\documentclass[12pt]{article}
\pdfoutput=1

\usepackage[a4paper, total={6.5in, 9.2in}]{geometry}
\usepackage{lipsum}
\usepackage{amsfonts}
\usepackage{graphicx}
\usepackage{epstopdf}
\usepackage{algorithmic}
\ifpdf
  \DeclareGraphicsExtensions{.eps,.pdf,.png,.jpg}
\else
  \DeclareGraphicsExtensions{.eps}
\fi

\title{Graphlets in multilayer networks}

\author{Sallamari Sallmen\thanks{Department of Computer Science, Aalto University School of Science, P.O.  Box 12200, FI-00076, Finland}
\and Tarmo Nurmi\footnotemark[1] \thanks{tarmo.nurmi@aalto.fi}
\and Mikko Kivel\"a\footnotemark[1]}

\usepackage{multirow}
\usepackage{colortbl}
\usepackage{float}
\usepackage{comment}
\usepackage{bbm}
\usepackage{physics}
\usepackage{diagbox}
\usepackage{tikz}
\usepackage{amsmath}
\usetikzlibrary{shapes.geometric, arrows, calc, positioning}

\usepackage{url}
\usepackage{hyperref}

\newcommand{\bs}{\boldsymbol}

\begin{document}

\maketitle

\begin{abstract}
Representing various networked data as multiplex networks, networks of networks and other multilayer networks can reveal completely new types of structures in these system. We introduce a general and principled graphlet framework for multilayer networks which allows one to break any multilayer network into small multilayered building blocks. 
These multilayer graphlets can be either analyzed themselves or used to do tasks such as comparing different systems. The method is flexible in terms of multilayer isomorphism, automorphism orbit definition, and the type of multilayer network. 
We illustrate our method for multiplex networks and show how it can be used to distinguish networks produced with multiple models from each other in an unsupervised way.
In addition, we include an automatic way of generating the hundreds of dependency equations between the orbit counts needed to remove redundant orbit counts. The framework introduced here allows one to analyze multilayer networks with versatile semantics, and these methods can thus be used to analyze the structural building blocks of myriad multilayer networks.
\end{abstract}

\section{Introduction}

Representing networked systems as graphs has been an extremely successful approach for analyzing the structure of various such systems, ranging from societies and transportation systems to brains and cellular regulation \cite{newman2010networks}.
One of the reasons for the rapid growth of structural network analysis is that,
building on the graph abstraction, it has been possible to analyze the structure of these otherwise disparate systems with the same methods and models.
Despite this success, there are several systems and research questions for which one needs to consider more general network structures, such as multilayer networks \cite{kivela2014multilayer, boccaletti2014structure}.
Several methods and concepts have been generalized in order to understand a wide class of multilayered networks, including community detection methods \cite{mucha2010community,magnani2019community,huang2021survey}, spreading processes \cite{salehi2015spreading}, centrality measures \cite{de2013centrality,de2013mathematical}, and local network features \cite{battiston2014structural,cozzo2015structure}. As multilayer networks are higher order structures than graphs, these generalizations typically have more degrees of freedom in their parametrization, and many concepts can be generalized in multiple different ways. In order to understand which of these generalizations to use, or how to navigate the new degrees of freedom, it is useful to take a principled approach where complicated concepts are built bottom-up from more fundamental ones.

Recently, the concept of graph isomorphism, a notion formalizing structural similarity, was extended to multilayer networks \cite{kivela2015isomorphisms}.
Graph isomorphism is a fundamental idea behind structural analytics such as graphlets, motifs, and network comparison. When generalized to multilayer networks, these concepts can then be used to analyze various types of multilayer networks coming from multiple application areas. Further, additional layers of methodology can be developed. 
Here we focus on one such methodology known as \emph{graphlets}, which have several applications in network data analysis. They reduce the topological information on single nodes or complete networks into vectorized format, which can be used, for example, in supervised and unsupervised learning tasks. A common use case is to compute topological similarity and dissimilarity of pairs of networks using graphlets.
This type of alignment-free methods \cite{Yaverouglu2015proper} are less computationally intensive than methods where the full network would need to be aligned \cite{wegner2017identifying,aliakbary2015distance}. Further, they can be used to group together networks that share common features, for example, because they are created via similar mechanisms \cite{prvzulj2007biological,yaverouglu2014revealing,ali2014alignment,wegner2017identifying}.
Graphlet-based methods are built on \emph{graphlet degree distributions}, which are generalizations of the degree distribution of a network \cite{prvzulj2007biological}. For example, they are used to assess the distance between two different networks \cite{yaverouglu2014revealing,wegner2017identifying,Yaverouglu2015proper}. Graphlet degrees for nodes can be defined based on the full network \cite{prvzulj2007biological,yaverouglu2014revealing,wegner2017identifying} or on the ego-networks of nodes \cite{ali2014alignment}. Mathematically, the graphlet degree of a node is defined by how many times it is found on a specific \emph{automorphism orbit}, a node equivalence class based on network automorphisms within the graphlets of the network (\emph{orbit} for short).

While graphlets and orbits have been defined and studied in graphs, analysis for multilayer networks is not fully developed. Graphlets in multiplex networks, which is an important special case of multilayer networks \cite{kivela2014multilayer}, have been studied recently \cite{dimitrova2020graphlets}. 
Orbits of nodes in multiplex graphlets were defined as "sub-orbits" of aggregated single-layer graphlet orbits, where the specific multiplex edge contents of the aggregated graphlets classify the nodes into the sub-orbits. Additionally, these multiplex graphlets can be grouped together using various reduction strategies in order to decrease the number of sub-orbits \cite{dimitrova2020graphlets}. However, the sub-orbit and reduction approach is not based on the explicit definition of orbits via multilayer network automorphism groups. This means that it will catch only a subset of all isomorphisms available for multiplex networks.
Multiplex graphlets have also been defined for three nodes in two-layer networks \cite{jiao2020sampling} and multiplex motifs (statistically enriched graphlets) have been defined for larger numbers of nodes and layers as manually constructed compositions of smaller subgraphs \cite{battiston2017multilayer}, using node isomorphism as the isomorphism type of choice explicitly or implicitly (node isomorphism is only one of the many types of isomorphisms possible in multiplex networks \cite{kivela2015isomorphisms}).

In this paper, we develop multilayer definition of graphlets and related concepts needed for comparing multilayer networks via graphlet degree distributions.
Our definition applies to general multilayer networks with any number of aspects \cite{kivela2014multilayer}. It is based on an explicit definition of multilayer network isomorphism \cite{kivela2015isomorphisms}, and we use a multilayer automorphism group as a basis of our definition of the multilayer orbits. We apply our framework to a special case of multiplex networks 
in order to illustrate the concept of multilayer graphlets and to compare it to the recently developed multiplex graphlets \cite{dimitrova2020graphlets}, which results in a different orbit definition (see Supplementary Materials).
In addition, we define a procedure to automatically produce multiplex dependency equations \cite{yaverouglu2014revealing} for the graphlet degrees for any choice of isomorphism, number of nodes, and number of layers.
In contrast to previous methods for analysing small substructures of multilayer networks \cite{boekhout2019efficiently,enright2017counting,takes2018multiplex,ren2019finding,paranjape2017motifs}, our definition of multilayer graphlets and orbits  captures graphlets with any number of nodes, layers and aspects in addition to all multilayer isomorphism types. That is, instead of suggesting a single-purpose method, our method includes  considerable and transparent freedom to choose how the graphlet analysis should be extended to various use cases.

\begin{figure}[!htb]
\centering
\includegraphics[width=\textwidth]{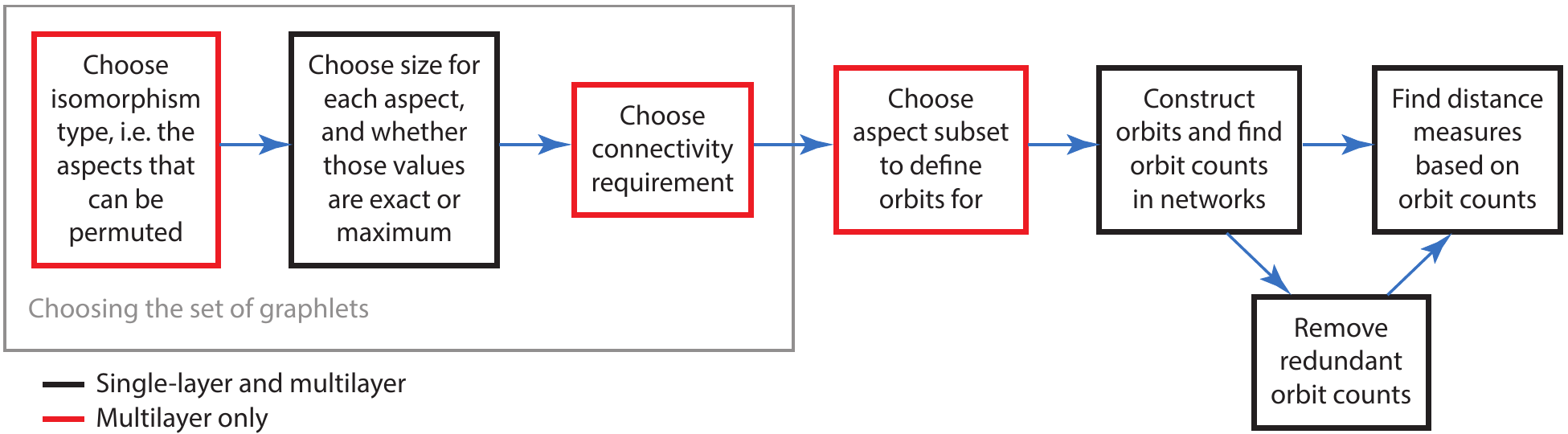}
\caption{The pipelines for single-layer and multilayer graphlet/orbit-based network distance measure calculation. The black boxes depict steps that have to be taken in both the single-layer and multilayer pipeline, while the red boxes depict steps unique to the multilayer pipeline. A typical pipeline for calculating graphlet/orbit-based distance measures in single-layer networks starts with deciding how many nodes the graphlets should have. %(here we consider node identities to be the sole "aspect" of single-layer networks), them having either \emph{exactly} that many nodes or \emph{at most} that many nodes. 
The graphlets are then constructed, orbits are found and enumerated from the networks in question, and the distance measures are calculated based on those orbit counts. The orbit counts contain redundancies because some graphlets contain other graphlets as sub-graphs; therefore, if desired, the level of redundancy can be reduced by removing some of the redundant orbit counts prior to distance measure calculation. In multilayer networks, the pipeline involves more choices. Now, we have to decide the type of multilayer isomorphism that is used to define the graphlets and orbits, the graphlet size for each aspect (and whether that size means exactly that many or at most that many), the definition of when a graphlet is "connected" in the multilayer setting, and the subset of aspects we want to find the orbits for.}
\label{fig:flowchart}
\end{figure}

The additional choices and multilayer extensions one needs to make when working with multilayer graphlets are illustrated clearly in the task of comparing multilayer networks (see Figure ~\ref{fig:flowchart}). This article is organised such that we build up this pipeline starting from the basics, and divide the text into the following aims: 1) Defining automorphism orbits in multilayer networks by extending the notion of conventional automorphism orbits; 2) Defining multilayer graphlets in order to apply the orbit definition to them; 3) Using the previous two to construct multilayer versions of graphlet/orbit-based alignment-free network distance measures; 4) Illustrating the multilayer pipeline on a simplified type of multiplex networks, and explicitly constructing dependency equations that quantify how orbit counts depend on other orbit counts in them (the removal of redundant orbit counts has been used before for single-layer networks, where it was found to improve the performance of a graphlet-based distance measure in some cases \cite{yaverouglu2014revealing}); Finally, 5) Evaluating different distance measures on test sets of multiplex network models.
The pipeline for creating multiplex orbit dependency equations, calculating multiplex graphlet degrees and creating Figures \ref{fig:pre-rec-combined}--\ref{AUPRs} is implemented by the authors and is publicly available \cite{pipeline-code}.

\section{Framework for multilayer network graphlet analysis}

We start by defining the multilayer graphlet framework. As the multilayer graphlets are an extension of graphlets in ordinary single-layer graphs, we first review how automorphism orbits are defined in graphs. Then, we extend this definition to multilayer networks by incorporating multilayer network isomorphisms in place of graph isomorphisms. Then, we define graphlets in multilayer networks and the automorphism orbits within them. Finally,  we generalize some existing graphlet-based distance measures for multilayer networks.

\subsection{Automorphism orbits} 

\subsubsection{Ordinary graphs}

Graph isomorphism is a concept defining when two graphs are structurally equivalent. Informally, one can think of two graphs  being isomorphic if they can be drawn in exactly the same way (while disregarding the node names). Formally, two graphs are isomorphic if one can transform one of the graphs into the other by relabeling the nodes.
A permutation of node labels that performs the transformation is called an isomorphism, and an isomorphism from a graph to itself is called an automorphism. In other words, if a relabeling $g$ assigns (maps) new identities to the nodes such that the set of edges in the network is the exact same before and after applying it, $g$ is an automorphism. The relabeling does not need to change the identity of each node: in fact, the relabeling that maps each node to itself is called the identity permutation.

Note that if you relabel nodes of a graph $G$ with a permutation $g$ that is an automorphism and then with another automorphism $f$, then this combination $fg$ is another permutation that is also an automorphism for $G$. That is, the automorphisms of a graph are related to each other. The structure of these relationships can be studied  algebraically by  forming a group of all automorphisms and the combination relation. This group is known as the \emph{automorphism group} of the graph denoted by $Aut(G)$ \cite{prvzulj2007biological}.
These automorphisms partition the nodes of the graph into equivalence classes called automorphism orbits. Formally, for a graph $G = (V,E)$, where $V$ and $E$ are the sets of nodes and edges, the automorphism orbit of node $u$ is defined as the set of all the nodes it can be mapped to with any automorphism \cite{prvzulj2007biological}
\begin{eqnarray}
Orb(u) & = & \{ v \ | \ v = g(u) \ \exists \ g \in Aut(G) \}\,.
\label{eq:singlelayernodeorbits}
\end{eqnarray}
One can interpret the automorphism orbits (or orbits for short) as sets of nodes that are structurally equivalent in the given graph.

\subsubsection{Multilayer networks}

A multilayer network is a quadruplet $M = (V_M, E_M, V, \mathbf{L})$ where $\mathbf{L} = \{L_a\}_{a=1}^d$ is a sequence of sets of elementary layers, $V_M$ is the set of node-layers, $E_M$ is the set of edges between them ($E_M \subseteq V_M \times V_M$), and $V$ is the set of nodes %(elementary layers of the zeroth aspect, node identity)
\cite{kivela2014multilayer}. The number of aspects in the network, $d$, corresponds to the "dimensionality" of the layers $\bs{\alpha} \in L_1 \times L_2 \times ... \times L_d$ in the network. Each node-layer is a combination of a node identity $v \in V$ and layer identity $\bs{\alpha}$, and $V_M \subseteq V \times L_1 \times L_2 \times ... \times L_d$. 
For notational convenience we will denote  $L_0 = V$, i.e. define the elementary layers of the 0th aspect as the node identities \cite{kivela2015isomorphisms}.
We will denote node-layers as $(v,\bs{\alpha}) \in V_M$ or $(\gamma_0, \dots, \gamma_d) \in V_M$.
The graph formed by the combination $(V_M,E_M)$ is called the underlying graph of the multilayer network.

In multilayer networks the isomorphisms and automorphisms depend on the aspects $p$ that are allowed to be permuted \cite{kivela2015isomorphisms}. For a network with $d$ aspects, $p \subseteq \{0,1,2,...,d\}$. The isomorphism of choice (the set of aspects to be permuted) affects the orbits in multilayer networks. The automorphisms of multilayer network $M$ with $p$ as the set of aspects that are allowed to be permuted form the $p$-automorphism group $Aut_p(M)$. Let $\bs{\zeta} \in Aut_p(M)$ be a $p$-automorphism (relabeling of nodes and elementary layers) of $M$. The total relabeling consists of a relabeling for each aspect: $\bs{\zeta} = (\zeta_0,\zeta_1,...,\zeta_d)$, where $\zeta_a = \mathbbm{1}_{L_a}$ if $a \notin p$ and $\mathbbm{1}_{L_a}$ is the identity permutation for the set of elementary layers $L_a$ of aspect $a$. For node $u$ in a multilayer network $M$ the $p$-automorphism orbit is defined analogous of Equation \ref{eq:singlelayernodeorbits} as
\begin{eqnarray}
Orb_p(u)  =  \{ v \ | \ v = \zeta_0(u) \ \exists \ \bs{\zeta} \in Aut_p(M) \} \text{, where } \bs{\zeta} = (\zeta_0,\zeta_1,...,\zeta_d) \,.
\label{eq:nodeorbits}
\end{eqnarray}
In addition to defining the orbits for nodes (as in ordinary graphs), in multilayer networks it is also possible to define the orbits for node-layers, layers, or any other subset of aspects. 
For node-layer $(u, \bs{\beta})$ the orbit is defined as
\begin{eqnarray}
Orb_p((u, \bs{\beta}))  =  \{ (v, \bs{\alpha}) \ | \ (v, \bs{\alpha}) = \bs{\zeta}(u, \bs{\beta}) \ \exists \ \bs{\zeta} \in Aut_p(M) \} \,.
\label{eq:nodelayerorbits}
\end{eqnarray}
We can define the orbits for any subset of aspects: let $\bs{\gamma} \in L_{a_1} \times L_{a_2} \times ... \times L_{a_k}$, where $a_1,...,a_k$ are the desired aspects. Then,
\begin{eqnarray}
Orb_p(\bs{\gamma}) & = & \{ \bs{\delta} \ | \ \delta_1 = \zeta_{a_1}(\gamma_1),  \delta_2 = \zeta_{a_2}(\gamma_2), ..., \delta_k = \zeta_{a_k}(\gamma_k) \ \exists \ \bs{\zeta} \in Aut_p(M) \} \ .
\label{eq:generalorbits}
\end{eqnarray}
Equation \ref{eq:generalorbits} becomes Equation \ref{eq:nodeorbits} when we choose $k = 1, a_1 = 0$ ($L_0 = V$), and Equation \ref{eq:nodelayerorbits} when we choose $k = d + 1, a_1 = 0, a_2 = 1, a_3 = 2, ..., a_k = d$.
As in ordinary graphs, the orbits in multilayer networks partition the nodes, node-layers, or the entities $\bs{\gamma}$ of any subset of aspects into disjoint equivalence classes. 
This follows from the fact that the application of an automorphism is a group action and we can apply the known result that group action induces equivalence relation \cite{cohn1982algebra} with the equivalence class of $\bs{\gamma}$ that is $Orb_p(\bs{\gamma})$.
For an explicit proof involving the properties of multilayer automorphisms, see Supplementary Materials.

Figure~\ref{fig:all_orbits_and_isoms_example} illustrates all the automorphisms, orbits and orbit equivalence classes in a small two-aspect multilayer network.

\begin{figure}[]
\begin{center}
\begin{minipage}{0.5\textwidth}
\begin{center}
\includegraphics[width=\textwidth]{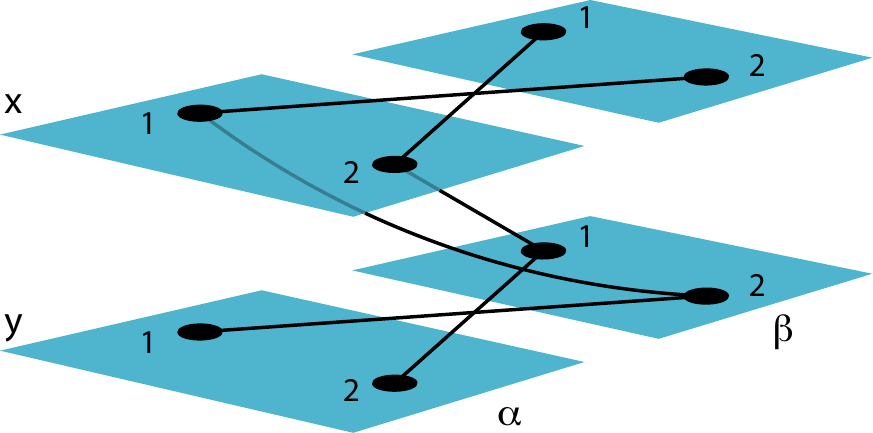}
\end{center}
\end{minipage}
%\hspace{0.1cm}
\begin{minipage}{0.45\textwidth}
\small
\begin{alignat*}{2}
    &Aut_{\{0\}} = &&\{\mathbbm{1},(1 \leftrightarrow 2,\mathbbm{1}_{L_1},\mathbbm{1}_{L_2})\} \\
    &Aut_{\{1\}} = &&\{\mathbbm{1}\} \\
    &Aut_{\{2\}} = &&\{\mathbbm{1}\} \\
    &Aut_{\{0,1\}} = &&\{\mathbbm{1},(1 \leftrightarrow 2,\mathbbm{1}_{L_1},\mathbbm{1}_{L_2})\} \\
    &Aut_{\{0,2\}} = &&\{\mathbbm{1},(1 \leftrightarrow 2,\mathbbm{1}_{L_1},\mathbbm{1}_{L_2})\} \\
    &Aut_{\{1,2\}} = &&\{\mathbbm{1},(\mathbbm{1}_{L_0},x \leftrightarrow y,\alpha \leftrightarrow \beta)\} \\
    &Aut_{\{0,1,2\}} = &&\{\mathbbm{1},(1 \leftrightarrow 2,\mathbbm{1}_{L_1},\mathbbm{1}_{L_2}), \\
    & && (\mathbbm{1}_{L_0},x \leftrightarrow y,\alpha \leftrightarrow \beta), \\
    & && (1 \leftrightarrow 2,x \leftrightarrow y,\alpha \leftrightarrow \beta)\}
\end{alignat*}
\end{minipage}
\\
\bigskip
\begin{minipage}{\textwidth}
\begin{center}
\tiny
\newcommand{\centered}[1]{\begin{tabular}{c} #1 \end{tabular}}
\begin{tabular}{c|@{}c@{}@{}c@{}@{}c@{}@{}c@{}@{}c@{}@{}c@{}@{}c@{}}
 \multicolumn{1}{c}{} & \multicolumn{7}{c}{$Orb_p(\bs{\gamma})$ equivalence classes ($\bs{\gamma} \in L_{a_1} \times ... \times L_{a_k}$)} \\
 \diagbox{$p$}{$a_1,...,a_k$} & 0 & 1 & 2 & 0, 1 & 0, 2 & 1, 2 & 0, 1, 2 \\
 \cline{1-8}
 \centered{$\{0\}$,\\ $\{0,1\}$,\\and $\{0,2\}$} & $\{1,2\}$ & \centered{$\color{red}\{x\}$,\\$\color{blue}\{y\}$} & \centered{$\color{red}\{\alpha\}$,\\$\color{blue}\{\beta\}$} & \centered{$\color{red}\{(1,x),(2,x)\}$,\\$\color{blue}\{(1,y),(2,y)\}$} & \centered{$\color{red}\{(1,\alpha),(2,\alpha)\}$,\\$\color{blue}\{(1,\beta),(2,\beta)\}$} & \centered{$\color{red}\{(x,\alpha)\}$,\\$\color{blue}\{(y,\alpha)\}$,\\$\color{olive}\{(x,\beta)\}$,\\$\color{purple}\{(y,\beta)\}$} & \centered{$\color{red}\{(1,x,\alpha),(2,x,\alpha)\}$,\\$\color{blue}\{(1,y,\alpha),(2,y,\alpha)\}$,\\$\color{olive}\{(1,x,\beta),(2,x,\beta)\}$,\\$\color{purple}\{(1,y,\beta),(2,y,\beta)\}$} \\[8mm]
 $\{1,2\}$ & \centered{$\color{red}\{1\}$,\\$\color{blue}\{2\}$} & $\{x,y\}$ & $\{\alpha,\beta\}$ & \centered{$\color{red}\{(1,x),(1,y)\}$,\\$\color{blue}\{(2,x),(2,y)\}$} & \centered{$\color{red}\{(1,\alpha),(1,\beta)\}$,\\$\color{blue}\{(2,\alpha),(2,\beta)\}$} & \centered{$\color{red}\{(x,\alpha),(y,\beta)\}$,\\$\color{blue}\{(x,\beta),(y,\alpha)\}$} & \centered{$\color{red}\{(1,x,\alpha),(1,y,\beta)\}$,\\$\color{blue}\{(1,y,\alpha),(1,x,\beta)\}$,\\$\color{olive}\{(2,x,\alpha),(2,y,\beta)\}$,\\$\color{purple}\{(2,y,\alpha),(2,x,\beta)\}$} \\[8mm]
 $\{0,1,2\}$ & $\{1,2\}$ & $\{x,y\}$ & $\{\alpha,\beta\}$ & \centered{$\{(1,x),(2,x),$\\$(1,y),(2,y)\}$} & \centered{$\{(1,\alpha),(2,\alpha),$\\$(1,\beta),(2,\beta)\}$} & \centered{$\color{red}\{(x,\alpha),(y,\beta)\}$,\\$\color{blue}\{(x,\beta),(y,\alpha)\}$} & \centered{$\color{red}\{(1,x,\alpha),(2,x,\alpha),$\\$\color{red}(1,y,\beta),(2,y,\beta)\}$,\\$\color{blue}\{(1,y,\alpha),(2,y,\alpha),$\\$\color{blue}(1,x,\beta),(2,x,\beta)\}$}
\end{tabular}
\end{center}
\end{minipage}
\caption{\textbf{Top left:} A two-aspect multilayer network with $L_0 = V = \{1,2\}, \; L_1 = \{x,y\}, \; L_2 = \{\alpha,\beta\}$. \textbf{Top right:} The automorphism groups of the network on the left, where $\mathbbm{1} = (\mathbbm{1}_{L_0},\mathbbm{1}_{L_1},\mathbbm{1}_{L_2})$ is the identity permutation in every aspect. In the non-identity permutations, $\square \leftrightarrow \Diamond$ denotes that $\square$ is relabeled to $\Diamond$ and $\Diamond$ is relabeled to $\square$. Note that in general the $p$-automorphism group cannot be inferred from the automorphism groups of subsets of $p$ \cite{kivela2015isomorphisms}: $Aut_{\{1\}}$ and $Aut_{\{2\}}$ contain only the identity permutation, but $Aut_{\{1,2\}}$ contains also another permutation. \textbf{Bottom:} The orbit equivalence classes for the network on the top left. When $p = \{1\}$ or $p = \{2\}$, each entity $\bs{\gamma}$ is alone in its own equivalence class regardless of $a_1,...,a_k$; therefore, these have been omitted from the table. If there is more than one equivalence class, different colors have been used to visually separate them. An entity is always in its own equivalence class, so the table can be used to find the orbit of each entity: for example, $Orb_{\{0\}}(1) = \{1,2\} = Orb_{\{0\}}(2)$ and $Orb_{\{0,1,2\}}((x,\alpha)) = \{(x,\alpha),(y,\beta)\} = Orb_{\{0,1,2\}}((y,\beta))$.}
\label{fig:all_orbits_and_isoms_example}
\end{center}
\end{figure}

\subsection{Graphlets and graphlet degrees}

Graphlets in ordinary single-layer networks are defined as small, connected, non-isomorphic induced subgraphs of a larger network \cite{prvzulj2004modeling, prvzulj2007biological}. A graphlet is thus an isomorphism class of connected induced subgraphs. Graphlet analysis is usually restricted to a subset of all possible graphlets, for example by looking only at graphlets with at most some number of nodes. While the graphlet definition is quite straight-forward in single-layer networks, it leads to different definitions of graphlets in multilayer networks based on how one defines isomorphism, connectivity, and size. Figure~\ref{fig:flowchart} illustrates the additional choices to be made in graphlet definition in the multilayer setting compared to the single-layer case. We now present a formal definition of multilayer graphlets, elaborate on the concepts of isomorphism, connectivity, and size, and define graphlet degrees in multilayer networks.

Formally, an induced subnetwork $M' = (V'_M, E'_M, V', \mathbf{L}')$ within a larger network $M = (V_M, E_M, V, \mathbf{L})$ is defined by $V' \subseteq V, \ L'_a \subseteq L_a \ \forall \ a \in \{1,2,...,d\}, \ V'_M = \{(v,\bs{\alpha}) \in V_M \ | \ (v,\bs{\alpha}) \in V' \times L'_1 \times L'_2 \times ... \times L'_d  \}, \ E'_M = \{((v,\bs{\alpha}),(u,\bs{\beta})) \in E_M \ | \ ((v,\bs{\alpha}),(u,\bs{\beta})) \in V'_M \times V'_M\}$. This definition fixes the elementary layer sets for each aspect, and then all the node-layers and edges that exist within the span of those elementary layer sets are included. $M'$ belongs in an isomorphism class with every other subnetwork it is isomorphic to. If the subnetworks in that isomorphism class are connected, we call that isomorphism class the graphlet that $M'$ corresponds to.

Unlike single-layer networks, multilayer networks have multiple possible types of isomorphism, one for each set $p \subseteq \{0,1,2,...,d\}$ of aspects that can be permuted. The choice of isomorphism affects the set of graphlets in a given network and influences which subnetworks correspond to the same graphlet. Which isomorphism is appropriate depends on the application \cite{kivela2015isomorphisms}.

When it comes to connectivity, one option is to require the underlying graph of the graphlet to be connected. However, this could lead to some entities in the network never participating in any graphlets, and one may wish to loosen the restriction to requiring only the layer-aggregated network to be connected. For example, if node-layer $(v,\bs{\alpha})$ is not connected to any other node-layer, then node $v$ does not participate in any graphlets in subnetworks that include layer $\bs{\alpha}$ if the underlying graph is required to be connected. However, if only the aggregated graph is required to be connected, $v$ can still participate in graphlets if there are sufficient connections on other layers.

The size of a multilayer graphlet can be defined based on, for example, the number of nodes, layers, or node-layers participating in the graphlet. A reasonable extension of the notion of graphlet size is that we give the size of the graphlet in every aspect (including the zeroth aspect of nodes), which means that in total we need $d+1$ numbers to define the graphlet size. In single-layer networks this would be just one number, the number of nodes, in one-aspect multilayer networks this would be the number of nodes and the number of layers, in two-aspect networks this would be the number of nodes, the number of elementary layers in the first aspect, and the number of elementary layers in the second aspect, and so on. According to this definition, the size of the graphlet that an induced subnetwork $M' = (V'_M, E'_M, V', \mathbf{L}')$ corresponds to is then $\abs{V'_M}$, $\abs{L'_1}$, $\abs{L'_2}$, ..., $\abs{L'_d}$. Notably, this definition of size does not fix the size of $V'_M$: two graphlets with the same size can contain a different number of node-layers.

\subsubsection*{Graphlet degrees}

\begin{figure}[!htb]
    \centering
    \includegraphics[width=0.4\textwidth]{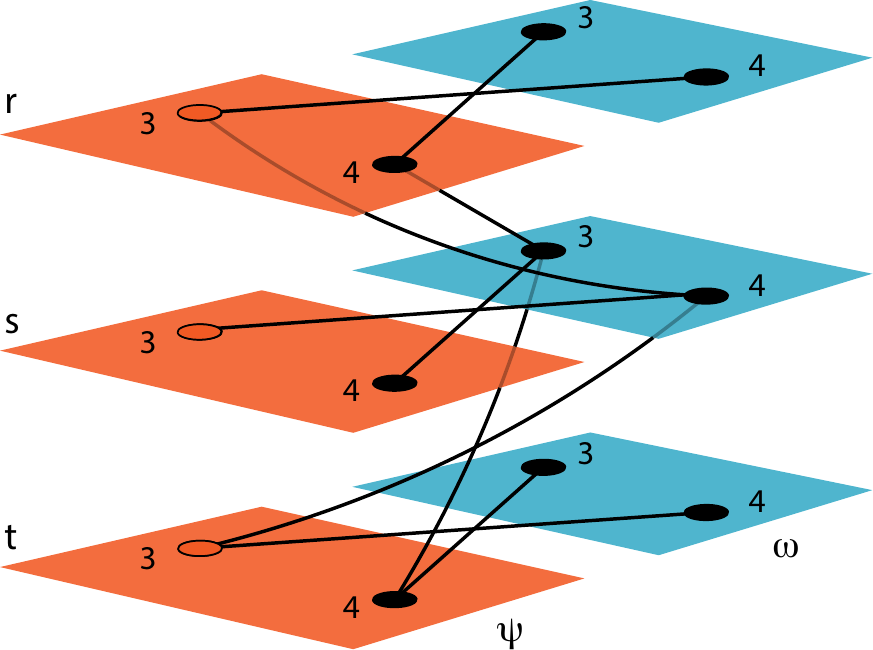}
    \caption{A two-aspect multilayer network with $L_0 = V = \{3,4\}, \; L_1 = \{r,s,t\}, \; L_2 = \{\psi,\omega\}$. We are interested in the graphlet degrees of $(3,\psi)$ (the corresponding node and the layers are shown in orange) with respect to graphlets of size (2, 2, 2). There are three induced subnetworks of that size, $(\{3,4\},\{r,s\},\{\psi,\omega\})$, $(\{3,4\},\{s,t\},\{\psi,\omega\})$, and $(\{3,4\},\{r,t\},\{\psi,\omega\})$. Depending on the connectivity requirements, $(\{3,4\},\{r,t\},\{\psi,\omega\})$ may or may not be connected and thus it may or may not be a graphlet according to the definition. If the set of aspects that can be permuted, $p$, is $\{1\}$, $\{0,1\}$, $\{1,2\}$ or $\{0,1,2\}$, then $(\{3,4\},\{r,s\},\{\psi,\omega\})$ and $(\{3,4\},\{s,t\},\{\psi,\omega\})$ are isomorphic and thus correspond to the same graphlet, and $(3,\psi)$ is in the same orbit in both of them. The graphlet degree of $(3,\psi)$ with respect to that orbit of that graphlet is then 2. If $(\{3,4\},\{r,t\},\{\psi,\omega\})$ is considered connected, it corresponds to a different graphlet (it is not isomorphic to either of the two other subnetworks) and thus the graphlet degree of $(3,\psi)$ with respect to the orbit it is on in that graphlet is 1. If $p \in \{\{0\},\{2\},\{0,2\}\}$, then none of the induced subnetworks are isomorphic to the other(s) and every one of them corresponds to a different graphlet. For each of these graphlets, $(3,\psi)$ is the only element in its orbit in that graphlet, and thus the graphlet degree of $(3,\psi)$ with respect to each of those orbits is 1. If $(\{3,4\},\{r,t\},\{\psi,\omega\})$ is considered connected, there are three different graphlet degrees equal to 1, and if it is not, there are two.}
    \label{fig:graphlet_degree_strange_orbit_example}
\end{figure}

Graphlets are intertwined with the concept of node roles and automorphism orbits in single-layer networks, such that the automorphism orbit of a node within a graphlet can be used to define the node's role in that graphlet \cite{yaverouglu2014revealing}. This concept can be immediately generalized to multilayer networks by applying the multilayer automorphism node orbit definition (Equation \ref{eq:nodeorbits}) in place of the single-layer automorphism orbit. The number of times a node is found on a specific orbit of a specific graphlet in a network is called the \emph{graphlet degree} %(not orbit degree) 
of that node with respect to that orbit \cite{prvzulj2007biological}. The distribution of graphlet degrees of a specific orbit over all the nodes in a network is called the \emph{graphlet degree distribution} of that orbit \cite{prvzulj2007biological} --- there is one such distribution for each orbit, both in an ordinary graph and in a multilayer network (naturally, the orbits themselves will be different in the two cases). In addition to defining graphlet degrees of nodes, in multilayer networks we can define a graphlet degree with respect to layers, node-layers, or any other subset of aspects, since orbits are defined for all of them (see the previous section). The graphlet degree of $\bs{\gamma} \in L_{a_1} \times L_{a_2} \times ... \times L_{a_k}$ w.r.t. an orbit of a graphlet in a multilayer network $M$, where $a_1,...,a_k$ are aspects of $M$, is then simply the number of times $\bs{\gamma}$ is found on that specific orbit in $M$. Figure \ref{fig:graphlet_degree_strange_orbit_example} illustrates graphlet degrees for a combination of a node and an elementary layer in a two-aspect multilayer network with respect to graphlets of certain size.
Because of the added degrees of freedom in multilayer networks compared to ordinary graphs, when talking about graphlet degrees one needs to specify which combination of aspects and which isomorphism is considered.

When finding the graphlets contained in a network, we need to choose the type of isomorphism, and when determining the orbits inside those graphlets, we need to choose the type of automorphism. Both of these require the choice of which aspects are allowed to be permuted, and it is reasonable to use the same set of aspects for both. However, this is not required and they can be different, in which case one needs to be careful of the interpretations of the real-world meaning of the graphlet degree distributions.

As is the case with single-layer graphlets, multilayer graphlets contain smaller graphlets as subnetworks, and therefore there are dependencies between the graphlet degrees of different orbits. Similar to single-layer networks \cite{yaverouglu2014revealing,sarajlic2016graphlet}, it is possible to construct orbit count equations that exactly determine these dependencies in multilayer networks. In the multiplex network case study in this paper, we describe a process of automatically generating dependency equations for single-aspect multiplex networks in detail.

\subsection{Graphlet-based methods and measures}

Now that the definitions for graphlets and their orbits are established one can compute the graphlet degrees of nodes, layers or any other entities $\bs{\gamma}$ in multilayer networks. After this step, one can in general use multilayer graphlets in a very similar way as ordinary graphlets, by simply substituting multilayer graphlet degrees in place of single-layer graphlet degrees in any graphlet-based methods. For example, the graphlet degree vectors \cite{prvzulj2007biological} and/or graphlet correlation matrices \cite{yaverouglu2014revealing} can be used as an input for supervised or unsupervised learning tasks or they can be visualised and interpreted in the context of the input multilayer network data.

Consider for example a task of computing distances between networks. An existing statistic called the graphlet correlation distance (GCD) \cite{yaverouglu2014revealing} is calculated based on graphlet correlation matrices, which are simply matrices regardless of if the orbits in question are single-layer or multilayer ones. The calculation of GCD is then exactly the same once graphlet degree vectors have been found in multilayer networks (the procedure for calculating GCD shown in Supplementary Materials). Besides GCD, there are also other graphlet/orbit-based measures that can similarly be generalized to multilayer networks. These include the graphlet degree distribution agreement (GDDA) \cite{prvzulj2007biological}, the relative graphlet frequency distance (RGFD) \cite{prvzulj2004modeling}, Netdis \cite{ali2014alignment} (requires the definition of multilayer ego-networks), and NetEmd \cite{wegner2017identifying}. In the next section, GCD is taken as the multilayer distance measure of choice, and in Supplementary Materials, multilayer NetEmd and multilayer GDDA are investigated in more detail.

\section{Comparing multiplex networks using graphlets}

The framework for multilayer network graphlets encompasses a large number of use cases and types of networks. This comes at a cost of the framework being relatively abstract. We will next focus on the special case of graphlets in node-aligned single-aspect multiplex networks \cite{kivela2014multilayer} in order to illustrate how the general multilayer network graphlet framework can be applied. This significantly simplifies many of the notions and serves as a concrete example for the various concepts we have introduced.
To further illustrate the applicability of the graphlet framework, we use it to  conduct a case study classifying multiplex network models in an unsupervised way using graphlet distance measures.

Multiplex networks are a type of multilayer networks that are diagonally coupled and where each layer shares at least one node with another layer \cite{kivela2014multilayer}. In other words, all interlayer edges are between a node on some layer and the same node on another layer (i.e. have the form $((v,\bs{\alpha}),(v,\bs{\beta}))$) and there are no layers where the set of nodes is completely disjoint from every other layer. In this section, we focus only on single-aspect multiplex networks where each layer contains the same set of nodes and every node is connected to all of its counterparts in other layers ($(v,\bs{\alpha}),(v,\bs{\beta}) \in V_M \implies ((v,\bs{\alpha}),(v,\bs{\beta})) \in E_M$). Such networks are called node-aligned with categorical couplings \cite{kivela2014multilayer}. For simplicity we refer to them as multiplex networks in this section.

\subsection{Graphlets in multiplex networks}

\begin{figure}[!htb]
\begin{center}
\includegraphics[width=\linewidth]{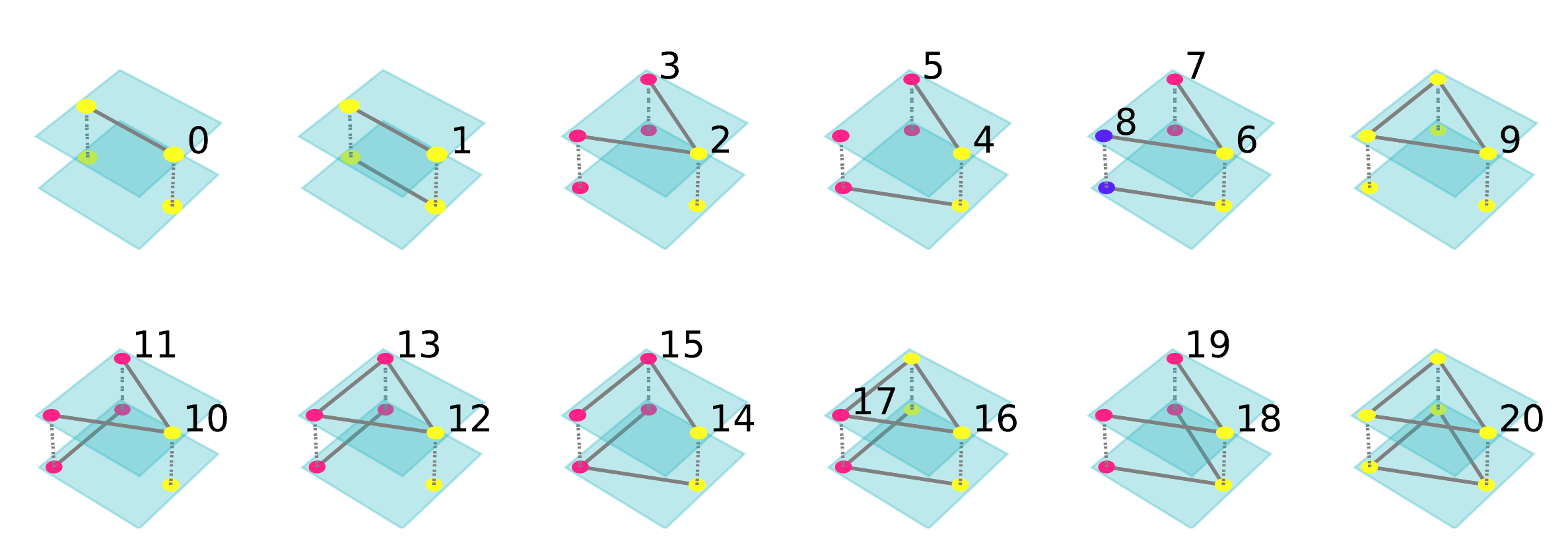}
\end{center}
\caption{Graphlets and their automorphism orbits computed for nodes when nodes and layers are allowed to be permuted $Orb_{\{0,1\}}(u)$ for single-aspect multiplex networks with two layers and up to three nodes (single-node graphlet omitted). The orbits are numbered from 0 to 20. Within each graphlet, nodes colored with the same color belong to the same orbit.}
\label{l2_n3}
\end{figure}

We can enumerate multiplex graphlets by finding the isomorphism classes of all possible connected multiplex networks of desired size. 
Unlike in general multilayer networks, in multiplex networks we only need to consider different variations of intralayer edge configurations and the combinations of these intralayer networks, because by the definition of multiplex networks
the same nodes are present in all of the layers and the interlayer edges are connecting the layers fully symmetrically.
Figure \ref{l2_n3} presents all multiplex graphlets with two nodes and two layers or three nodes and two layers when the isomorphism allows the permutation of both node and layer labels ($p = \{0,1\}$). Graphlets with four nodes and two layers are illustrated in Supplementary Materials. The two notions of connectivity discussed before (underlying network is connected, aggregated network is connected) are equivalent in our multiplex networks, since every node is connected to its counterparts in the other layers \cite{mucha2010community,kivela2014multilayer}.
Note that if we didn't allow the permutation of both node and layer labels in the isomorphism, the number of graphlets of given size would depend on the number of distinct nodes (if $p \in \{\emptyset,\{1\}\}$) and layers (if $p \in \{\emptyset,\{0\}\}$) in our network. In real-world applications this could correspond to a case where the nodes/layers contain some semantics that we want to preserve in our analysis \cite{kivela2015isomorphisms}.

Figure \ref{l2_n3} also illustrates all the node orbits within the graphlets. The automorphism orbits are numbered starting from zero (we skip the arbitrary single-node graphlet and orbit), so in total there are 21 different automorphism orbits in the graphlets in Figure \ref{l2_n3}. When finding the graphlet degree distributions within a larger network, each orbit corresponds to one distribution. The node orbits of graphlets with four nodes and two layers are shown in Supplementary Materials. The number of node orbits grows quickly as the number of nodes and layers grows; Table \ref{table:orbits_and_independent_equations} lists the number of orbits in graphlets with up to four nodes and three layers when node and layer labels are allowed to be permuted and when only node labels are allowed to be permuted (in the latter case, every graphlet is assumed to contain the same set of layer labels as the number of graphlets would grow with the set of possible layer labels).

As previously explained, we can consider layer and node-layer orbits in addition to node orbits. However, we limit our attention to node orbits and their graphlet degree distributions, and in the following text \emph{orbit} refers to node orbit and \emph{graphlet degree} refers to node graphlet degree, unless specified otherwise.

\subsection{Dependency equations for multiplex networks}

We define an automatic process to construct equations that encode dependencies between orbit counts of different node orbits in multiplex networks. Each equation (for a pair of distinct orbits $x_1 \neq x_2$) has the form 
\begin{equation}
\binom{C_{x_1}}{c_1} \binom{C_{x_2} - b}{c_2} = a_1 C_{y_1} + \ldots + a_k C_{y_k}\,,
\label{eq:two-orbit-dependency}
\end{equation}
where the left side of the equation represents all possible combinations of two orbits $x_1$ and $x_2$ with respect to a node, and the right side represents all the possible orbits $y_j$ these combinations could generate. In other words, for each way two orbits can be combined, there must be some larger orbit that matches this combination (and so this larger orbit is one of $y_1,y_2,...,y_k$). $C_{x_i}$ and $C_{y_j}$ are the counts of orbits $x_i$ and $y_j$, respectively, $c_1$ and $c_2$ are the numbers of times orbits $x_1$ and $x_2$ are included in the combination, respectively, $b$ is the number of times orbit $x_2$ is included in orbit $x_1$, and $a_j$ is the number of ways a node can touch orbit $y_j$ when it touches the combined orbits $x_1$ and $x_2$. We can also combine multiple instances of a single orbit, which results in an equation of slightly different form:
\begin{equation}
\binom{C_{x_1}}{c_1} = a_1 C_{y_1} + \ldots + a_k C_{y_k}\,,
\label{eq:single-orbit-dependency}
\end{equation}
where the elements are the same as before, except now only orbit $x_1$ is included $c_1$ times in the combination.

The process of constructing the equations is essentially the process of determining the orbits $y_j$ and the coefficients $a_j$. The discovery of the orbits $y_j$ that contribute to the counts of orbits $x_i$ can be divided into three phases: combining the orbits, merging nodes and adding links.
When combining orbits $o_1$ and $o_2$ (which can correspond to different orbits $x_1$ and $x_2$, as in Equation \ref{eq:two-orbit-dependency}, or to the same orbit $x_1$, as in Equation \ref{eq:single-orbit-dependency}) residing in graphlets $g_1$ and $g_2$ respectively, one of the nodes representative of orbit $o_1$ and one of the nodes representative of orbit $o_2$ are merged into a single node $v_0$ which will be connected to all the node-layers the two merged nodes were connected to.
Otherwise the connections between nodes will remain unchanged.
If layer labels are allowed to be permuted in the isomorphism, all possible mergings with respect to layer combinations are done (for example, if in $g_1$ there are layers 1 and 2 and in $g_2$ there are also layers 1 and 2, there's two possible mergings: matching layers 1 to 1 and 2 to 2, and matching layer 1 of one graphlet to layer 2 of the other graphlet and vice versa.) If layer labels are not allowed to be permuted, then the layers where the two nodes are found should match.

In the second phase, nodes originating from different graphlets in the resulting networks of the first phase can be merged while the network has at least $max(n_1, n_2) + 1$ nodes, where $n_1$ and $n_2$ are the numbers of nodes in graphlets $g_1$ and $g_2$, respectively.
For the merge to be allowed, the two nodes to be merged need to have (possible) edges connecting them to $v_0$ on exactly the same layers.

In the final phase, links can be added between node-layers that belonged to different graphlets (i.e. one belonged to $g_1$ and the other to $g_2$), since $v_0$ will still touch both orbits $o_1$ and $o_2$ in the resulting network.
All the possible edge combinations should be added to all the obtained networks from the previous two phases. Since in our multiplex networks the interlayer edges are already specified, we can only add intralayer edges in this phase.

The orbits $y_j$ can then be obtained by checking in which orbit node $v_0$ resides in each of the resulting networks.
The coefficients $a_j$ for the orbit counts $C_{y_j}$ are obtained by computing the number of ways in which the graphlets $g_1$ and $g_2$ can be embedded into the resulting network such that $v_0$ touches orbit $y_j$.
The subtrahend $b$ in the left side of the equation is determined by calculating how many times a node in orbit $o_1$ touches orbit $o_2$ assuming graphlet $g_1$ has more nodes than $g_2$.

For example, if we combine two orbit 0s in multiplex networks with two layers (see Figure~\ref{l2_n3}), the edges can be either in the same layer, resulting in orbit 2, or in different layers, resulting in orbit 4, when both node and layer labels are allowed to be permuted.
The wedge-end nodes in these graphlets can be connected either in neither of the layers, only one of the layers, or both layers.
Therefore, combining two orbit 0s can also result in orbits 9, 10, 11, 12 and 14, and the dependencies can be expressed with the equation $\binom{C_0}{2} = C_2 + C_4 + C_9 + C_{10} + C_{11} + C_{12} + C_{14}$, where $C_i$ denotes the graphlet degree of a node with respect to orbit $i$. The dependency equations for up to 4-node graphlets in multiplex networks with 2 layers are listed in Supplementary Materials, and a computer program that produces these equations for any choice of parameters is provided as part of our multiplex graphlet analysis pipeline \cite{pipeline-code}.

We could combine more than two orbits to create more complex dependency equations with the same process. However, these equations can be derived from the equations where only two orbits are combined (considering also that the orbit can be combined with itself), as shown in Supplementary Materials.
Therefore, for the goal of discovering which equations are independent it is enough to consider equations where two orbits are combined. 

A similar process applied here for multiplex networks could be constructed for general multilayer networks as well. However, there are further considerations that need to be taken into account in the general case. When merging two nodes in the first phase, we need to make sure that they appear on the same layers and have the same connectivity to their other instances on other layers, considering also all layer permutations where the aforementioned is true if layer labels are allowed to be permuted.  When merging nodes in the second phase, we need to make sure they also appear on the same layers and have the same interlayer connectivity to their other instances, and also have the same connectivity to $v_0$. In the third phase, we must add all possible interlayer edges in addition to intralayer edges. In the general case, we can also consider any aspect or combination of aspects in addition to nodes for which to construct the equations.

\subsection{Reducing multiplex orbit counts based on dependencies between them}

Once we have generated a set of orbit dependency equations, we can discover which equations in the set are independent of the others. For each independent equation, we can calculate the value of one of the orbit count variables appearing in that equation based on the others. Therefore, for each independent equation, one orbit count is "redundant" and can be removed from the graphlet degree vectors of nodes in a network, to obtain reduced graphlet degree vectors. The process of finding these independent equations is described in detail in Supplementary Materials.

Table \ref{table:orbits_and_independent_equations} describes the number of orbits (and therefore the lengths of unreduced graphlet degree vectors) and the number of independent equations for graphlets up to four nodes and three layers, when both node and layer labels and only node labels are allowed to be permuted. For example, if we construct graphlet degree vectors of nodes in a network for orbits of graphlets with two layers and up to four nodes and $p = \{0,1\}$, each vector will contain $2+19+391=412$ orbit counts. We can then reduce this vector to $412-(0+3+36)=373$ orbit counts by removing one orbit count for each independent equation. As the number of nodes and layers grows, the number of orbits rapidly increases and the relative number of independent equations diminishes.

Finding a set of independent equations among a set of equations is not a process dependent on the multiplexity of the graphlets. Thus, the orbit count reduction method can also be applied to more general multilayer dependency equations, as long as such equations are first constructed.

\begin{table}[ht]
\centering
\caption{The number of all possible node orbits and independent orbit count equations for multiplex graphlets up to four nodes and three layers, when orbits are defined using either node-layer isomorphism or node isomorphism (in the latter case, each graphlet is assumed to contain the same set of layer labels, otherwise the number of graphlets and orbits would be infinite).}
\begin{tabular}{c||ccc|ccc||ccc|ccc} 
 \multicolumn{1}{c}{} & \multicolumn{6}{c}{{\bf Orbits}} & \multicolumn{6}{c}{{\bf Independent equations}} \\
 %\cline{2-14}
 Isomorphism type &
\multicolumn{3}{c|}{Node-layer} & \multicolumn{3}{c||}{Node} & \multicolumn{3}{c|}{Node-layer} & \multicolumn{3}{c}{Node} \\
 \cline{1-13}
 \diagbox{Layers}{\emph{Nodes}} & \emph{2} & \emph{3} & \emph{4} & \emph{2} & \emph{3} & \emph{4} & \emph{2} & \emph{3} & \emph{4} & \emph{2} & \emph{3} & \emph{4} \\
 \cline{1-13}
 1 & 1 & 3 & 11 & 1 & 3 & 11 & 0 & 1 & 3 & 0 & 1 & 3 \\
 2 & 2 & 19 & 391 & 3 & 33 & 751 & 0 & 3 & 36 & 0 & 6 & 91 \\
 3 & 3 & 67 & 8121 & 7 & 273 & 45311 & 0 & 6 & 193 & 0 & 28 & 1827 \\
\end{tabular}
\label{table:orbits_and_independent_equations}
\end{table}

\subsection{Multiplex network models}

To assess the performance of the different distance measures and multilayer orbit definition approaches, we apply them to three single-aspect test sets of multiplex networks. In the first test set, we compare different random multiplex network models, some of which contain relevant edge correlations and overlaps between layers. This set consists of the eight models "BA-ind" -- "WS", presented below. Between all networks, the average intralayer degrees of nodes are kept approximately constant. In the second set, we have the same models, but now we increase the average intralayer degree of networks for each model in a steadily progressing fashion, such that for every model there is the same mixture of different average intralayer degrees. In the third test set, we compare network groups which have different graphlets \emph{purposefully inserted} into them, to mimic a real-world situation where 
some graphlets are enriched in the networks, for example because they are important for the function of the network or there is a particular mechanism creating such structures. This insertion method is described in "Graphlet insertion" subsection.

In the first and second set, we have eight different models, translating into eight different classes of networks. In the third set, we fix the number of classes to be five (different class means different graphlets inserted into the networks). In each set, we have 30 networks per class (the first and second set thus have $30 \times 8 = 240$ and the third has $30 \times 5 = 150$ networks in total). Each network has 1000 nodes and 3 layers. In the first set, for most of the networks, the average intralayer degree of nodes is approximately equal to 4 (the average intralayer degree is approximately equal to $2m$, where $m$ is the parameter of the Barab\'asi-Albert (BA) networks. We set $m = 2$ in the first set.). In the second set, the average intralayer degrees progress as 2, 4, 6, 8, 10, and 12 ($m=1,2,3,4,5,6$, respectively), such that for each model there are five networks for each average intralayer degree. In the third set, the average intralayer degree of nodes is exactly 4 in every network.

\subsubsection*{Independent Barab\'asi-Albert models (BA-ind)}

A multiplex network is constructed by generating a Barab\'asi-Albert random network \cite{barabasi1999emergence} for each layer independently of each other.
First, a (complete) seed network with $m$ nodes is created, and after that each new node is attached to $m$ existing nodes with probabilities proportional to the degrees of the nodes.
Each layer will have a power law degree distribution, but the degrees of a node between layers do not correlate and there is little overlap in the edges between layers. The average intralayer degree of a node is approximately $2m$.
The node names are randomized for each layer separately and they do not follow the time when the nodes are introduced to the network.

\subsubsection*{Interdependent Barab\'asi-Albert models (BA-dep)}

Otherwise the same model as above, but new nodes are attached to the existing nodes with probabilities proportional to the sum of the degrees of a node over all the layers \cite{kim2013coevolution}.
The model produces high interlayer degree correlations, but the overlap of edges between layers remains quite low. 

\subsubsection*{Independent configuration models (Conf-ind)}

Each layer of the multiplex network is constructed as a configuration model random network \cite{newman2010networks} independently of each other.
The degree distributions of BA-ind networks are used to generate the Conf-ind networks.
The model introduces low interlayer degree correlations and edge overlaps.

\subsubsection*{Interdependent configuration models (Conf-dep)}

For each type of edge overlap (including having an edge only on a single layer), a configuration model \cite{newman2010networks} is generated. The final multiplex network is constructed by adding the edges from each configuration model to the layers corresponding to that overlap. The generated network approximately matches the degree correlations and edge overlaps of the network used as the basis. The multiplex degree distributions and edge overlaps of the BA-dep networks were used to generate the Conf-dep networks.

\subsubsection*{Zero-overlap Erd\H{o}s-R\'enyi networks (ER-0)}

Each layer is constructed as a separate Erd\H os-R\'enyi random network \cite{erdHos1959random} with the restriction that an edge cannot exist in multiple layers.
The model produces networks with zero edge overlap and interlayer degree correlations close to zero. The average intralayer degree is $2m$, as in the Barab\'asi-Albert networks, and when the network is aggregated into a single layer, the average degree is $2m$ times the number of layers (since there is no overlap between layers).

\subsubsection*{Overlapping Erd\H{o}s-R\'enyi networks (ER-20)}

Similarly as in the Conf-dep model, each type of overlap is generated as a regular Erd\H os-R\'enyi random network with the restriction that an edge can exist in at most one "overlap layer/network".
The final multiplex network is then constructed by adding links of one of the networks to all the layers, links of another layer to all but one of the layers and so forth such that for each combination of layers we use the links of one generated network.
With this model we generate networks with equal edge densities to ER-0 in the aggregated networks (average degree in an aggregated network is $2m$ times the number of layers), but with 20 \% of the edges overlapping between every pair of layers. Because of the overlap, the average intralayer degree is therefore higher than $2m$.

\subsubsection*{Random geometric graphs with shared node location (GEO)}

In the soft geometric random graph model \cite{penrose2016connectivity}, each node is randomly assigned a position in the unit square. Nodes within a threshold distance $r$ are connected by an edge with probability $e^{-d}$, where $d$ is their Euclidean distance. For the multiplex network, the nodes are positioned in the same locations in all the layers, but the edges between the nodes are added independently in each layer.
The model produces networks with high edge overlap and interlayer degree correlations.
The threshold distance is set to $r = \sqrt{\frac{2.2 \times m}{\pi \times (n-1)}}$, which produces average intralayer degrees of approximately $2m$.

\subsubsection*{Watts-Strogatz models with same initial lattice (WS)}

The Watts-Strogatz model \cite{watts1998collective} starts with a ring and connects each node to its $2m$ nearest neighbors in all the layers.
Then for each layer edges are rewired independently with probability $p=0.3$.
The model generates networks with high edge overlap, but with low interlayer degree correlations. The average intralayer degree is $2m$.

\subsubsection*{Graphlet insertion}

To insert a graphlet with $n$ nodes and $l$ layers into a network, we randomly sample $k$ instances of all possible combinations of $n$ nodes and $l$ layers in the network such that no two instances overlap for more than one node if they share at least one layer. We then change the edge configurations at those locations to match the inserted graphlet. We then add or remove edges outside those locations as needed to match the average intralayer degree on each layer with that of the original network, by either choosing random edges to be deleted (if there are more edges than in the original network) or by choosing two random nodes and adding an edge between them (if there are fewer edges than in the original network).

To construct a test set of networks with graphlet insertions, we repeat the following three steps five times:
(1)    We generate 30 layer-independent Erd\H{o}s-R\'enyi (ER) multiplex networks with three layers, 1000 nodes and average intralayer degree 4,
(2)    we randomly pick 20 different graphlets with $n = 4$ nodes and $l = 2$ layers from all such graphlets,
(3)    for each network in the ER networks, we insert $k = 3$ instances of each of the chosen graphlets into that network.
As a result, we get five sets of 30 networks with different inserted graphlets. In Supplementary Materials, the graphlet insertion method is repeated for graphlets with 3 nodes and 3 layers.

\subsection{Evaluation of measures}

The GCDs for the network test sets were computed using graphlets with one, two and three layers.
For the one-layer and two-layer cases, the orbits were counted for up to 3- and 4-node graphlets, and for the 3-layer case, the orbits were counted for only up to 3-node graphlets. The number of nodes here denotes the maximum number of nodes: for example, for four nodes and two layers, all graphlets with two layers and two, three, or four nodes were included. The orbits were computed for all the nodes in the networks using node-layer isomorphism.
For the calculation of the one-layer orbits, the networks were layer-aggregated and treated as if they were single-layer networks to model the typical analysis procedure when ordinary graphs are used.
For the 2-layer measures, the orbit counts were computed for each of the 2-layer combinations in the networks and the counts from all of the combinations were summed together to obtain the final orbit counts.
The test set networks had three layers, so there was only one 3-layer combination to consider for the 3-layer graphlets.
For each pair of number of nodes and number of layers, the distances were computed both including and excluding the redundant orbits.

The performance of each measure is evaluated by the area under the precision-recall curve (AUPR).
Precision is defined as the fraction of true positives out of all positives and recall is the fraction of true positives and the sum of true positives and false negatives.
Here, true positives are pairs of networks generated using the same model that have a distance smaller than the threshold $\epsilon$ at which the precision and recall values are evaluated.
False positives are pairs of networks generated from different models but have a distance smaller than $\epsilon$, and false negatives are the pairs of networks generated from the same model but have a distance larger than $\epsilon$.

\subsection{Comparison results}

For the measures we use the following naming convention.
The first number denotes the number of layers and the second is the maximum number of nodes in the graphlets.
'R' in the end denotes that redundant orbits have been removed.
For example GCD-2-4R denotes GCD computed using orbits of multiplex graphlets with two layers and up to four nodes excluding the redundant orbits. 

For comparison purposes, we also calculated GCDs following an alternative orbit definition GCD-DPK (named here after Dimitrova, Petrovski, Kocarev)~\cite{dimitrova2020graphlets} based on the implementation the authors have provided \cite{rovski-github}. In the implementation, graphlets of two and three nodes are taken into account, so the performance is similar but not identical to our graphlets with three layers and up to three nodes. In general, our method produces different orbits than the alternative method as shown in Supplementary Materials.

Comparison results for NetEmd and GDDA distances are shown in Supplementary Materials, and they support the conclusions drawn from the GCD results.

\subsubsection{Precision-recall curves}

We plot the precision-recall curves and AUPRs of all three multiplex network test sets in Figure \ref{fig:pre-rec-combined}.
When separating different models with constant average intralayer degrees (Figure \ref{fig:pre-rec-combined} (a)), GCD-2-3R has the highest AUPR, and all the multiplex graphlet methods perform better than the single-layer graphlet methods GCD-1-4(R) and GCD-1-3(R), showing that considering multiplex structure is indeed important when constructing graphlet-based distance measures for multiplex networks. When each model includes networks with progressing average intralayer degrees (Figure \ref{fig:pre-rec-combined} (b)), the separation task is much more difficult. However, multilayer measures still perform much better than single-layer ones and GCD-3-3 has the highest AUPR. When separating graphlet insertion networks (Figure \ref{fig:pre-rec-combined} (c)), GCD-2-4 achieves perfect separation. There is a clear difference between the performance of GCD-2-4(R) and GCD-2-3(R) compared to the other methods (and the other multilayer methods perform comparably to the single-layer methods). The difference is explained by the fact that the inserted graphlets had two layers and four nodes, which GCD-2-4(R) specifically finds in the networks, and that they therefore also contain some two-layer-three-node graphlets as subnetworks. When looking to separate real-world networks which are expected to contain different kinds of graphlets, one then has to choose a distance measure that contains the expected graphlets. The precision-recall curves illustrate that 1) methods created specifically with multiplex isomorphisms in mind are necessary for handling multiplex networks, and 2) just applying any multiplex method in an unsupervised context may not be good enough, instead the method should be adapted to whatever kind of graphlets the data contains (or is expected to contain). 

\definecolor{blue}{HTML}{0000FF}
\definecolor{cyan}{HTML}{00FFFF}
\definecolor{purple}{HTML}{800080}
\definecolor{magenta}{HTML}{FF00FF}
\definecolor{pink}{HTML}{FFC0CB}
\definecolor{orange}{HTML}{FFA500}
\definecolor{yellow}{HTML}{FFFF00}
\definecolor{green}{HTML}{008000}
\definecolor{black}{HTML}{000000}
\definecolor{gray}{HTML}{808080}

\begin{figure}[]
\begin{center}
\begin{minipage}{\textwidth}
\begin{center}
\begin{tikzpicture}
\node [anchor=south west] (image) at (0, 0) {
\includegraphics[width=0.95\textwidth]{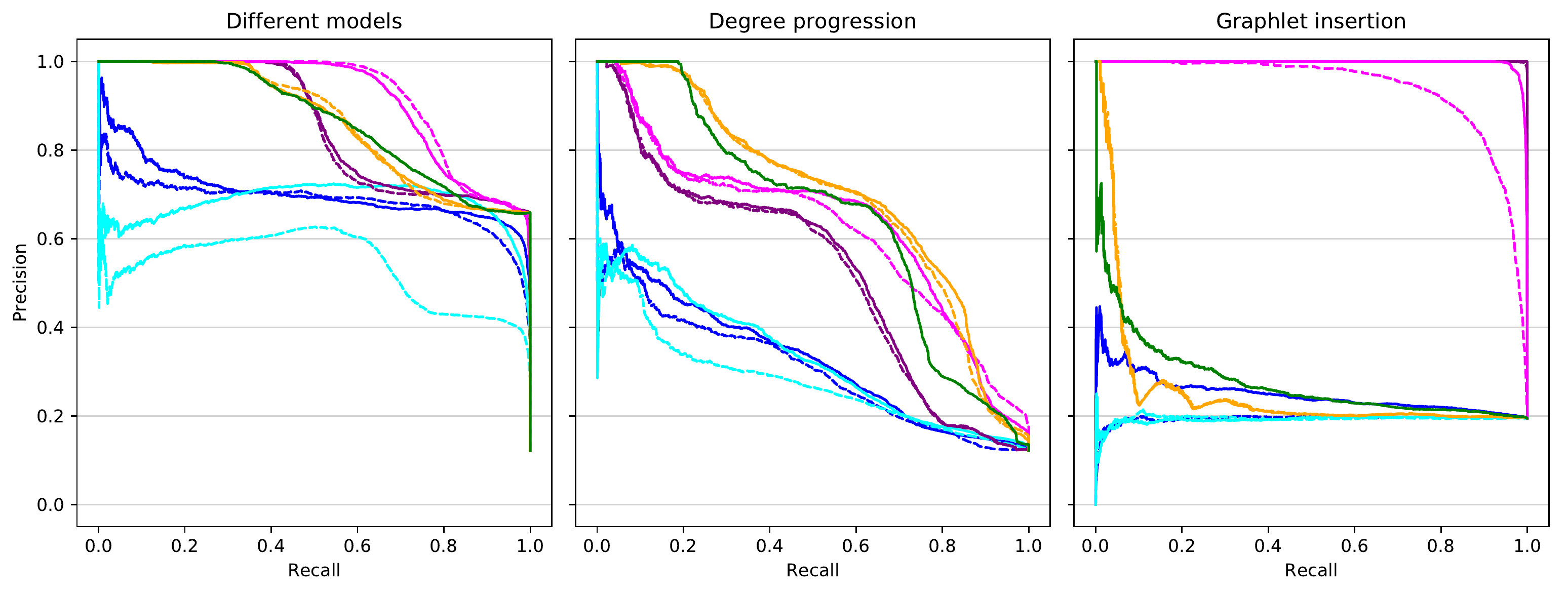}
};
\begin{scope}[x={(image.south east)},y={(image.north west)}]
\node[label={center, black}:{\footnotesize (a)}] at (0.11,0.22) {};
\node[label={center, black}:{\footnotesize (b)}] at (0.42,0.22) {};
\node[label={center, black}:{\footnotesize (c)}] at (0.73,0.22) {};
\end{scope}
\end{tikzpicture}
\end{center}
\end{minipage}
\\
\begin{minipage}{\textwidth}
{\tiny
\begin{tabular}{c|ccccccccccc}
& \cellcolor{blue} & \cellcolor{blue} & \cellcolor{cyan} & \cellcolor{cyan} & \cellcolor{purple} & \cellcolor{purple} & \cellcolor{magenta} & \cellcolor{magenta} & \cellcolor{orange} & \cellcolor{orange} & \cellcolor{green} \\
Distance (GCD) & 1-4 & 1-4R & 1-3 & 1-3R & 2-4   & 2-4R  & 2-3 & 2-3R   & 3-3   & 3-3R  & DPK \\
\hline
AUPR (a) & 0.7050 & 0.6848 & 0.6816 & 0.5360 & 0.8568 & 0.8531 & 0.9150 & \textbf{0.9226} & 0.8608 & 0.8629 & 0.8672 \\
AUPR (b) & 0.3309 & 0.3038 & 0.3304 & 0.2752 & 0.5305 & 0.5238 & 0.6370 & 0.6211 & \textbf{0.7031} & 0.6936 & 0.6539 \\
AUPR (c) & 0.2481 & 0.1938 & 0.1924 & 0.1930 & \textbf{1.0000} & 0.9999 & 0.9969 & 0.9395 & 0.2506 & 0.2535 & 0.2759
\end{tabular}}
\end{minipage}
\caption{Precision-recall curves and AUPRs for different distance measures. The dashed lines depict the measures where redundant orbits have been removed. (a) Test set with eight different multiplex network models, all networks have similar approximate intralayer degrees. (b) Test set with eight different multiplex network models, and for each model the average intralayer degrees progressed from 2 to 12 in steps of 2. (c) Test set with graphlets with four nodes and two layers inserted.}
\label{fig:pre-rec-combined}
\end{center}
\end{figure}

Removing redundant orbits from the orbit counts slightly increases or decreases the AUPR in the two- and three-layer measures. On the other hand, there is a mostly performance-reducing effect with the single-layer measures, which might be explained by the smaller number of single-layer orbits compared to two- and three-layer orbits.
The GCD-3-3(R) curves are close to the GCD-DPK curves, with GCD-3-3(R) performing slightly better in networks with degree progression, and slightly worse in constant degree and graphlet insertion networks.

For single-layer GCDs going from GCD-1-3 to GCD-1-4 increases the AUPR, whereas for the two-layer GCD increasing the graphlet size from three to four nodes decreases the performance except in the graphlet insertion networks.
This could be explained by the fact that there are only few single-layer-three-node orbits and the number of orbits increases much faster as the number of nodes is increased when there are more layers (see Table~\ref{table:orbits_and_independent_equations}).
Thus, a greater proportion of the orbits in the two-layer four-node case are more likely to be completely non-existent or have a lot of counts close to zero especially if the networks are not very dense.
For these orbits the ranks of the nodes used to compute the correlations between orbits can appear quite random.

\subsubsection{Multidimensional scaling embeddings}

\begin{figure}[]
\begin{center}
\begin{tikzpicture}
\node [anchor=south west] (image) at (0, 0) {
\includegraphics[width=0.9\textwidth]{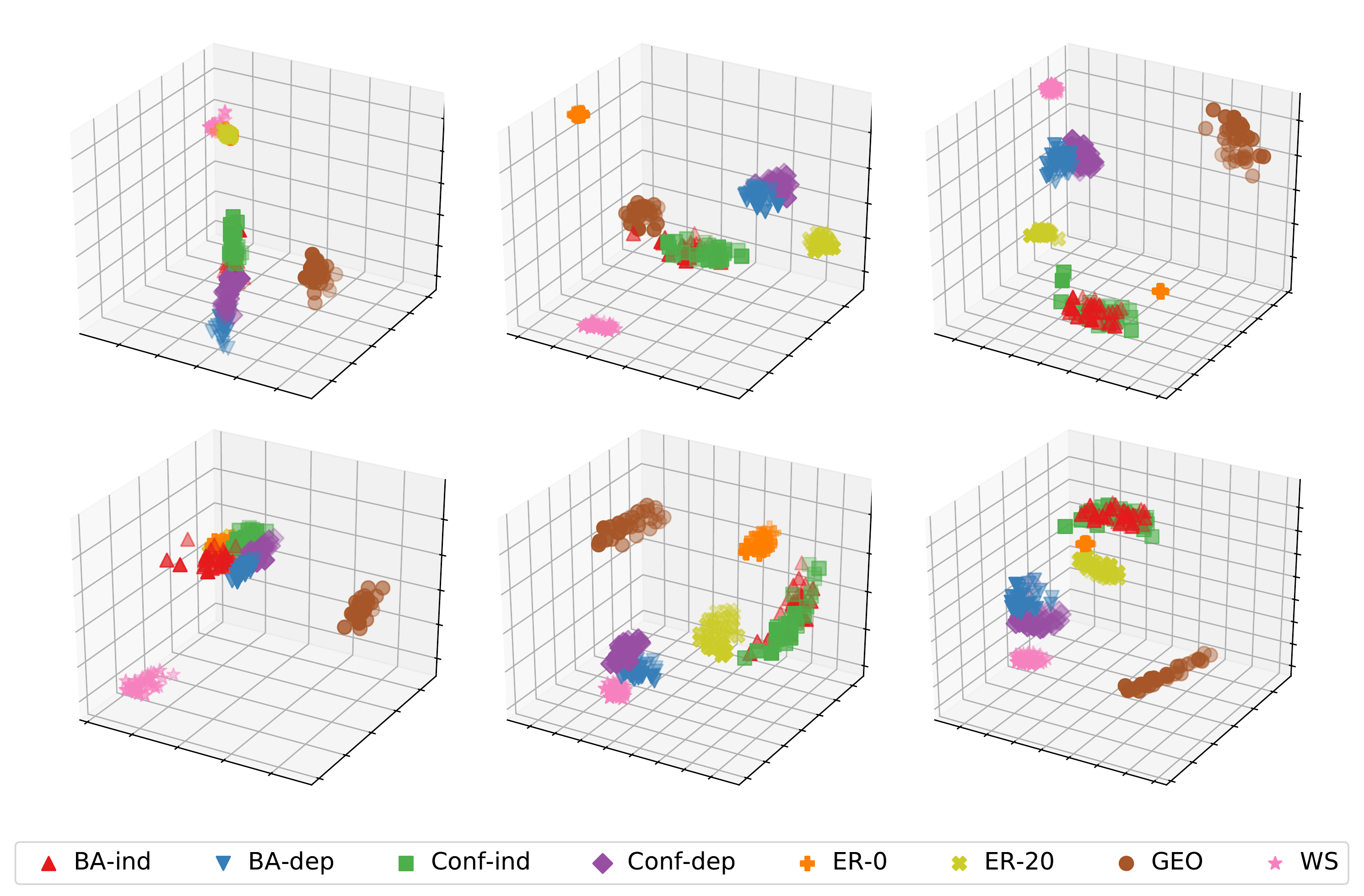}
};
\begin{scope}[x={(image.south east)},y={(image.north west)}]
\node[label={center, black}:{\footnotesize (a)}] at (0.07,0.9) {};
\node[label={center, black}:{\footnotesize (b)}] at (0.38,0.9) {};
\node[label={center, black}:{\footnotesize (c)}] at (0.69,0.9) {};
\node[label={center, black}:{\footnotesize (d)}] at (0.07,0.48) {};
\node[label={center, black}:{\footnotesize (e)}] at (0.38,0.48) {};
\node[label={center, black}:{\footnotesize (f)}] at (0.69,0.48) {};
\end{scope}
\end{tikzpicture}
\end{center}
\caption{Networks from different multiplex models with constant average intralayer degrees embedded into 3-dimensional space using MDS while preserving different distance measures. (a) GCD-1-3, (b) GCD-2-3, (c) GCD-3-3, (d) GCD-1-4, (e) GCD-2-4, (f) GCD-DPK. The multiplex measures generally separate the the models better than the single-layer models do. Separating BA-ind from Conf-ind and BA-dep from Conf-dep is difficult, while the other models form clearly separate clusters with the multiplex measures.}
\label{fig:MDS_3D}
\end{figure}
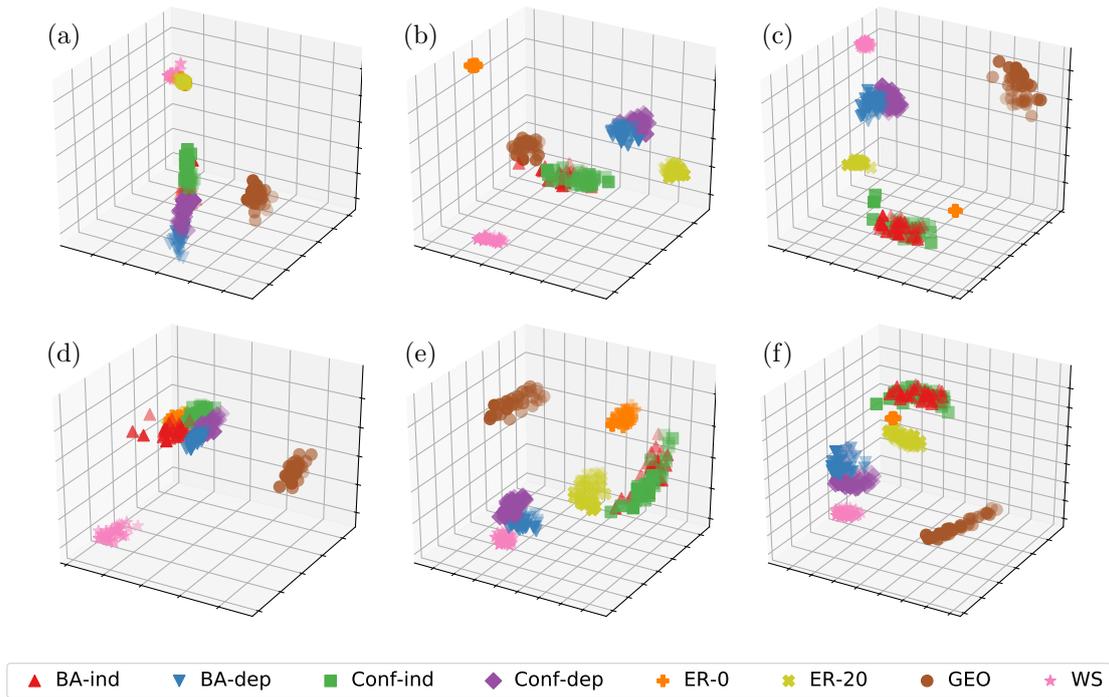

To see how well the different models are grouped by different distance measures, the networks are embedded into 3-dimensional space using the multidimensional scaling method (MDS) \cite{cox2000multidimensional} which preserves the different distances as well as possible. Figure \ref{fig:MDS_3D} shows the embeddings for the different multiplex models with constant average intralayer degree; the embeddings for progressing intralayer degree and graphlet insertion networks are shown in Supplementary Materials. All the multiplex methods separate the network classes somewhat clearly, except for the BA-ind and Conf-ind, and BA-dep and Conf-dep models. These models have pairwise matched edge overlaps between layers, i.e. their multiplex structure is very similar, which may explain the difficulty in separating them.

\subsubsection{Pairwise AUPRs}

\begin{figure}[]
\begin{center}
\begin{tikzpicture}
\node [anchor=south west] (image) at (0, 0) {
\includegraphics[width=0.9\textwidth]{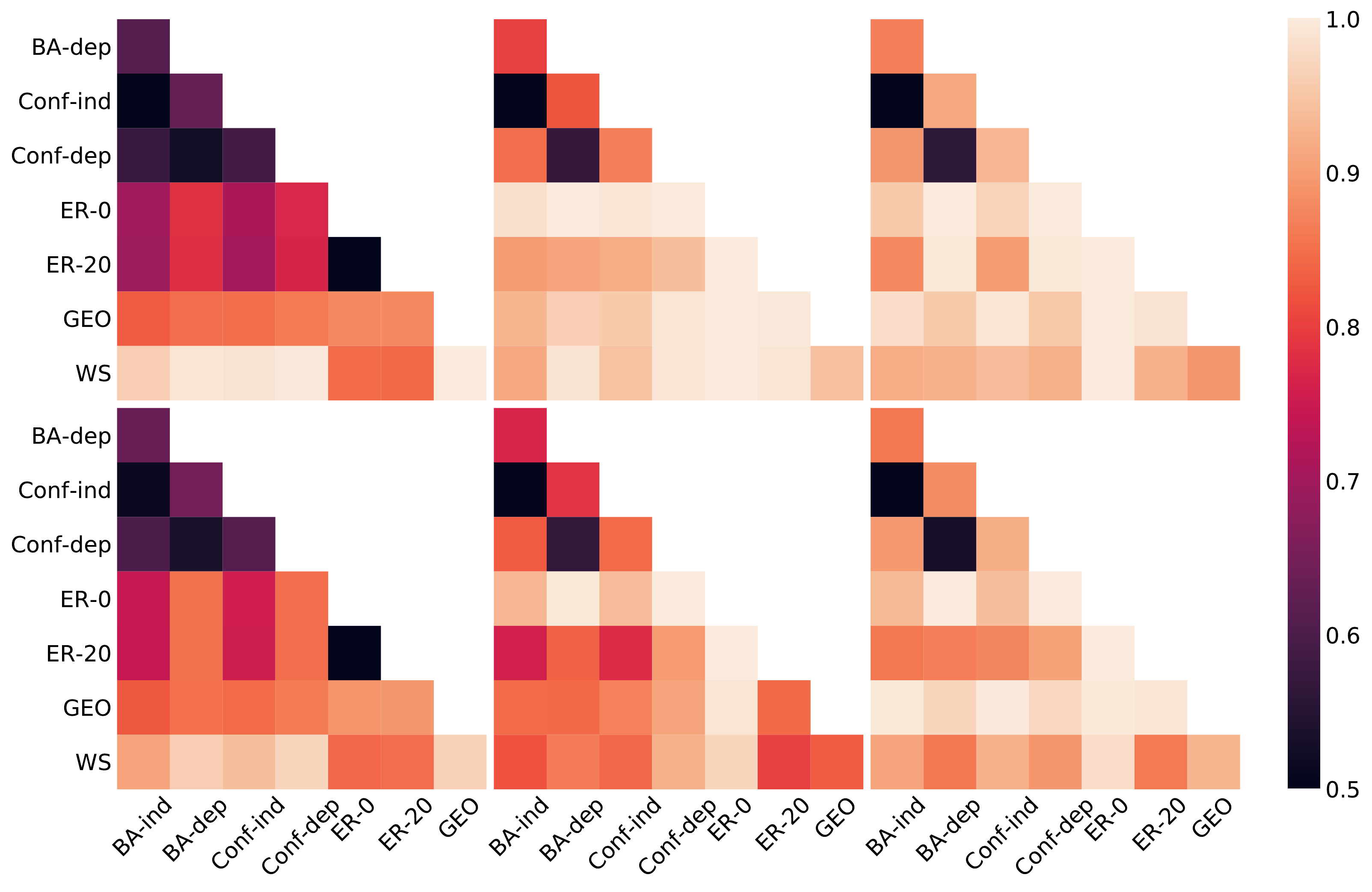}
};
\begin{scope}[x={(image.south east)},y={(image.north west)}]
\node[label={center, black}:{\footnotesize (a)}] at (0.25,0.9) {};
\node[label={center, black}:{\footnotesize (b)}] at (0.52,0.9) {};
\node[label={center, black}:{\footnotesize (c)}] at (0.79,0.9) {};
\node[label={center, black}:{\footnotesize (d)}] at (0.25,0.48) {};
\node[label={center, black}:{\footnotesize (e)}] at (0.52,0.48) {};
\node[label={center, black}:{\footnotesize (f)}] at (0.79,0.48) {};
\end{scope}
\end{tikzpicture}
\end{center}
\caption{Pairwise AUPRs when separating networks with average intralayer degree progression from different models using different distance measures. (a) GCD-1-3, (b) GCD-2-3, (c) GCD-3-3, (d) GCD-1-4, (e) GCD-2-4, (f) GCD-DPK.}
\label{AUPRs}
\end{figure}
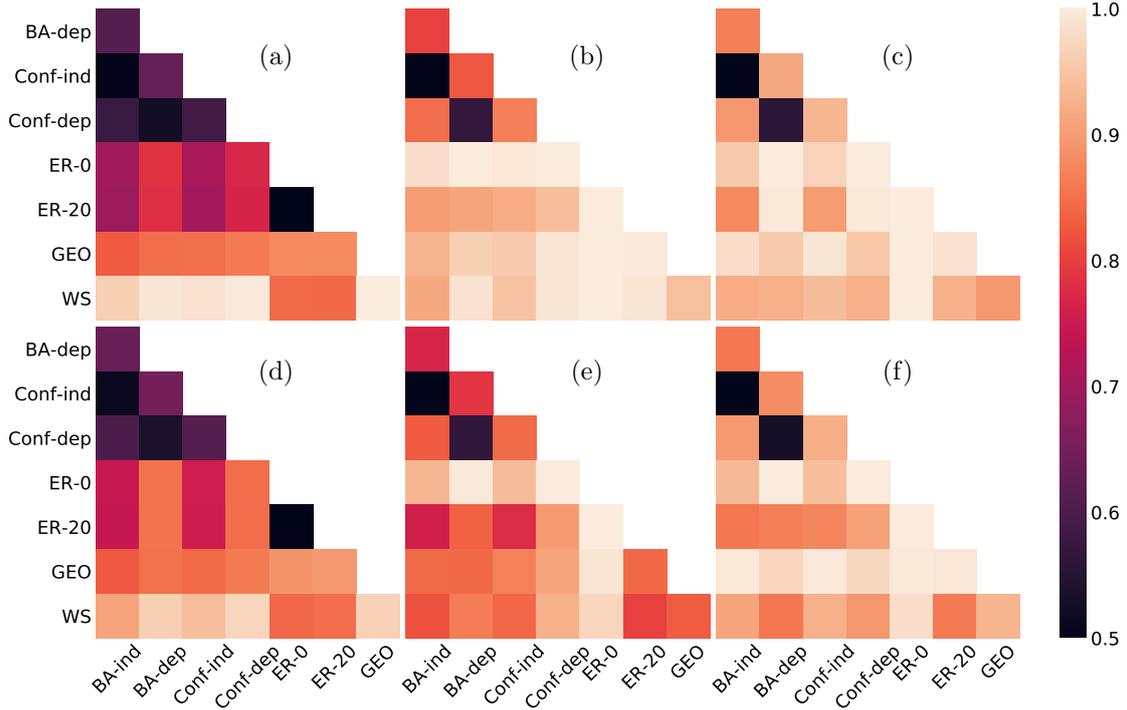

Figure~\ref{AUPRs} illustrates the pairwise AUPR values when clustering networks from different random network models with average intralayer degree progression. The pairwise AUPRs for the random networks with constant intralayer degrees and graphlet insertion networks are shown in Supplementary Materials.
The single-layer measures GCD-1-3 and GCD-1-4 perform worse than the multilayer measures overall, and as expected they are especially unsuited for cases, such as the two ER models, that are designed to be statistically indistinguishable at the aggregated network level.
All the measures have trouble distinguishing BA-ind, BA-dep, Conf-ind, and Conf-dep from one another, especially in BA-ind--Conf-ind and BA-dep--Conf-dep pairs.
None of the multilayer methods is the best in every situation, however GCD-3-3 seems to have similar but overall slightly better performance than GCD-DPK in most pairs. The results highlight that multilayer methods are better than single-layer ones also in pairwise separation tasks.

\section{Discussion}

To answer the need for graphlet-based tools for investigating the structure of multilayer networks, we have created a systematic and principled framework for multilayer network graphlet analysis, starting from the definition of isomorphism and automorphism orbits in multilayer networks. 
The framework can be used with any kind of multilayer network with any number of aspects. We have illustrated the usefulness of the framework with test sets of multiplex networks, showing that 1) multilayer graphlet correlation distance performs considerably better in grouping networks from different multiplex random models than single-layer graphlet correlation distance, and that 2) when there is graphlet structure in the networks, choosing the right multilayer graphlet size in the correlation distance measure is highly important, and that 3) the number of redundancy equations is relatively small compared to the number of orbits in multiplex networks and using them does not lead to significant improvements in the unsupervised prediction tasks we constructed.

We have presented the graphlet analysis pipeline in the light of comparing a set of networks to one another and in a general format, leaving for example the choice of isomorphism to the reader. Naturally, the choices made in the pipeline (Figure \ref{fig:flowchart}) affect the end results, and it might not be immediately clear what the correct ones for a given situation are. Especially when there are graphlets of a specific size in the network(s) we are investigating and we include graphlets of the wrong size in our graphlet-based measures, the results can be devastating in unsupervised learning tasks (see Figure \ref{fig:pre-rec-combined} (c)).

Once one obtains the graphlet degree vectors or the graphlet correlation matrix, they can be used as feature vectors/matrices in a variety of ways \cite{tantardini2019comparing,milenkovic2010optimal,milenkovic2008uncovering,vijayan2015magna++,lyu2017enhancing,hayes2013graphlet}.
The analysis presented here on clustering multiplex networks using node orbits and node-layer isomorphism barely scratches the surface of the possibilities given by multilayer graphlet analysis. Clearly, other types of multilayer networks can be analyzed with different isomorphisms. Additionally, the possibility to define orbits for layers, and combinations of layers, in addition to nodes could result in interesting new ways of analyzing multilayered systems.
Further, in addition to unsupervised learning problems (such as clustering networks), supervised learning models can be constructed to e.g. predict which group a network or an element of a network belongs in based on the graphlet degrees. 
Even though such methods have been previously applied to single-layer networks \cite{zhang2020panda,bouritsas2020improving,shervashidze2009efficient,li2020improving}, there is a clear avenue for further research in the application of machine learning to multilayer graphlets.
Presumably, the problem of choosing the correct set of graphlets is less severe in supervised problem in which the method should be able to find the important orbits based on the training data.

In this article, we applied the framework to multiplex networks generated from different multiplex network random models. Such synthetic networks are convenient for establishing a "ground truth" for evaluation of network measures, but they do not necessarily reflect well the properties of real-world multiplex and especially multilayer networks. Since there are already multiple well-labeled multilayer/plex data sets available (e.g. \cite{DickisonMagnaniRossi2016}), the usefulness of the multilayer graphlet framework should be established on them. Further, the exact choice of the models will probably reflect on the results greatly. For example, apart from the graphlet insertion method, our models did not include complex interlayer relations, which probably would highlight the importance of larger graphlets.

The generalization of single-layer graphlet and orbit concepts and methods to multilayer networks massively expands the possibilities of network analysis with different types of data, relationships, and hierarchies present in real-world networked systems. Starting from the mathematical foundation of multilayer network automorphisms provides a solid theoretical basis for multilayer orbits, on which advanced concepts and theory can be built in the future. The freedom gained in the leap from single-layer to multilayer networks enables more accurate representation and structural analysis of network-like systems, opening the door for a multitude of applications in a wide variety of fields where multilayer networks appear ranging from natural sciences and engineering to social sciences and humanities.

\section*{Acknowledgments}
MK acknowledges support from the Academy of Finland project ECANET (grant number: 320781).

%\pagebreak

\bibliographystyle{siamplain}
\bibliography{arxiv}

\pagebreak

%\usepackage{subcaption}
%usepackage{colortbl}
%\usepackage{multirow}
%\usepackage{float}
%\usepackage[section]{placeins}

%begin{document}
%\maketitle

\newcommand{\nn}{\nonumber}
\renewcommand\thesection{\Alph{section}}

\section*{Supplementary Materials}

\appendix 

\flushbottom

\thispagestyle{empty}

\definecolor{blue}{HTML}{0000FF}
\definecolor{cyan}{HTML}{00FFFF}
\definecolor{purple}{HTML}{800080}
\definecolor{magenta}{HTML}{FF00FF}
\definecolor{pink}{HTML}{FFC0CB}
\definecolor{orange}{HTML}{FFA500}
\definecolor{green}{HTML}{008000}
\definecolor{yellow}{HTML}{FFFF00}
\definecolor{black}{HTML}{000000}
\definecolor{gray}{HTML}{808080}

\section{Comparison of orbit definition methods}

We compare our method to that of Dimitrova et al. \cite{dimitrova2020graphlets}, and show that the two methods do not yield the same orbits by a counter-example.

\begin{figure}[!htb]
    \centering
    \includegraphics[width=0.3\textwidth]{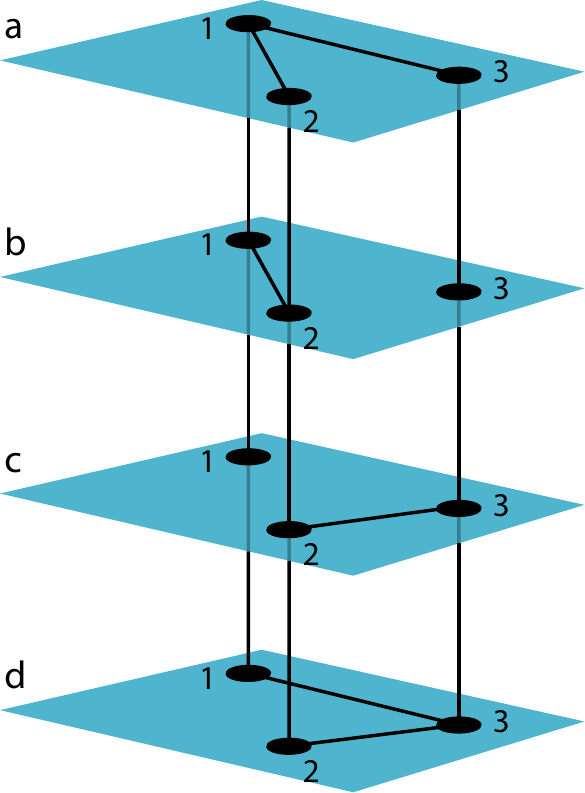}
    \caption{Example multiplex network with $V = \{1,2,3\}, L_1 = \{a,b,c,d\}$.}
    \label{fig:orbit-definitions-example}
\end{figure}

Consider the network in Figure \ref{fig:orbit-definitions-example}. Intuitively, the "roles" of nodes 1 and 3 are similar in the network, and the role of 2 is different from them. Now, $\bs{\zeta}: 1 \leftrightarrow 3, a \leftrightarrow d, b \leftrightarrow c$ is a node-layer automorphism. According to our definition, $Orb_{\{0,1\}}(1) = \{1,3\} = Orb_{\{0,1\}}(3)$ and $Orb_{\{0,1\}}(2) = \{2\}$, corresponding to the intuition.

However, using the definitions of Dimitrova et al., the sub-orbits of the nodes are 1: $3_{ab.ad.cd}$, 2: $3_{ab.cd.ad}$, 3: $3_{ad.cd.ab}$, i.e. they are all different. The first reduction method gives orbits 1: $3_{2.2.2}$, 2: $3_{2.2.2}$, 3: $3_{2.2.2}$, i.e. they are all the same. The second reduction method gives orbits 1: $3_{2x.2y.2z}$, 2: $3_{2x.2y.2z}$, 3: $3_{2x.2y.2z}$, i.e. they are all the same. Therefore, the orbits of this method are different from the orbits defined by our method, and the underlying intuition is also different.

{}

\subsection{Number of graphlet degree distributions and the size of the network}

With our method, for graphlets of specific size, the number of graphlet degree distributions does not depend on the size of the network when all aspects of the network are allowed to be permuted in the isomorphism. That is, the number of orbits for a specific graphlet size is fixed. With the Dimitrova et al. method, the number of orbits very quickly explodes if not reduced, scaling exponentially with the number of layers \cite{dimitrova2020graphlets}. The size of the graphlet correlation matrix (GCM) is orbits times orbits, so computing it quickly becomes prohibitively expensive. With the Dimitrova et al. method, in our test networks with 3 layers, the number of orbits (dimension of GCM) was 280. In networks with 4 layers, it was 2160. In networks with 5 layers, it was 16864. With 5 layers, each GCM (16864 times 16864) already consumed 7.3-7.4 GB of disk space when pickle-serialized from a numpy array, the total coming up to 1.1 TB for the whole 150-network set. This made it practically impossible to calculate the distance measures with Dimitrova et al. orbits for networks with more than 5 layers.

\begin{figure}[H]
\begin{center}
\includegraphics[width=0.8\textwidth]{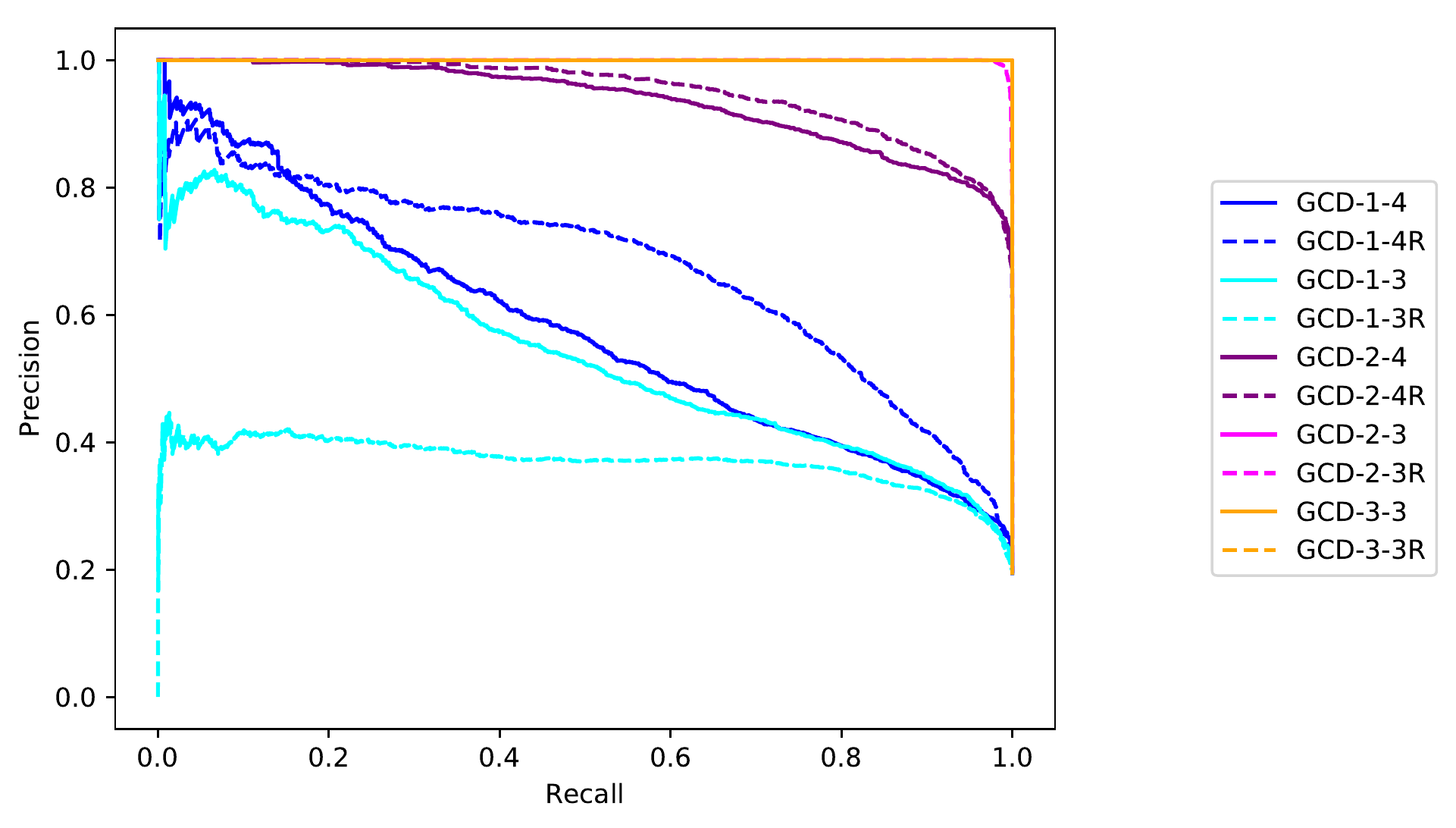}
\end{center}
\caption{10-layer 3-node-3-layer graphlet insertion precision-recall}
\label{fig:big-graphlet33-prerec}
\end{figure}

\begin{figure}[H]
\begin{center}
\includegraphics[width=0.8\textwidth]{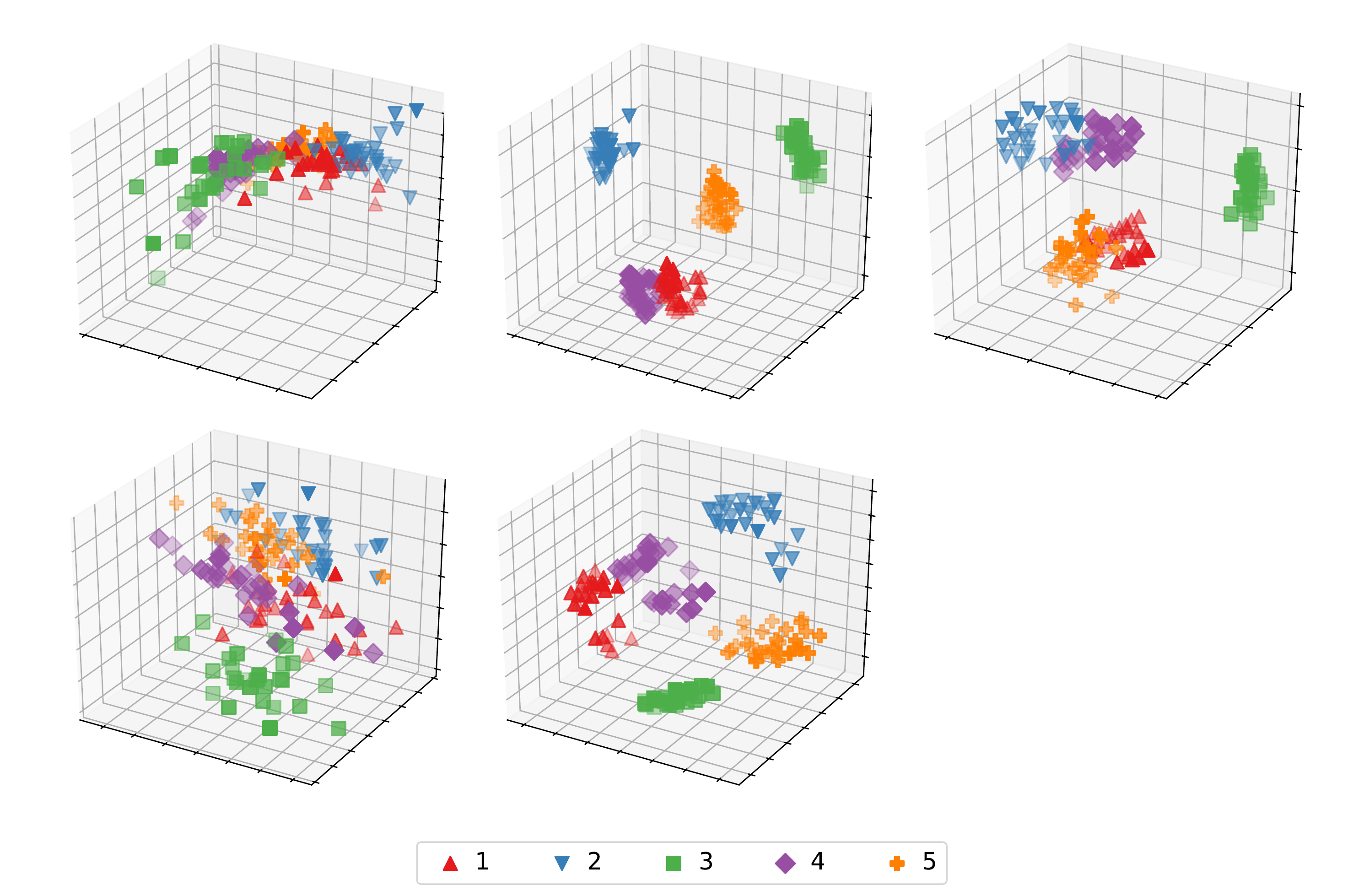}
\end{center}
\caption{10-layer 3-node-3-layer graphlet insertion MDS}
\label{fig:big-graphlet33-mds}
\end{figure}

\begin{figure}[H]
\begin{center}
\includegraphics[width=0.8\textwidth]{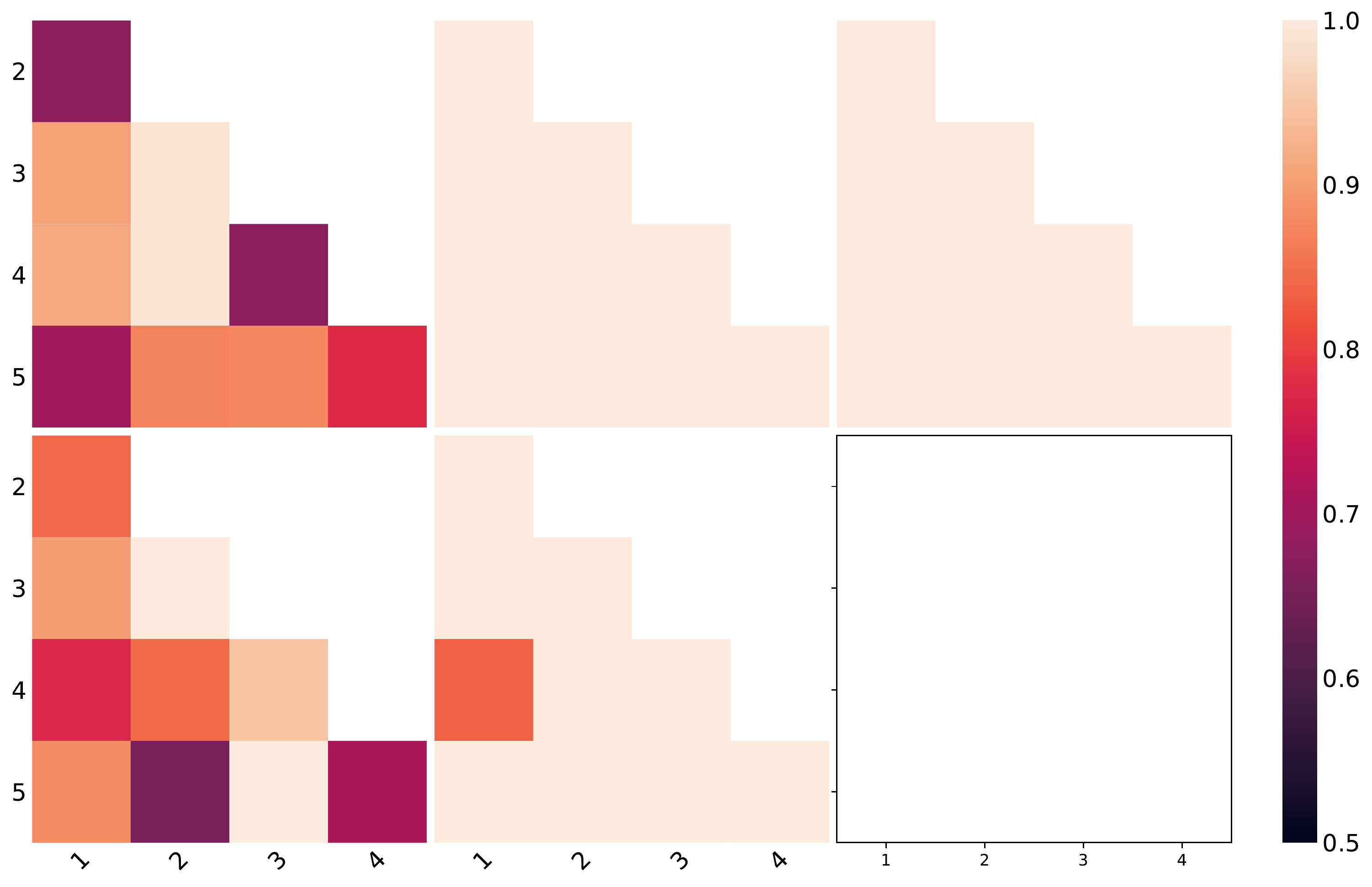}
\end{center}
\caption{10-layer 3-node-3-layer graphlet insertion pairwise AUPRs}
\label{fig:big-graphlet33-auprs}
\end{figure}

On the other hand, we calculated the precision-recall curves (Figure \ref{fig:big-graphlet33-prerec}), MDS embeddings (Figure \ref{fig:big-graphlet33-mds}), and pairwise AUPRs (Figure \ref{fig:big-graphlet33-auprs}) for a set of 3-node-3-layer graphlet insertion networks with 10 layers (10 graphlets inserted, 350 each) with our method. In Figures \ref{fig:big-graphlet33-mds} and \ref{fig:big-graphlet33-auprs} we use the following convention for subfigures: \textbf{top left:} GCD-1-3; \textbf{top middle:} GCD-2-3; \textbf{top right:} GCD-3-3; \textbf{bottom left:} GCD-1-4; \textbf{bottom middle:} GCD-2-4. The number of different orbits is the same as with any other number of layers, and applicability of the method is only limited by computational time required to find the graphlet degree vectors.

\clearpage

\section{Detailed proof for multilayer automorphism orbits induce equivalence classes}

Let $\sim$ be a relation with $\bs{\gamma} \sim \bs{\delta} \Leftrightarrow \bs{\delta} \in Orb_p(\bs{\gamma})$. Now, we need to prove that $\sim$ is an equivalence relation, i.e. that it is reflexive, symmetric, and transitive. The identity permutation is always an automorphism, and therefore $\exists \ \bs{\zeta}$ such that $\bs{\zeta}(\bs{\gamma}) = \bs{\gamma} \ \forall \ \bs{\gamma}$. Then, $\bs{\gamma} \sim \bs{\gamma}$ and $\sim$ is reflexive. If $\bs{\gamma} \sim \bs{\delta}$, $\exists \ \bs{\zeta}$ such that $\bs{\delta} = \bs{\zeta}(\bs{\gamma})$. Every automorphism $\bs{\zeta}$ has an inverse $\bs{\zeta}^{-1}$ that's also an automorphism, because $\bs{\zeta}$ maps a multilayer network to itself and we can therefore reverse the relabeling done in $\bs{\zeta}$ to also map the network to itself. Then, $\bs{\delta} = \bs{\zeta}(\bs{\gamma}) \Leftrightarrow \bs{\zeta}^{-1}(\bs{\delta}) = \bs{\zeta}^{-1}(\bs{\zeta}(\bs{\gamma})) \Leftrightarrow \bs{\zeta}^{-1}(\bs{\delta}) = \bs{\gamma}$ which means that $\bs{\gamma} \in Orb_p(\bs{\delta})$ and $\bs{\delta} \sim \bs{\gamma}$. Therefore, $\bs{\gamma} \sim \bs{\delta} \Leftrightarrow \bs{\delta} \sim \bs{\gamma}$ and $\sim$ is symmetric. If we combine two automorphisms, the result is also an automorphism, because each automorphism maps a multilayer network back to itself. If $\bs{\gamma}_1 \sim \bs{\gamma}_2$ and $\bs{\gamma}_2 \sim \bs{\gamma}_3$, then $\exists \ \bs{\zeta}_1$ such that $\bs{\zeta}_1(\bs{\gamma}_1) = \bs{\gamma}_2$ and $\exists \ \bs{\zeta}_2$ such that $\bs{\zeta}_2(\bs{\gamma}_2) = \bs{\gamma}_3$. Now, $\bs{\zeta}_2\bs{\zeta}_1$ is also an automorphism, and $\bs{\zeta}_2\bs{\zeta}_1(\bs{\gamma}_1) = \bs{\zeta}_2(\bs{\zeta}_1(\bs{\gamma}_1)) = \bs{\gamma}_3$, so $\bs{\gamma}_3 \in Orb_p(\bs{\gamma}_1)$ and $\bs{\gamma}_1 \sim \bs{\gamma}_3$. Therefore, $\bs{\gamma}_1 \sim \bs{\gamma}_2, \bs{\gamma}_2 \sim \bs{\gamma}_3 \implies \bs{\gamma}_1 \sim \bs{\gamma}_3$ and $\sim$ is transitive. $\square$

\clearpage

\section{Orbits for graphlets with two layers and four nodes}

\begin{figure}[!htb]
\begin{center}
\includegraphics[width=0.95\textwidth]{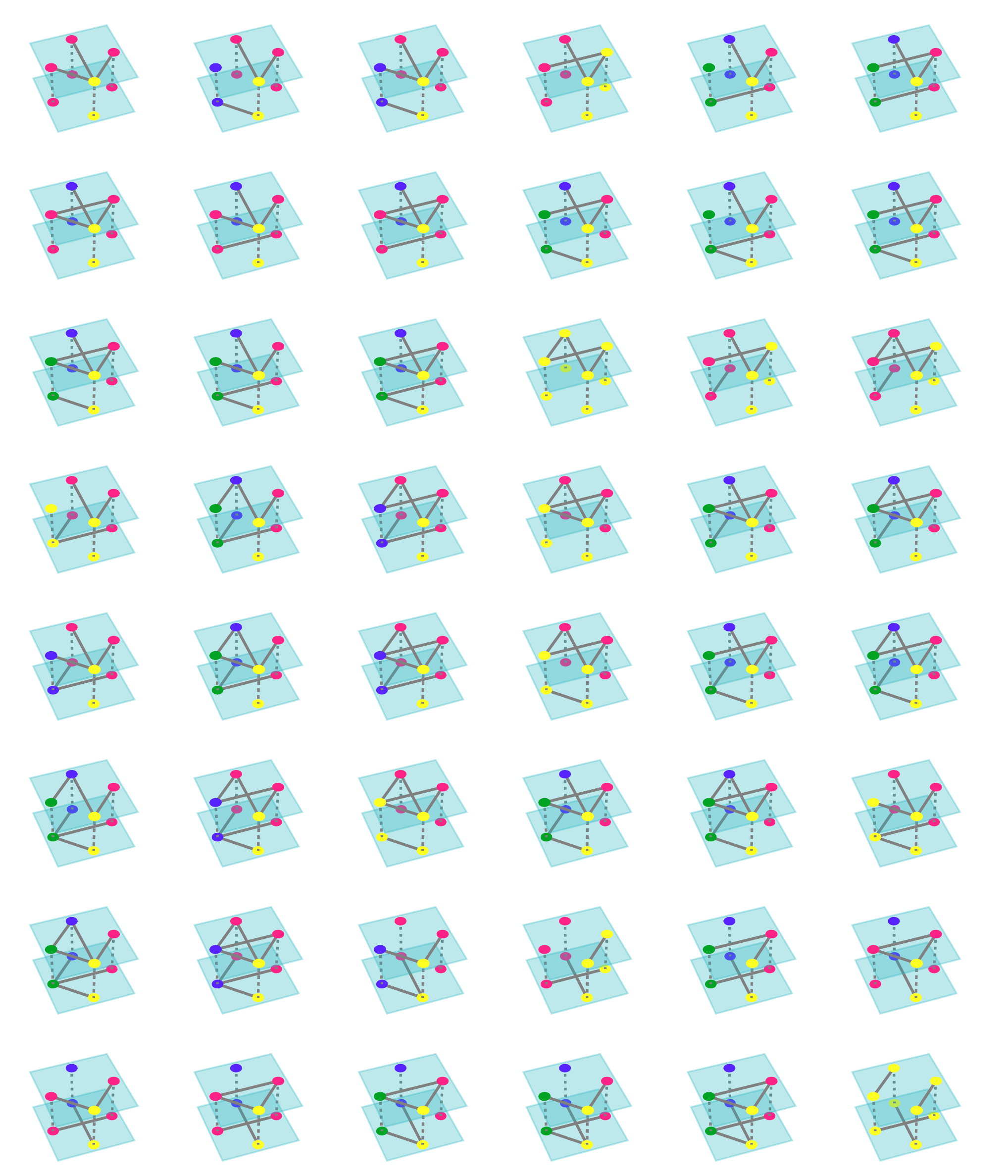}
\end{center}
\caption{2-layer-4-node multiplex graphlets 1-48 out of 137 and their orbits for nodes using node-layer isomorphism $Orb_{\{0,1\}}(u)$.
Within a graphlet nodes colored with the same color belong to the same orbit.}
\label{liitekuva1}
\end{figure}

\begin{figure}[!htb]
\begin{center}
\includegraphics[width=0.95\textwidth]{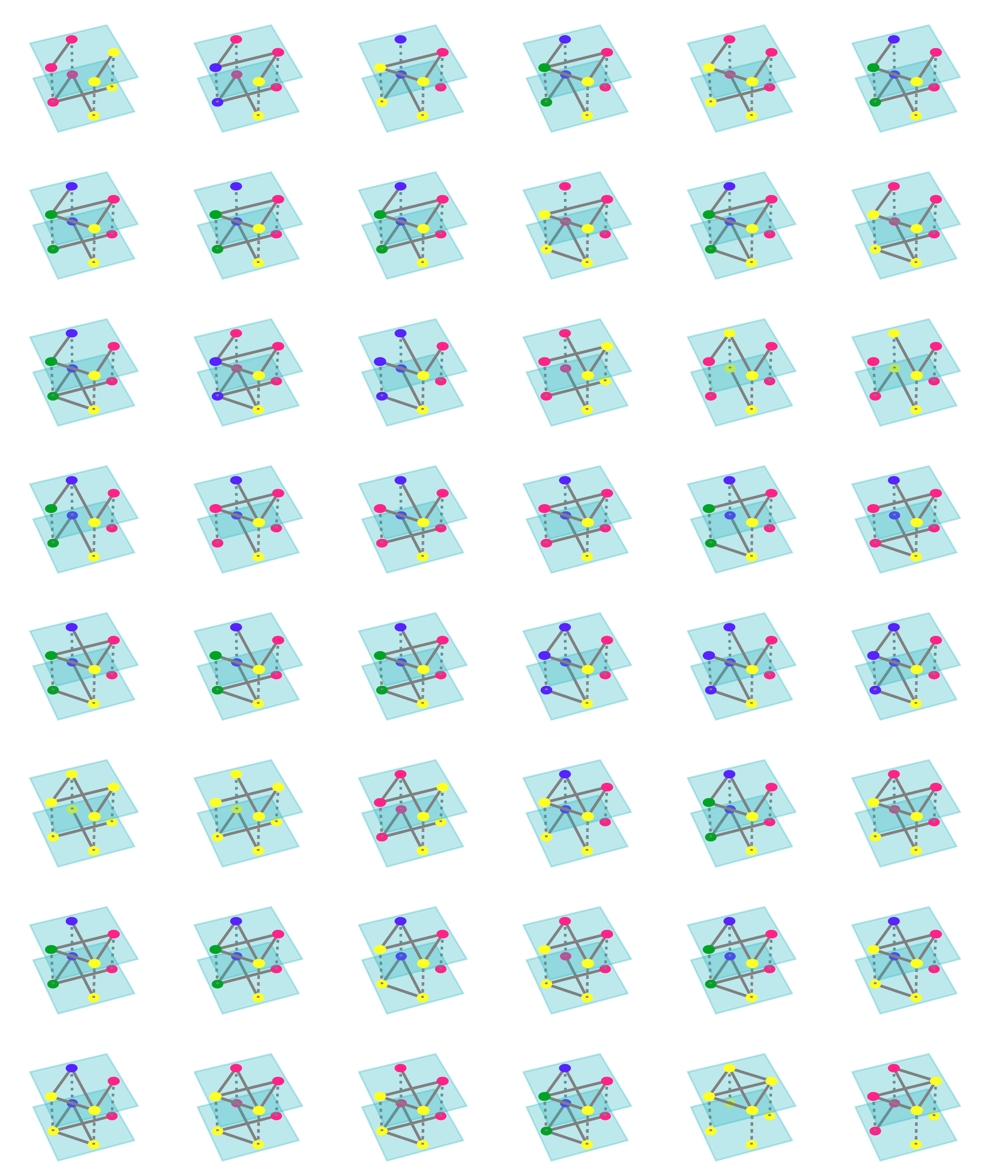}
\end{center}
\caption{2-layer-4-node multiplex graphlets 49-96 out of 137 and their orbits for nodes using node-layer isomorphism $Orb_{\{0,1\}}(u)$.
Within a graphlet nodes colored with the same color belong to the same orbit.}
\label{liitekuva2}
\end{figure}

\begin{figure}[!htb]
\begin{center}
\includegraphics[width=0.95\textwidth]{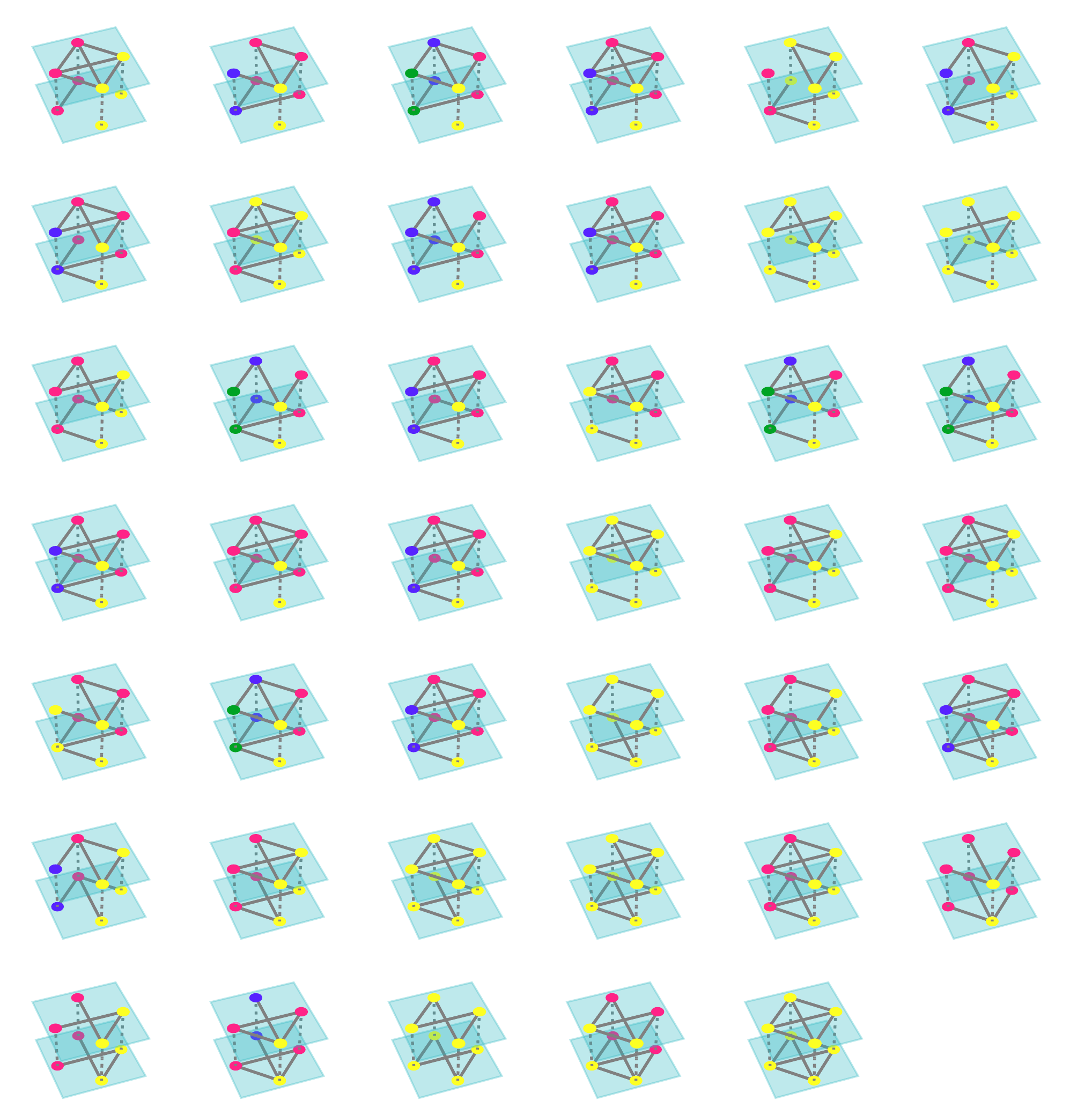}
\end{center}
\caption{2-layer-4-node multiplex graphlets 97-137 out of 137 and their orbits for nodes using node-layer isomorphism $Orb_{\{0,1\}}(u)$.
Within a graphlet nodes colored with the same color belong to the same orbit.}
\label{liitekuva3}
\end{figure}

We list the multiplex graphlets with two layers and four nodes and their orbits ($p = \{0,1\}$) in Figures \ref{liitekuva1}--\ref{liitekuva3}.

\clearpage

\section{Orbit equations for up to 4-node graphlets in multiplex networks with 2 layers}

The first three equations are obtained when combining two graphlets that produce graphlets with 3 nodes.
The graphlets combined in the rest of the equations produce graphlets with 4 nodes.
With the algorithm described in section \ref{section:independent_eqs}, we obtain that equations~\ref{eq_12} and \ref{eq_13} can be derived from the rest of the equations.

{
\tiny
\begin{eqnarray}
\binom{C_{0}}{2} & = & C_{2} + C_{4} + C_{9} + C_{10} + C_{11} + C_{12} + C_{14} \\ 
\binom{C_{1}}{2} & = & C_{17} + C_{18} + C_{20} \\
\binom{C_{1}}{1} \binom{C_{0}}{1} & = & C_{6} + C_{13} + C_{15} + C_{16} \\
\binom{C_{2}}{1} \binom{C_{0} - 2}{1} & = & 3 C_{21} + C_{23} + 2 C_{39} + 2 C_{42} + 2 C_{45} + C_{48} + C_{52} + C_{56} + C_{86} + C_{88} + C_{92} + C_{96} + C_{98} +  \\
 & & C_{99} + C_{103} + C_{106} + C_{108} + C_{112} + C_{116} + C_{120} \nn \\ 
\binom{C_{3}}{1} \binom{C_{0} - 1}{1} & = & C_{29} + C_{32} + C_{40} + C_{43} + C_{49} + C_{53} + C_{61} + C_{65} + 2 C_{72} + C_{73} + C_{74} + C_{75} + 2 C_{78} + C_{80} +  \\
 & & 2 C_{87} + C_{89} + C_{90} + C_{93} + 2 C_{97} + C_{100} + 2 C_{107} + C_{109} + C_{110} + C_{113} + C_{117} + 2 C_{124} + \nn \\
 & & C_{126} + C_{127} + C_{130} + 2 C_{134} + C_{136} \nn\\
\binom{C_{4}}{1} \binom{C_{0} - 2}{1} & = & 2 C_{23} + C_{48} + C_{52} + C_{56} + C_{91} + C_{111} + 2 C_{151} + 2 C_{154} + 2 C_{157} + C_{178} + C_{181} + C_{185} +  \\
 & & C_{187} + C_{191} + C_{195} + C_{199} \nn \\ 
\binom{C_{5}}{1} \binom{C_{0} - 1}{1} & = & C_{31} + C_{51} + C_{55} + C_{73} + C_{74} + 2 C_{77} + C_{79} + C_{89} + C_{90} + C_{109} + C_{110} + C_{126} + C_{127} +  \\
 & & C_{145} + C_{152} + C_{155} + C_{161} + C_{165} + 2 C_{172} + C_{173} + 2 C_{179} + 2 C_{180} + C_{182} + 2 C_{186} + \nn \\
 & & C_{188} + C_{193} + C_{197} + 2 C_{204} + C_{206} + 2 C_{210} + C_{212} \nn\\ 
\binom{C_{6}}{1} \binom{C_{0} - 1}{1} & = & 2 C_{26} + C_{60} + C_{64} + C_{68} + C_{95} + C_{102} + C_{115} + C_{119} + 2 C_{142} + C_{160} + C_{164} + C_{168} + \\
 & & C_{184} + C_{190} + C_{194} + C_{198} + 2 C_{231} + 2 C_{234} + 2 C_{237} + 2 C_{240} + 2 C_{244} + C_{272} + C_{275} + \nn \\
 & & C_{278} + C_{279} + C_{281} + C_{284} + C_{285} + C_{289} + C_{292} + C_{294} \nn\\
\binom{C_{7}}{1} \binom{C_{0} - 1}{1} & = & C_{35} + C_{75} + C_{79} + C_{80} + 2 C_{83} + C_{93} + C_{100} + C_{113} + C_{117} + C_{130} + C_{136} + C_{147} +  \\
 & & C_{173} + 2 C_{175} + C_{182} + C_{188} + C_{193} + C_{197} + C_{206} + C_{212} + C_{232} + C_{235} + C_{241} + C_{243} + \nn \\
 & & C_{248} + C_{252} + 2 C_{273} + 2 C_{276} + 2 C_{290} + 2 C_{299} + 2 C_{302} \nn \\
\binom{C_{8}}{1} \binom{C_{0}}{1} & = & C_{63} + C_{67} + C_{76} + C_{81} + C_{82} + C_{94} + C_{101} + C_{114} + C_{118} + C_{131} + C_{137} + C_{163} + C_{167} +  \\
 & & C_{174} + C_{183} + C_{189} + C_{192} + C_{196} + C_{207} + C_{213} + C_{223} + C_{225} + C_{261} + C_{264} + C_{268} + \nn \\
 & & C_{269} + C_{280} + C_{282} + C_{283} + C_{293} + C_{305} + C_{307} \nn \\
\binom{C_{9}}{1} \binom{C_{0} - 2}{1} & = & C_{39} + 2 C_{86} + C_{88} + C_{91} + C_{92} + C_{151} + C_{178} + C_{181} + 3 C_{312} + 2 C_{313} + C_{314} + 2 C_{315} + \\
 & & C_{317} + C_{318} + C_{320} + C_{321} + C_{324} + C_{327} + C_{329} + C_{332} \nn \\
\binom{C_{10}}{1} \binom{C_{0} - 2}{1} & = & C_{42} + C_{88} + 2 C_{96} + C_{99} + C_{111} + C_{154} + C_{185} + C_{187} + C_{313} + 2 C_{317} + C_{319} + C_{320} + \\
 & & 3 C_{328} + 2 C_{337} + C_{340} + C_{343} + C_{344} + C_{345} + C_{347} + C_{351} \nn \\
\binom{C_{11}}{1} \binom{C_{0} - 2}{1} & = & C_{48} + C_{52} + C_{91} + 2 C_{98} + 2 C_{106} + 2 C_{108} + C_{111} + C_{112} + C_{116} + C_{178} + C_{185} + C_{191} +  \\
 & & C_{195} + 2 C_{314} + 2 C_{318} + 2 C_{319} + C_{321} + 2 C_{327} + C_{329} + 2 C_{338} + 2 C_{343} + 2 C_{344} + C_{345} + \nn \\
 & & C_{347} + 2 C_{348} + 2 C_{355} + C_{357} + C_{361}\nn
 \end{eqnarray}
}

{
\tiny
\begin{eqnarray}
\binom{C_{12}}{1} \binom{C_{0} - 2}{1} & = & C_{45} + C_{92} + C_{99} + 2 C_{103} + C_{157} + C_{191} + C_{195} + C_{199} + C_{315} + C_{320} + 2 C_{324} + C_{337} + \label{eq_12} \\
 & & C_{338} + 2 C_{340} + C_{348} + C_{355} + C_{357} + C_{361} + 3 C_{367} + C_{369} \nn  \\
\binom{C_{13}}{1} \binom{C_{0} - 1}{1} & = & C_{60} + C_{95} + C_{102} + 2 C_{123} + C_{125} + C_{128} + C_{129} + C_{160} + C_{194} + C_{198} + 2 C_{203} + C_{205} +  \label{eq_13} \\
 & & C_{279} + C_{284} + 2 C_{316} + C_{322} + C_{323} + C_{325} + 2 C_{339} + C_{341} + C_{349} + C_{350} + 2 C_{354} + \nn \\
 & & C_{356} + C_{362} + 2 C_{372} + C_{373} + C_{374} + C_{375} + 2 C_{378} + C_{380} \nn \\
\binom{C_{14}}{1} \binom{C_{0} - 2}{1} & = & C_{56} + C_{112} + C_{116} + 2 C_{120} + C_{181} + C_{187} + C_{199} + C_{321} + C_{329} + 2 C_{332} + C_{345} + C_{347} +  \\
 & & 2 C_{351} + C_{357} + C_{361} + 2 C_{369} \nn \\
\binom{C_{15}}{1} \binom{C_{0} - 1}{1} & = & C_{64} + C_{115} + C_{119} + C_{125} + C_{128} + 2 C_{133} + C_{135} + C_{164} + C_{184} + C_{190} + 2 C_{209} + C_{211} +  \\
 & & C_{281} + C_{292} + C_{322} + C_{323} + 2 C_{330} + 2 C_{331} + C_{333} + 2 C_{346} + C_{349} + C_{350} + C_{352} + \nn \\
 & & C_{358} + C_{360} + C_{373} + C_{374} + 2 C_{377} + C_{379} + 2 C_{386} + C_{387} \nn \\
\binom{C_{16}}{1} \binom{C_{0} - 1}{1} & = & C_{68} + C_{129} + C_{135} + 2 C_{139} + C_{168} + C_{205} + C_{211} + 2 C_{215} + C_{272} + C_{275} + C_{278} + C_{285} +  \\
 & & C_{289} + C_{294} + C_{325} + C_{333} + 2 C_{335} + C_{341} + C_{352} + C_{356} + C_{358} + C_{360} + C_{362} + 2 C_{364} + \nn \\
 & & 2 C_{365} + C_{375} + C_{379} + C_{380} + 2 C_{383} + C_{387} + 2 C_{389} \nn \\ 
\binom{C_{17}}{1} \binom{C_{0}}{1} & = & C_{132} + C_{138} + C_{208} + C_{214} + C_{259} + C_{262} + C_{304} + C_{306} + C_{326} + C_{334} + C_{342} + C_{353} +  \\
 & & C_{359} + C_{363} + C_{376} + C_{381} + C_{382} + C_{388} + C_{397} + C_{398} \nn \\ 
\binom{C_{18}}{1} \binom{C_{0}}{1} & = & C_{105} + C_{122} + C_{202} + C_{218} + C_{247} + C_{251} + C_{255} + C_{288} + C_{297} + C_{395} \\
\binom{C_{19}}{1} \binom{C_{0}}{1} & = & C_{84} + C_{104} + C_{121} + C_{140} + C_{176} + C_{200} + C_{201} + C_{216} + C_{227} + C_{250} + C_{254} + C_{270} +  \\
 & & C_{286} + C_{295} + C_{309} + C_{392} \nn \\
\binom{C_{20}}{1} \binom{C_{0}}{1} & = & C_{265} + C_{298} + C_{301} + C_{308} + C_{368} + C_{370} + C_{371} + C_{384} + C_{390} + C_{399} \\  
\binom{C_{2}}{1} \binom{C_{1}}{1} & = & C_{26} + C_{60} + C_{64} + C_{68} + C_{123} + C_{125} + C_{129} + C_{133} + C_{135} + C_{139} \\
\binom{C_{3}}{1} \binom{C_{1}}{1} & = & C_{36} + C_{46} + C_{57} + C_{69} + C_{76} + C_{81} + C_{84} + C_{94} + C_{101} + C_{104} + C_{114} + C_{118} + C_{121} + \\
 & & C_{131} + C_{137} + C_{140} \nn  \\
\binom{C_{4}}{1} \binom{C_{1}}{1} & = & C_{128} + C_{142} + C_{160} + C_{164} + C_{168} + C_{203} + C_{205} + C_{209} + C_{211} + C_{215} \\
\binom{C_{5}}{1} \binom{C_{1}}{1} & = & C_{59} + C_{82} + C_{148} + C_{158} + C_{169} + C_{174} + C_{176} + C_{183} + C_{189} + C_{192} + C_{196} + C_{200} +  \\
 & & C_{201} + C_{207} + C_{213} + C_{216} \nn \\
\binom{C_{6}}{1} \binom{C_{1} - 1}{1} & = & C_{132} + C_{138} + C_{208} + C_{214} + 2 C_{218} + C_{247} + C_{251} + C_{255} + 2 C_{259} + 2 C_{262} + 2 C_{265} + \\
 & & C_{298} + C_{301} + C_{304} + C_{306} + C_{308} \nn \\
\binom{C_{7}}{1} \binom{C_{1}}{1} & = & C_{221} + C_{238} + C_{245} + C_{256} + C_{268} + C_{269} + C_{270} + C_{280} + C_{282} + C_{283} + C_{286} + C_{293} + \\
 & & C_{295} + C_{305} + C_{307} + C_{309} \nn \\ 
\binom{C_{8}}{1} \binom{C_{1} - 1}{1} & = & C_{71} + 2 C_{85} + C_{171} + 2 C_{177} + C_{229} + C_{267} + C_{271} + 2 C_{274} + 2 C_{277} + C_{287} + 2 C_{291} +  \\
 & & C_{296} + 2 C_{300} + 2 C_{303} + C_{310} \nn
\end{eqnarray}
}

{
\tiny
\begin{eqnarray}
\binom{C_{9}}{1} \binom{C_{1}}{1} & = & C_{95} + C_{184} + C_{231} + C_{272} + C_{316} + C_{322} + C_{325} + C_{330} + C_{333} + C_{335} \\ 
\binom{C_{10}}{1} \binom{C_{1}}{1} & = & C_{119} + C_{194} + C_{234} + C_{275} + C_{331} + C_{350} + C_{354} + C_{356} + C_{360} + C_{364} \\ 
\binom{C_{11}}{1} \binom{C_{1}}{1} & = & C_{102} + C_{115} + C_{190} + C_{198} + C_{240} + C_{278} + C_{289} + C_{323} + C_{339} + C_{341} + C_{346} + C_{349} +  \\
 & & C_{352} + C_{358} + C_{362} + C_{365} \nn \\
\binom{C_{12}}{1} \binom{C_{1}}{1} & = & C_{237} + C_{279} + C_{281} + C_{285} + C_{372} + C_{373} + C_{375} + C_{377} + C_{379} + C_{383} \\
\binom{C_{13}}{1} \binom{C_{1} - 1}{1} & = & 2 C_{105} + C_{132} + C_{202} + C_{208} + C_{247} + C_{288} + C_{298} + 2 C_{326} + 2 C_{342} + C_{359} + C_{363} +  \\
 & & 2 C_{368} + C_{370} + C_{376} + C_{381} + C_{384} \nn \\ 
\binom{C_{14}}{1} \binom{C_{1}}{1} & = & C_{244} + C_{284} + C_{292} + C_{294} + C_{374} + C_{378} + C_{380} + C_{386} + C_{387} + C_{389} \\ 
\binom{C_{15}}{1} \binom{C_{1} - 1}{1} & = & 2 C_{122} + C_{138} + C_{202} + C_{214} + C_{251} + C_{297} + C_{301} + 2 C_{334} + 2 C_{353} + C_{359} + C_{363} + \\
 & & C_{370} + 2 C_{371} + C_{382} + C_{388} + C_{390} \nn  \\
\binom{C_{16}}{1} \binom{C_{1} - 1}{1} & = & C_{255} + C_{288} + C_{297} + C_{304} + C_{306} + C_{308} + C_{376} + C_{381} + C_{382} + C_{384} + C_{388} + C_{390} +  \\
 & & 2 C_{395} + 2 C_{397} + 2 C_{398} + 2 C_{399} \nn \\ 
\binom{C_{17}}{1} \binom{C_{1} - 2}{1} & = & 2 C_{141} + 2 C_{217} + C_{311} + 3 C_{336} + 3 C_{366} + 2 C_{385} + 2 C_{391} + C_{393} + C_{400} \\
\binom{C_{18}}{1} \binom{C_{1} - 2}{1} & = & C_{141} + C_{217} + C_{311} + 2 C_{393} + 3 C_{401} + 2 C_{405} + C_{409} \\  
\binom{C_{19}}{1} \binom{C_{1} - 1}{1} & = & C_{258} + C_{271} + C_{287} + C_{296} + C_{310} + 2 C_{396} + C_{403} + C_{406} + 2 C_{408} + 2 C_{410} \\   
\binom{C_{20}}{1} \binom{C_{1} - 2}{1} & = & C_{311} + C_{385} + C_{391} + 2 C_{400} + C_{405} + 2 C_{409} + 3 C_{411}      
\end{eqnarray}
}

\clearpage

\section{Number of generated orbit equations}

Table \ref{table:no_of_eqs_nl_isom} shows the number of orbit equations generated by our method for graphlets with a specific number of layers and nodes when using node-layer isomorphism or node isomorphism. These set the upper bounds for how many \emph{independent} equations can be found.

\begin{table}[h!]
\centering
\caption{Number of generated equations for graphlets with given number of layers and nodes with node-layer and node isomorphisms.}
\begin{tabular}{cc|ccc|ccc}
 & \multicolumn{1}{c}{} & \multicolumn{3}{c}{Node-layer} & \multicolumn{3}{c}{Node} \\
 & Nodes & 2 & 3 & 4 & 2 & 3 & 4\\
\cline{2-8}
\multirow{3}{*}{\rotatebox[origin=c]{90}{Layers}} & 1 & 0 & 1 & 3 & 0 & 1 & 3 \\
 & 2 & 0 & 3 & 38 & 0 & 6 & 99\\
 & 3 & 0 & 6 & 201 & 0 & 28 & 1911
\end{tabular}
\label{table:no_of_eqs_nl_isom}
\end{table}

\section{Combining more than two orbits}

In the main article, we have only discussed how to generate equations where two orbits are combined, even though one could form equations where three or more orbits are combined.
However, these equations depend on the equations where two orbits are combined and can be derived from them.
Thus, they cannot be used to remove redundant orbits from the graphlet degree vectors.

In general, the equations combining any number of orbits can be expressed in the following form
\begin{eqnarray}
\binom{C_{x_1}}{c_1} \binom{C_{x_2} - b_2}{c_2} \cdots \binom{C_{x_l} - b_l}{c_l} & = & a_1 C_{y_1} + \ldots + a_k C_{y_k} \ . \label{eq_general}
\end{eqnarray}

In three-orbit equations, i.e. equations where three orbits are combined, all the combined orbits can be different, two of the orbits can be the same or all of them can be the same.
When all the orbits are the same, the three-orbit equation $\binom{C_x}{3}$ can be derived from equation $\binom{C_x}{2}$ by multiplying both sides of the equation by $(C_x - 2) / 3$.
Equation $\binom{C_{x_1}}{1} \binom{C_{x_2}-b}{2}$ can be derived by multiplying equation $\binom{C_{x_1}}{1} \binom{C_{x_2}-b}{1}$ by $(C_{x_2}-b-1)/2$, and equation $\binom{C_{x_1}}{1} \binom{C_{x_2}-b_2}{1} \binom{C_{x_3}-b_3}{1}$ by multiplying $\binom{C_{x_1}}{1} \binom{C_{x_2}-b_2}{1}$ by $C_{x_3}-b_3$.
The left sides of the equations become of the desired form, and the right sides of the equations become of the form $a_1 C_{y_1} (C_x-b) + \ldots + a_k C_{y_k} (C_x - b)$.
The terms $C_y (C_x - b)$ correspond to two-orbit equations that have been defined already and can be replaced by the right sides of the equations to get the three-orbit equation in the form of equation~\ref{eq_general}.

For the case $\binom{C_x}{3}$, there is only one equation it can be derived from, $\binom{C_x}{2}$.
In addition to that equation, it will also depend on the equations emerging in the right side of the multiplied equation.
For the case $\binom{C_{x_1}}{1} \binom{C_{x_2}-b}{2}$, there are two possible equations it can be derived from, equation $\binom{C_{x_1}}{1} \binom{C_{x_2}-b}{1}$ or equation $\binom{C_{x_2}}{2}$.
When $b>0$, the equation is not in form~\ref{eq_general}, but with some rearranging we obtain
\begin{eqnarray}
\binom{C_{x_2}}{2} \binom{C_{x_1}}{1} & = & \frac{1}{2} C_{x_1} C_{x_2} (C_{x_2} - 1) \nn \\
& = & \frac{1}{2} C_{x_1} C_{x_2} (C_{x_2} - b) + \frac{1}{2} (b-1) C_{x_1} C_{x_2} \nn \\
& = & \frac{1}{2} C_{x_1} (C_{x_2} - b-1)(C_{x_2} - b) + \frac{1}{2} (b+1) C_{x_1} (C_{x_2} - b) \nn \\
& & + \frac{1}{2} (b-1) C_{x_1} (C_{x_2} - b) + \frac{1}{2} b (b-1) C_{x_1} \nn \\
& = & \binom{C_{x_1}}{1} \binom{C_{x_2} - b}{2} + b \binom{C_{x_1}}{1} \binom{C_{x_2} - b}{1} + \frac{1}{2} b (b-1) C_{x_1} \ . \nn
\end{eqnarray}
Therefore, the equation $\binom{C_{x_1}}{1} \binom{C_{x_2} - b}{2}$ depends on equation $\binom{C_{x_1}}{1} \binom{C_{x_2} - b}{1}$ in addition to equation $\binom{C_{x_2}}{2}$, and the equations obtained by multiplying the right side of that equation by $C_{x_1}$.
For the equation $\binom{C_{x_1}}{1} \binom{C_{x_2}-b_2}{1} \binom{C_{x_3}-b_3}{1}$, there are three possibilities.

Equations combining more than three orbits can be derived in a similar manner from equations combining fewer nodes.
These equations will not affect the independency of two-orbit equations, and thus will not be discussed in more detail.

When inspecting which equations can be used to derive two-orbit equations, one can search for one of the two orbits from the right sides of the equations and multiply this equation by the other orbit count.
Then the equation depends on all the other two-orbit equations emerging on the right side of the equation and the three-orbit equation appearing on the left side of the equation.
The other option would be to search one of the orbit counts from the left sides of the equation and analogously multiply that equation by the other orbit count.
The formed equation could then be divided by the third orbit count to achieve the desired form on the left side of the equation.
However, on the right side of the equation there would be divisions by orbit counts for which we do not have defined equations.

\section{Finding independent equations}
\label{section:independent_eqs}

There are two-orbit equations that can be set independent for certain.
The equations that contain orbit counts that do not exist in any other equation cannot be derived from the other equations, and thus are independent.
Moreover, the equations where the orbit combination produces graphlets of size at most three nodes can be set independent.
These equations do not depend on each other, since each resulting two-star will represent a different graphlet, and one can investigate which of the equations can be derived from these equations.

To discover a set of independent equations, a network is formed, where the equations represent the nodes and there is a directed edge from equation $e_1$ to equation $e_2$ if equation $e_2$ is required for the derivation of equation $e_1$.
Next, a directed acyclic graph (DAG) is formed out of the strongly connected components of the network.
The components are processed in the linearized order of the DAG starting from sinks and proceeding to sources.
This way the equations will only depend on equations within the same component and equations in the earlier components of the linearized order.

The networks in sinks are first set independent.
Within other components, an equation can be set dependent if the equations it has links to have already been determined either independent or dependent.
If the component forms a complete graph, one of the equations can be set dependent and the rest are set independent.
If the component contains a three-orbit equation, it should be set dependent, since those equations are not explicitly generated.

If the component does not form a complete graph, one of the equations, $e_0$, in that component is selected and set dependent.
If the component contains three-orbit equations, these should be selected first.
Next, another equation, $e_1$, is selected.
If it can be derived from the other equations in the component and equations in the earlier components without the first selected equation $e_0$, then the equation $e_1$ is also set dependent.
Otherwise, $e_1$ is left in the component and another equation is selected as $e_1$.
One continues this way selecting equations and determining whether they can be derived from the remaining equations to find the largest set of equations that can be set dependent in the component.

\clearpage

\section{Calculating graphlet correlation distance (GCD)}

Consider for example a task of computing distances between networks based on their node-based orbits \cite{prvzulj2007biological}. In this task, 
the $i$th graphlet degree of a node depicts the number of graphlets in the network where the node is in orbit $i$.
These graphet degrees can be collected in a vector called the graphlet degree vector (GDV) \cite{prvzulj2007biological}.
One can reduce the number of orbits to be included in the computation of these measures by using dependency equations (demonstrated in the multiplex network comparison section). For each independent equation, one orbit can be ignored altogether, reducing the size of the GDVs.
The graphlet correlation matrix of a network (GCM) \cite{yaverouglu2014revealing} is obtained by computing Spearman's correlation coefficients between vectors containing the orbit counts for all the nodes in the network. The dimensions of the GCM are therefore \textit{number of orbits} $\times$ \textit{number of orbits}, enabling the comparison of networks of different sizes.
The correlation coefficient cannot be computed if all the counts for a certain orbit are equal, therefore a 'dummy node' with all orbit counts equal to one is added to the set of GDVs.
These matrices for different networks can then be compared with a statistic called the graphlet correlation distance (GCD) \cite{yaverouglu2014revealing} which is obtained by taking the Euclidean distance between the upper triangles of the GCMs,
\begin{equation}
GCD(G, H) = \sqrt{\sum_{i=1}^{n_{orb}} \sum_{j=i+1}^{n_{orb}} (GCM_G(i,j) - GCM_H(i,j))^2} \,,
\end{equation}
where $n_{orb}$ is the number of orbits used to compute the GCMs.
The value of GCD is dependent on the set of orbits we choose to include in the GCMs. One common way of choosing orbits is to select a maximum size for the graphlets and include the orbits within all graphlets of at most that size. In multilayer networks, as discussed before, the graphlet size can be defined by giving a size for each aspect. The more aspects there are, the more choices have to be made to define the set of orbits to be used in the measure.

\section{Additional figures for GCD}

\begin{figure}[H]
\begin{center}
\includegraphics[width=0.8\textwidth]{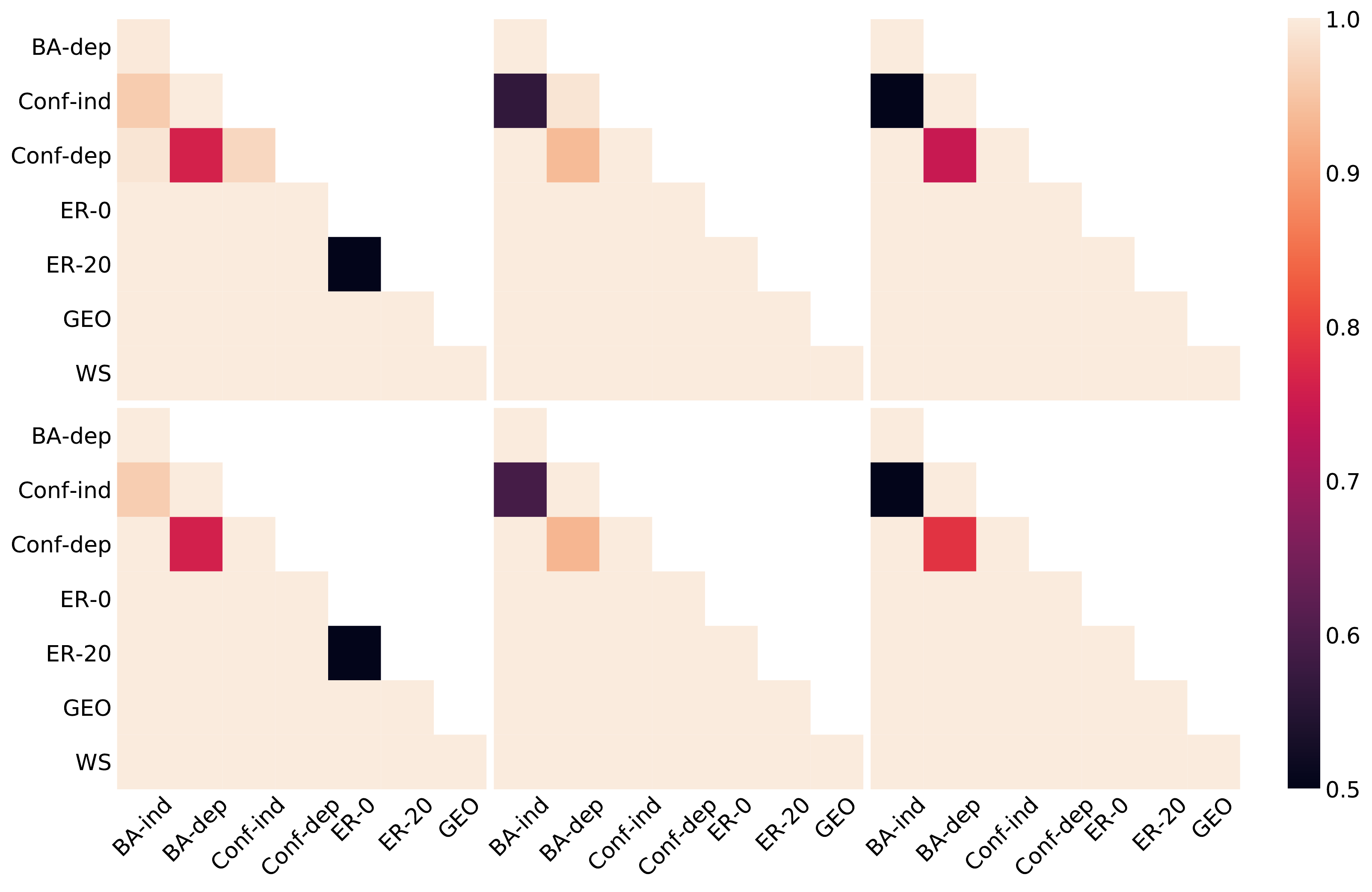}
\end{center}
\caption{Constant degree pairwise AUPRs}
\label{fig:fixed-auprs}
\end{figure}

\begin{figure}[H]
\begin{center}
\includegraphics[width=0.8\textwidth]{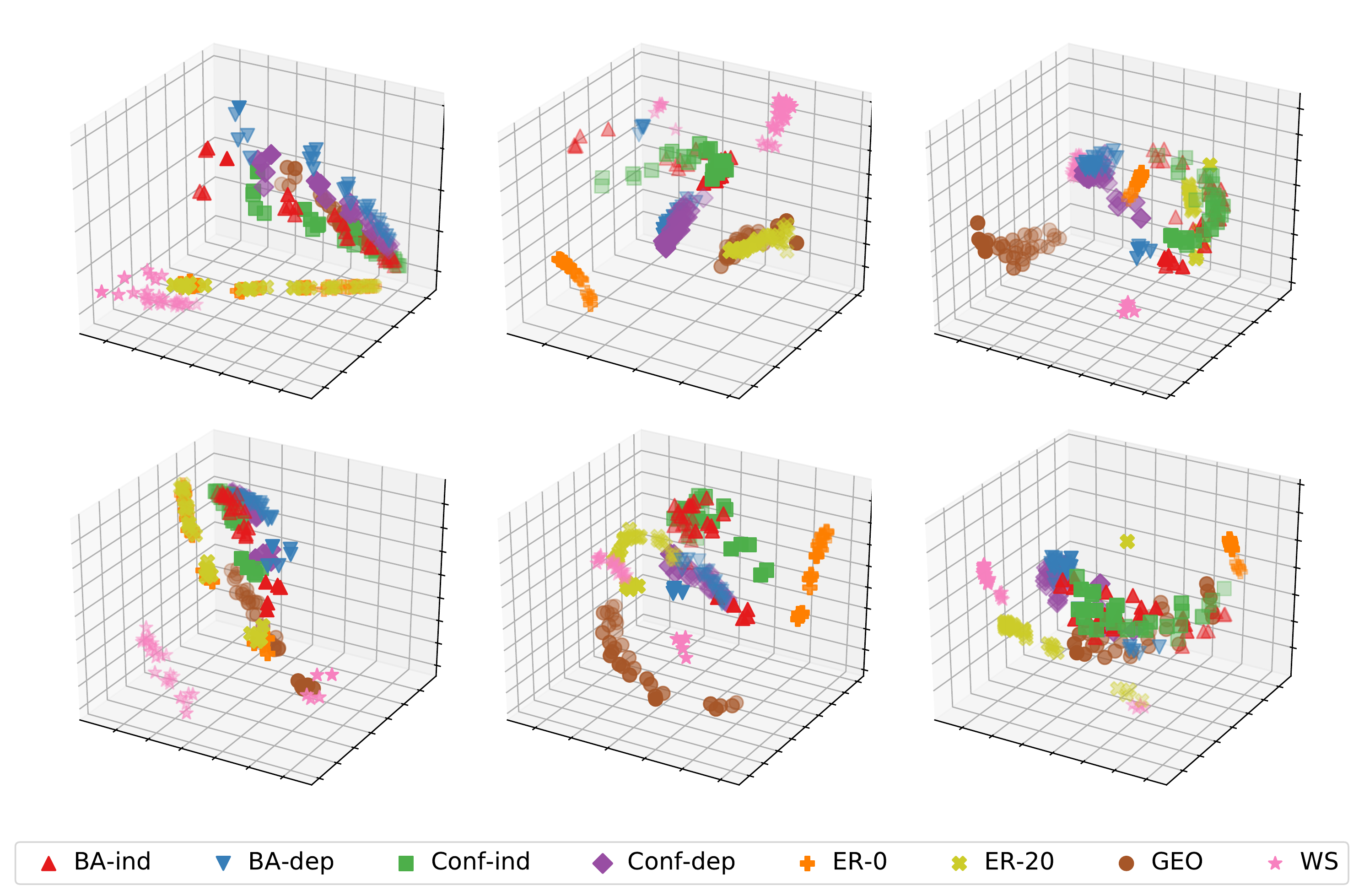}
\end{center}
\caption{Degree progression MDS}
\label{fig:degprog-mds}
\end{figure}

\begin{figure}[H]
\begin{center}
\includegraphics[width=0.8\textwidth]{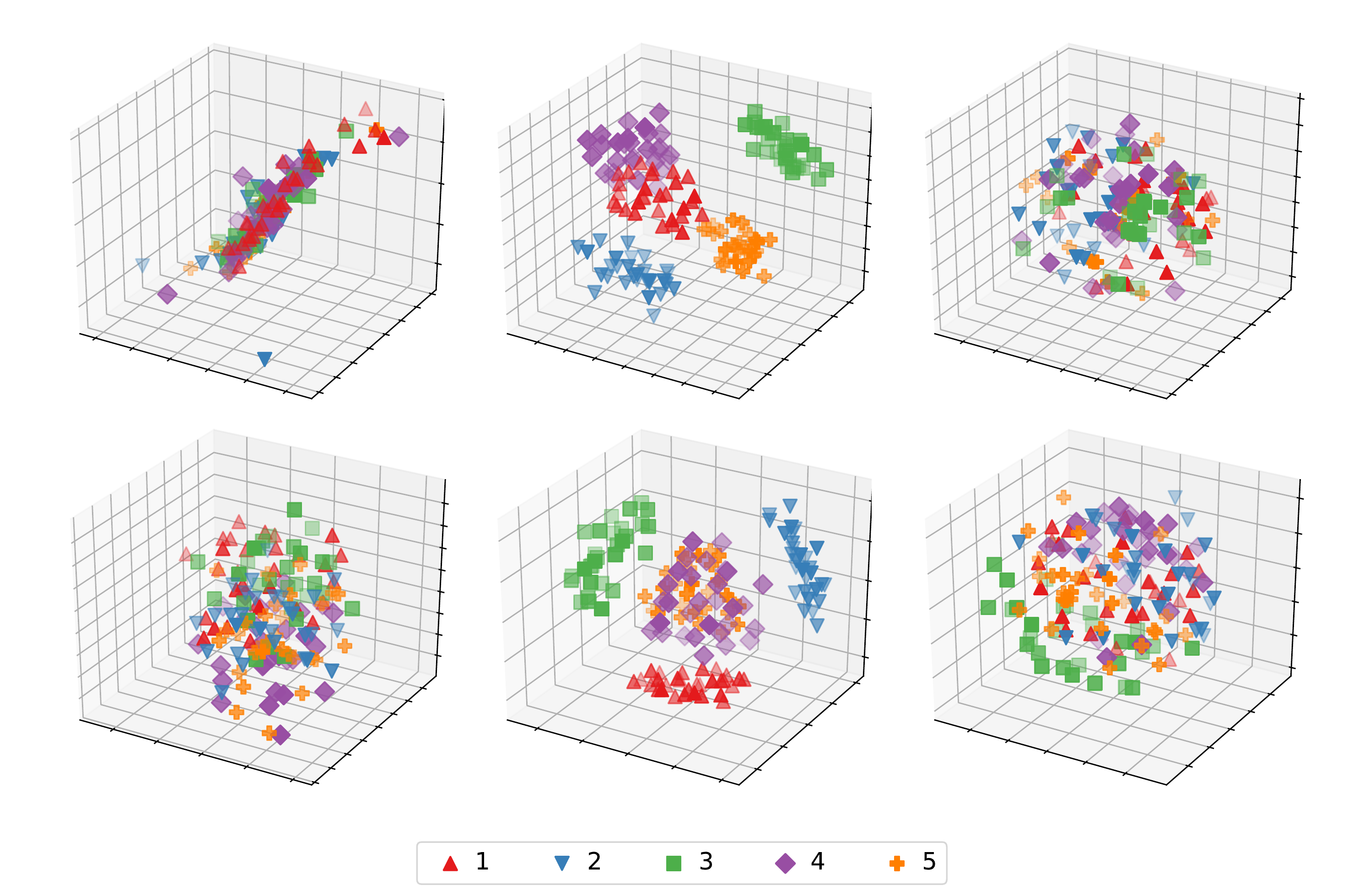}
\end{center}
\caption{4-node-2-layer graphlet insertion MDS}
\label{fig:graphlet42-mds}
\end{figure}

\begin{figure}[H]
\begin{center}
\includegraphics[width=0.8\textwidth]{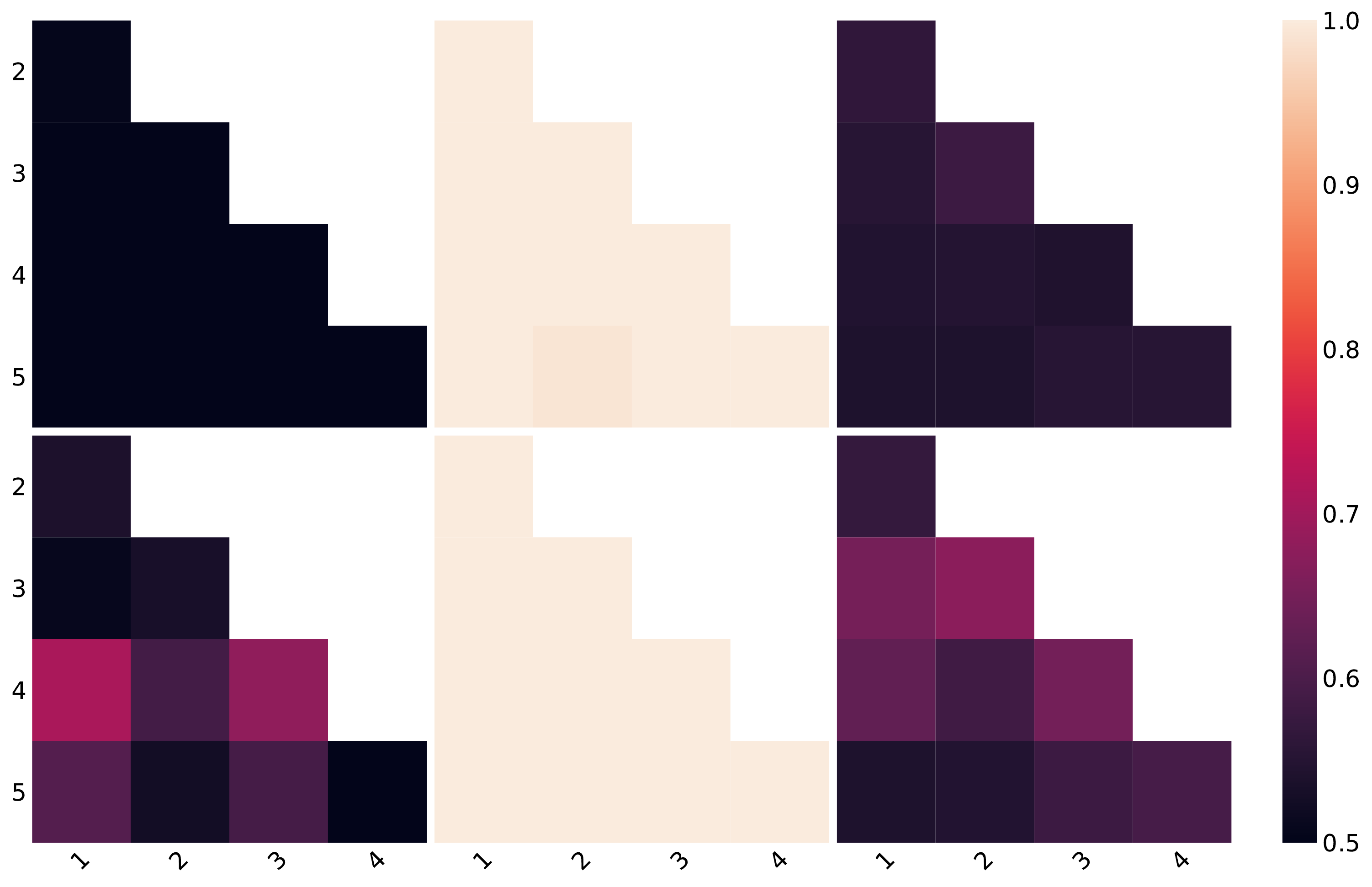}
\end{center}
\caption{4-node-2-layer graphlet insertion pairwise AUPRs}
\label{fig:graphlet42-auprs}
\end{figure}

In the main article, the precision-recall curves were shown for the constant degree random models, degree progression random models, and 4-node-2-layer graphlet insertion networks, with GCD as the distance measure. Additionally, MDS embedding for the constant degree random models and pairwise AUPRs for the degree progression random models were shown, also with GCD. Here, we list the rest of the figures with GCD that were not included in the main article: pairwise AUPRs for the constant degree random models (Figure \ref{fig:fixed-auprs}); MDS embedding for the degree progression random models (Figure \ref{fig:degprog-mds}); and MDS embedding (Figure \ref{fig:graphlet42-mds}) and pairwise AUPRs (Figure \ref{fig:graphlet42-auprs}) for the 4-node-2-layer graphlet insertion networks.

\begin{figure}[H]
\begin{center}
\includegraphics[width=0.8\textwidth]{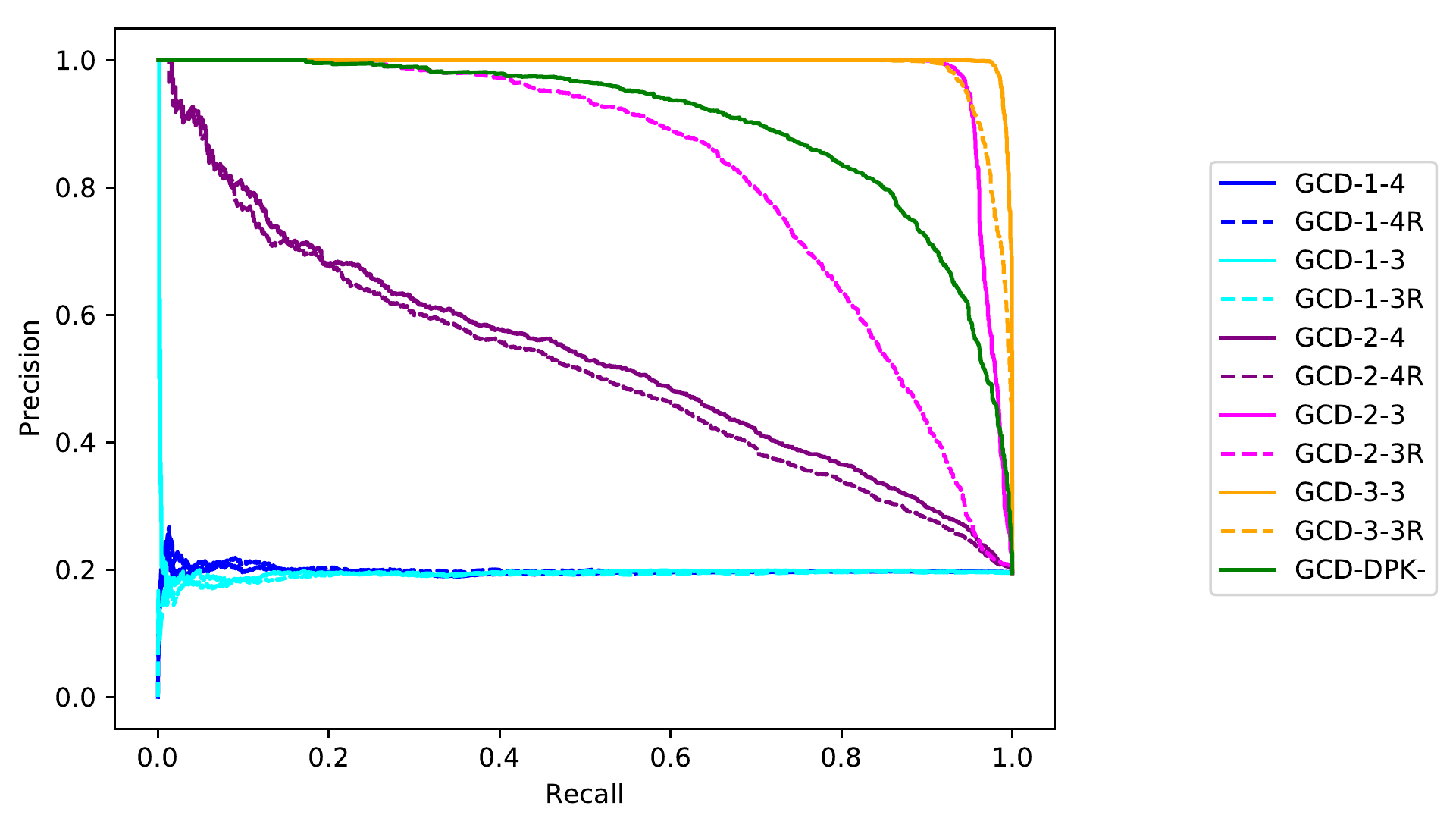}
\end{center}
\caption{3-node-3-layer graphlet insertion precision-recall}
\label{fig:graphlet33-prerec}
\end{figure}

\begin{figure}[H]
\begin{center}
\includegraphics[width=0.8\textwidth]{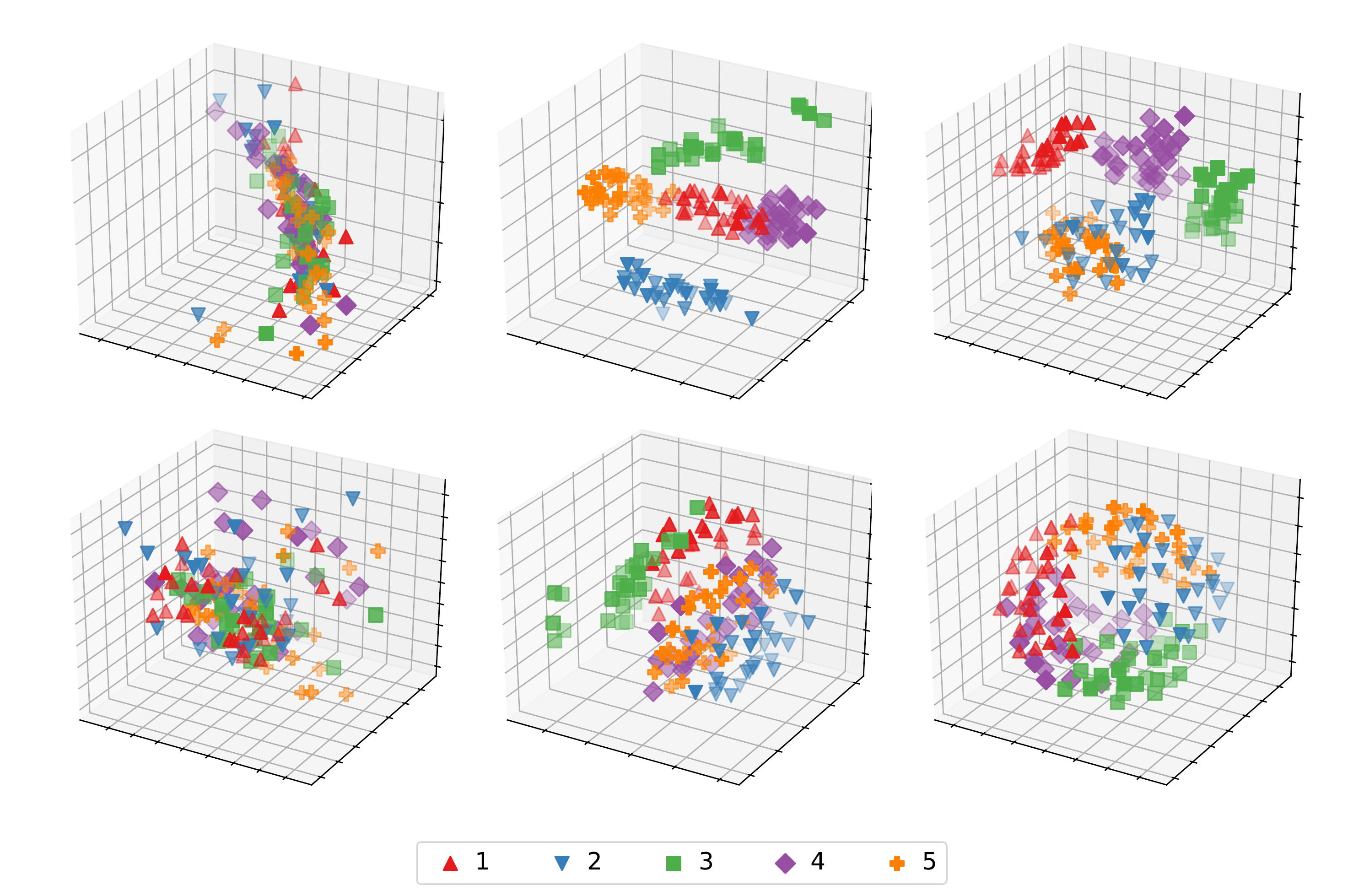}
\end{center}
\caption{3-node-3-layer graphlet insertion MDS}
\label{fig:graphlet33-mds}
\end{figure}

\begin{figure}[H]
\begin{center}
\includegraphics[width=0.8\textwidth]{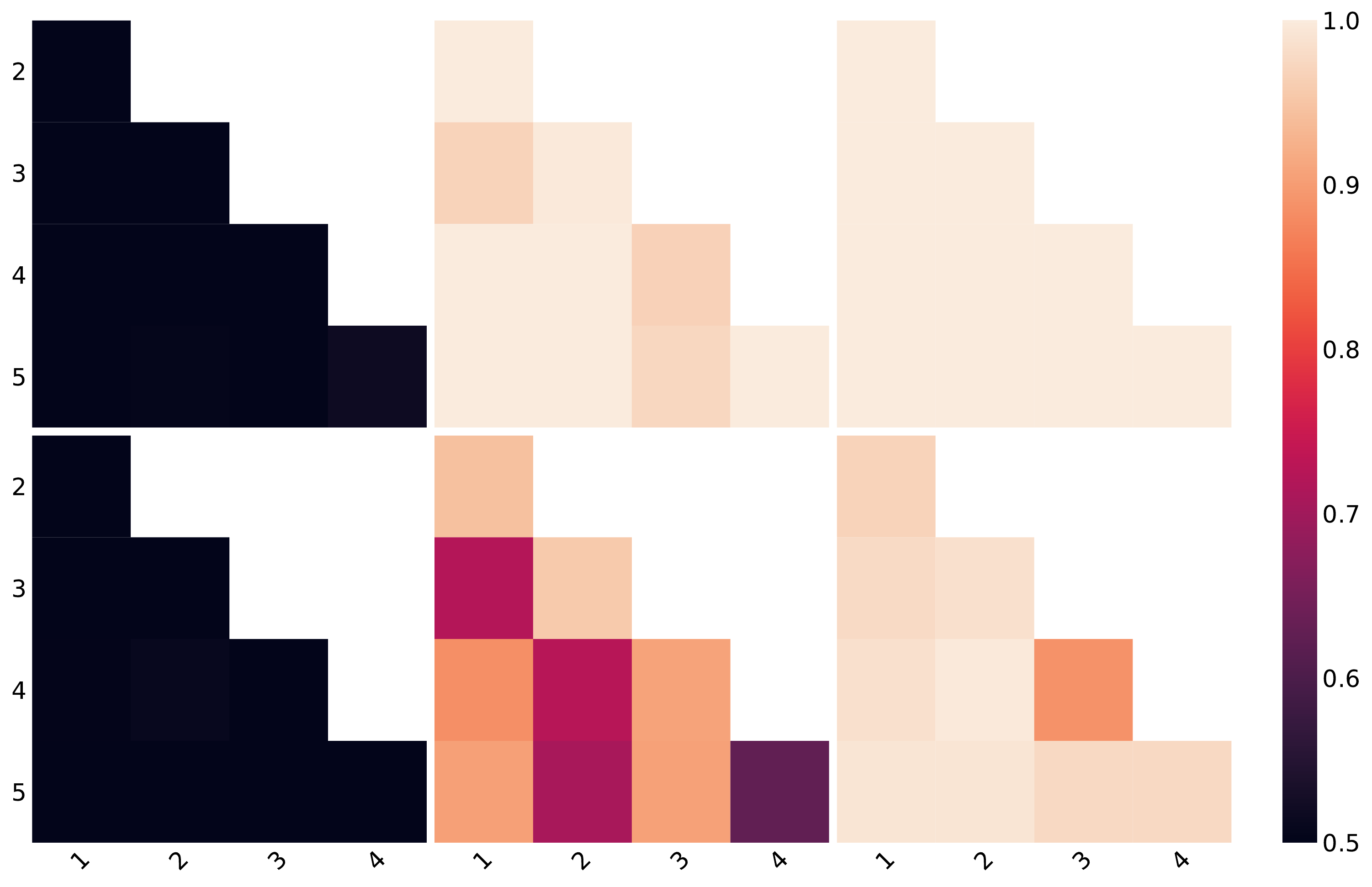}
\end{center}
\caption{3-node-3-layer graphlet insertion pairwise AUPRs}
\label{fig:graphlet33-auprs}
\end{figure}

Additionally, we run the analysis for graphlet insertion networks where 3-node-3-layer graphlets were inserted. The procedure is otherwise the same as in the 4-node-2-layer case, except now we only insert 10 different graphlets (instead of 20), because there are much fewer 3-node-3-layer graphlets than 4-node-2-layer graphlets. Three instances of each graphlet were inserted. The resulting precision-recall curves (Figure \ref{fig:graphlet33-prerec}), MDS embedding (Figure \ref{fig:graphlet33-mds}) and pairwise AUPRs (Figure \ref{fig:graphlet33-auprs}) show that, as expected, GCD-3-3 performs the best in this case.

In Figures \ref{fig:fixed-auprs}, \ref{fig:degprog-mds}, \ref{fig:graphlet42-mds}, \ref{fig:graphlet42-auprs}, \ref{fig:graphlet33-mds}, and \ref{fig:graphlet33-auprs} we use the following convention for subfigures: \textbf{top left:} GCD-1-3; \textbf{top middle:} GCD-2-3; \textbf{top right:} GCD-3-3; \textbf{bottom left:} GCD-1-4; \textbf{bottom middle:} GCD-2-4; \textbf{bottom right:} GCD-DPK.

\section{Additional network distance measures}

\subsection{NetEmd}
NetEmd \cite{wegner2017identifying} is a distance measure which is based on calculating earth mover's distance (EMD) between network feature distributions. Here, we use the graphlet degree distributions \cite{prvzulj2007biological} of the networks as the features. For a set $T = \{t_1,t_2,...,t_m\}$ of network measures, NetEmd between networks $G$ and $G'$ is defined as follows:
\[
NetEmd_T(G,G') = \frac{1}{m}\sum_{j=1}^m NetEmd_{t_j}(G,G')
\]
where
\[
NetEmd_t(G,G') = EMD^*(p_t(G),p_t(G'))
\]
where $p_t(G)$ and $p_t(G'))$ are the distributions of feature $t$ on $G$ and $G'$, respectively, and
\[
EMD^*(p,q) = \inf_{c \in \mathbb{R}}(EMD(\tilde{p}(\cdot+c),\tilde{q}(\cdot)))
\]
where $\tilde{p}$ and $\tilde{q}$ are obtained by rescaling $p$ and $q$ to have variance 1, and
\[
EMD(p,q) = \int_{-\infty}^{\infty} |F(x) - G(x)|
\]
where $F(x)$ and $G(x)$ are the cumulative distribution functions of $p$ and $q$, respectively.

Here, $p_t(G)$ is the $t$th graphlet degree distribution of $G$. Despite the name, the distribution corresponds to the $t$th automorphism orbit distribution within the graphlet set considered.

Here, NetEmd is implemented using pyemd \cite{pele2008,pele2009} for Python for EMD calculation and $scipy.optimize.minimize\_scalar$ \cite{scipy} for minimization in $EMD^*(p,q)$ calculation. Pyemd requires the construction of a histogram for each of the distributions to be compared, and as such, a choice of the number of histogram bins has to be made. The number 10 for the number of bins is chosen, since it is the default for $numpy.histogram$.

\begin{figure}[H]
\begin{center}
\includegraphics[width=0.8\textwidth]{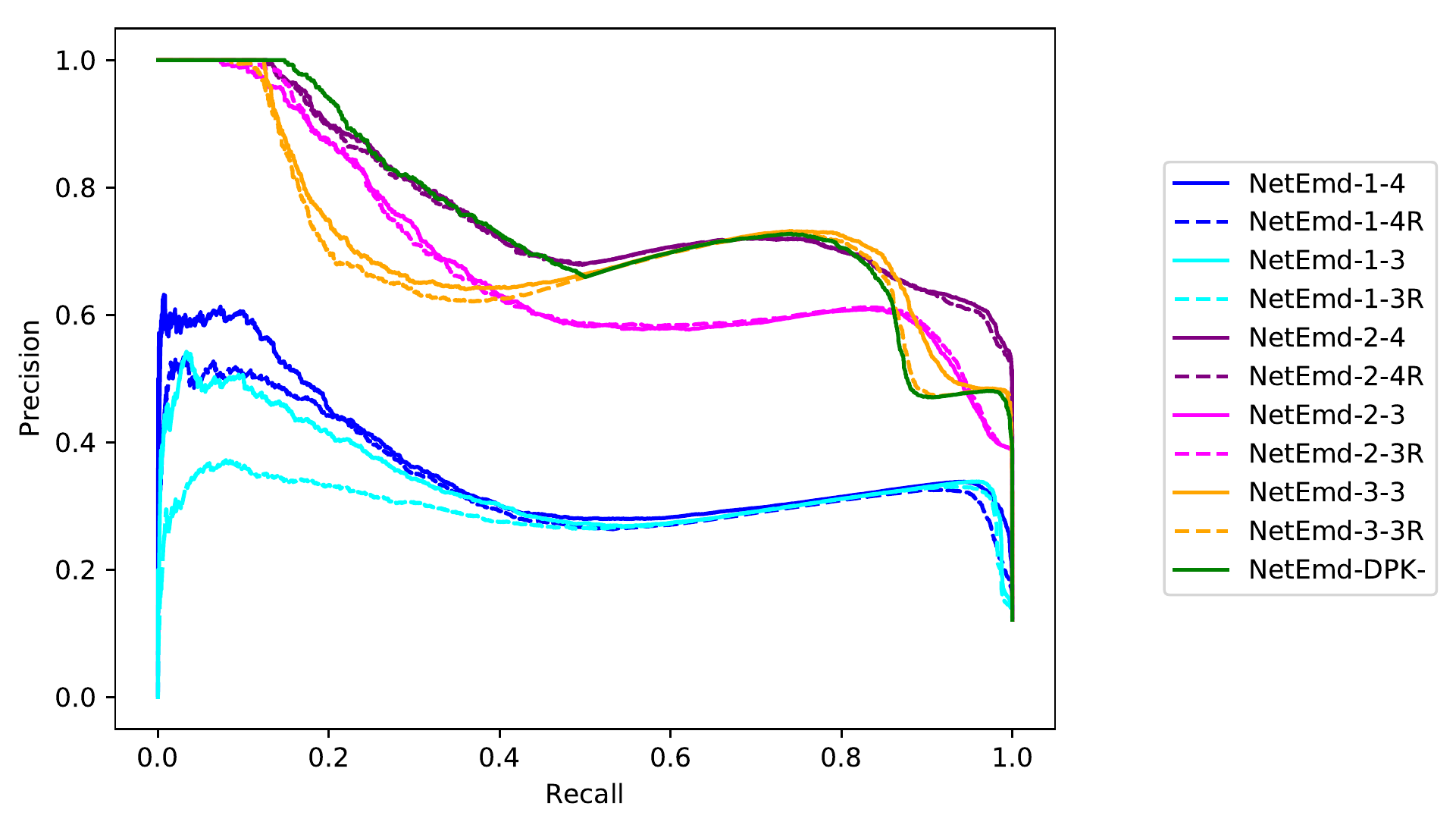}
\end{center}
\caption{NetEmd Constant degree precision-recall}
\label{fig:netemd-fixed-prerec}
\end{figure}

\begin{figure}[H]
\begin{center}
\includegraphics[width=0.8\textwidth]{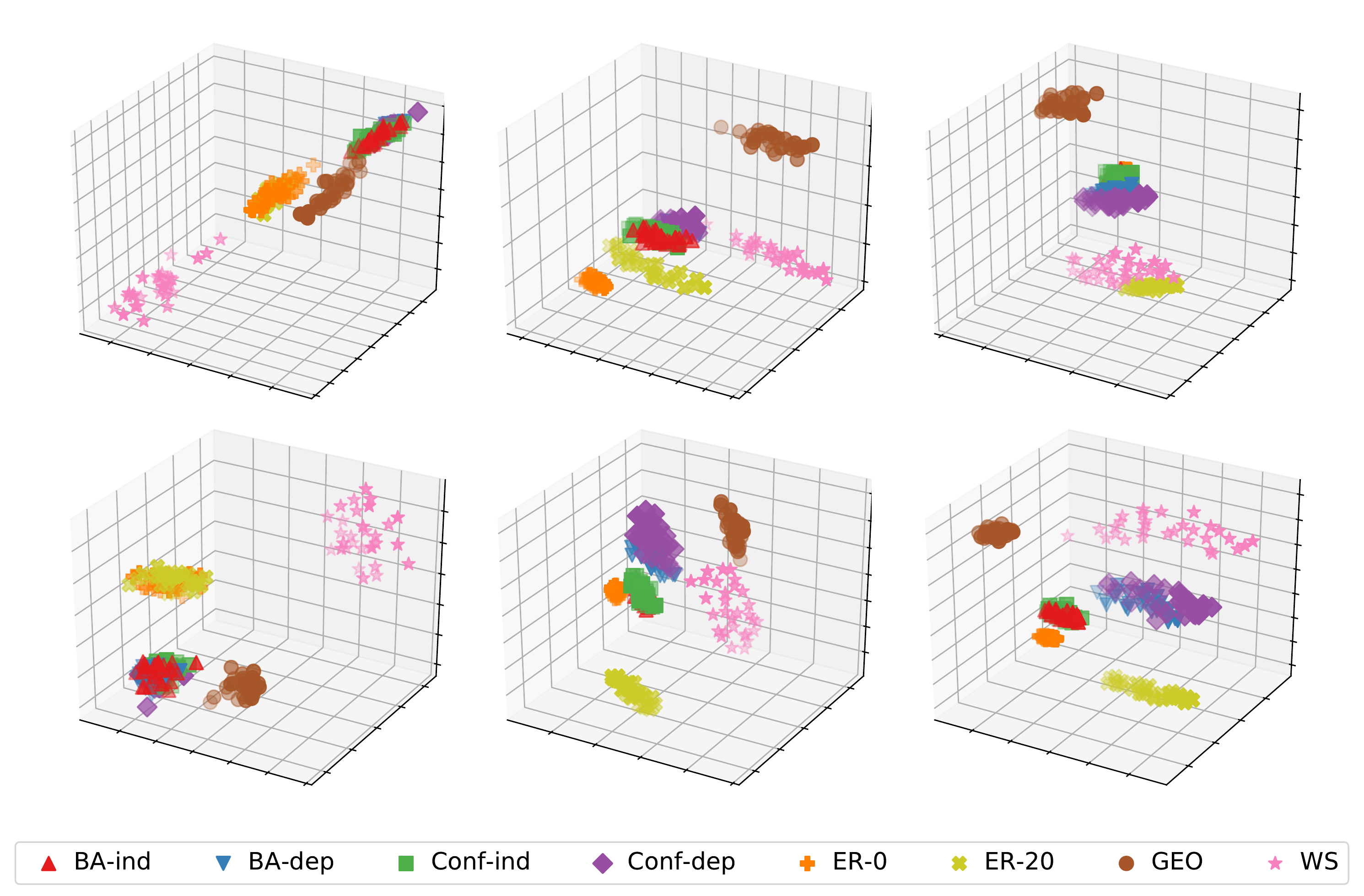}
\end{center}
\caption{NetEmd Constant degree MDS}
\label{fig:netemd-fixed-mds}
\end{figure}

\begin{figure}[H]
\begin{center}
\includegraphics[width=0.8\textwidth]{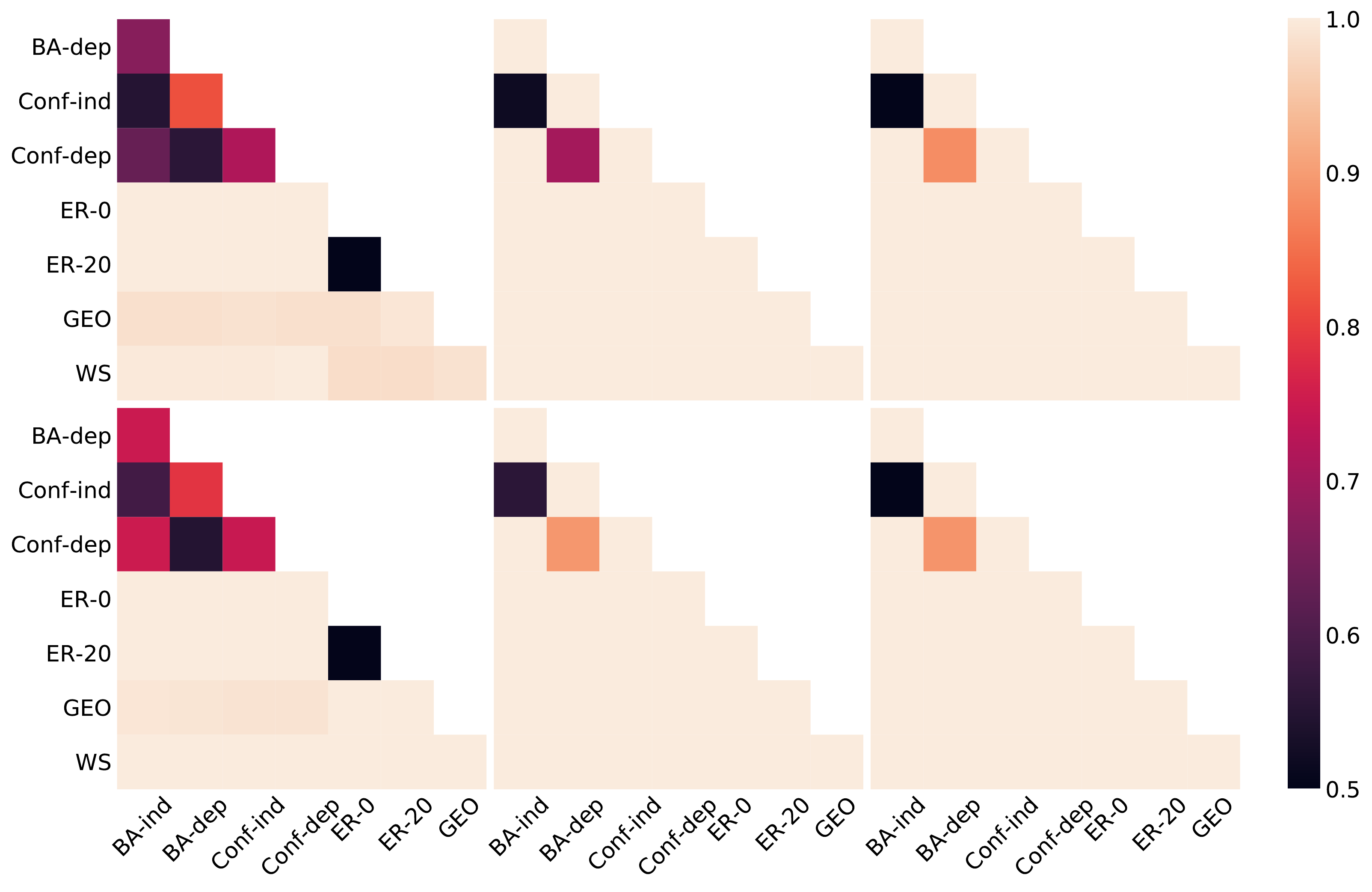}
\end{center}
\caption{NetEmd Constant degree pairwise AUPRs}
\label{fig:netemd-fixed-auprs}
\end{figure}

\begin{figure}[H]
\begin{center}
\includegraphics[width=0.8\textwidth]{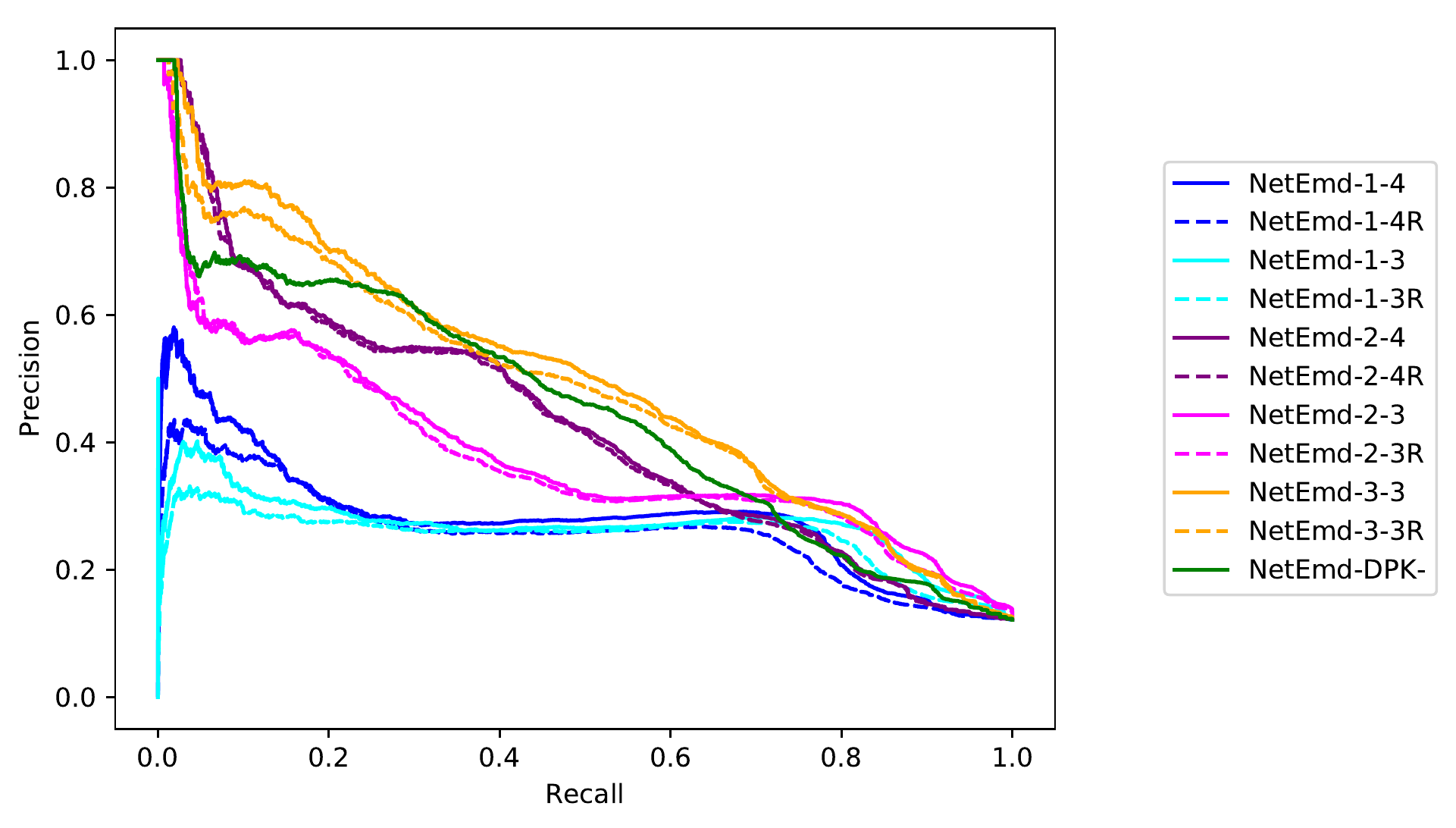}
\end{center}
\caption{NetEmd Degree progression precision-recall}
\label{fig:netemd-degprog-prerec}
\end{figure}

\begin{figure}[H]
\begin{center}
\includegraphics[width=0.8\textwidth]{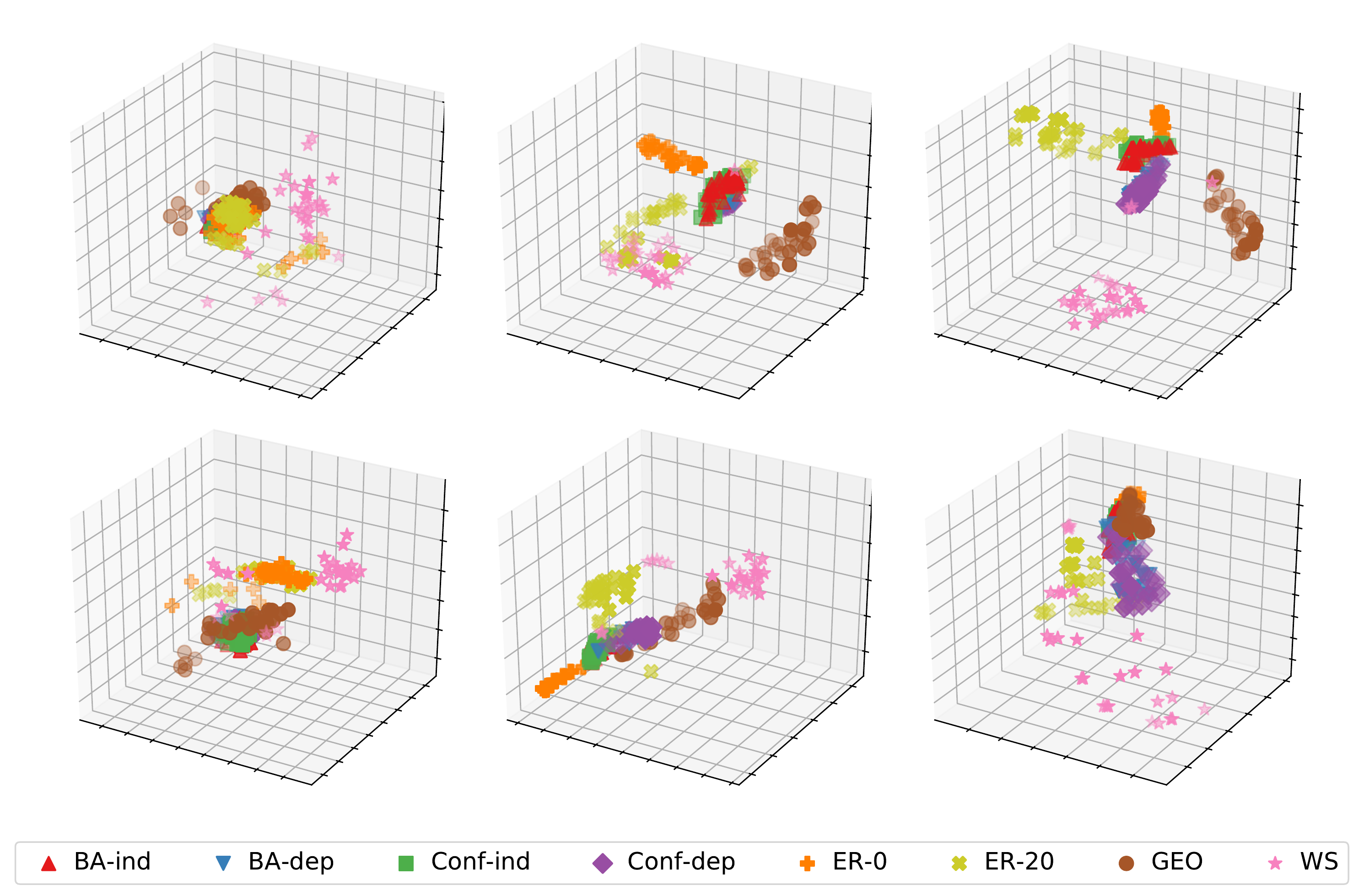}
\end{center}
\caption{NetEmd Degree progression MDS}
\label{fig:netemd-degprog-mds}
\end{figure}

\begin{figure}[H]
\begin{center}
\includegraphics[width=0.8\textwidth]{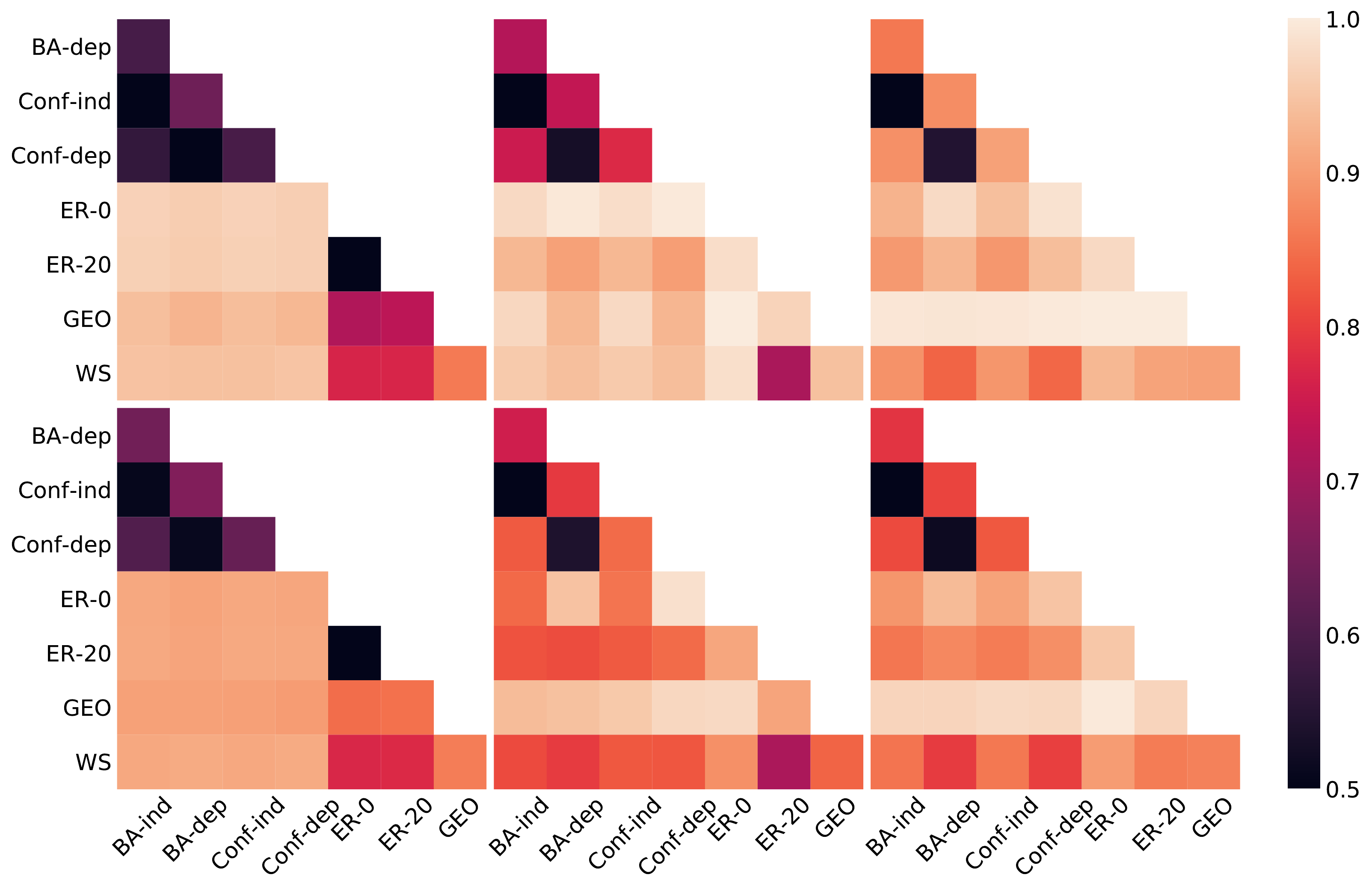}
\end{center}
\caption{NetEmd Degree progression pairwise AUPRs}
\label{fig:netemd-degprog-auprs}
\end{figure}

\begin{figure}[H]
\begin{center}
\includegraphics[width=0.8\textwidth]{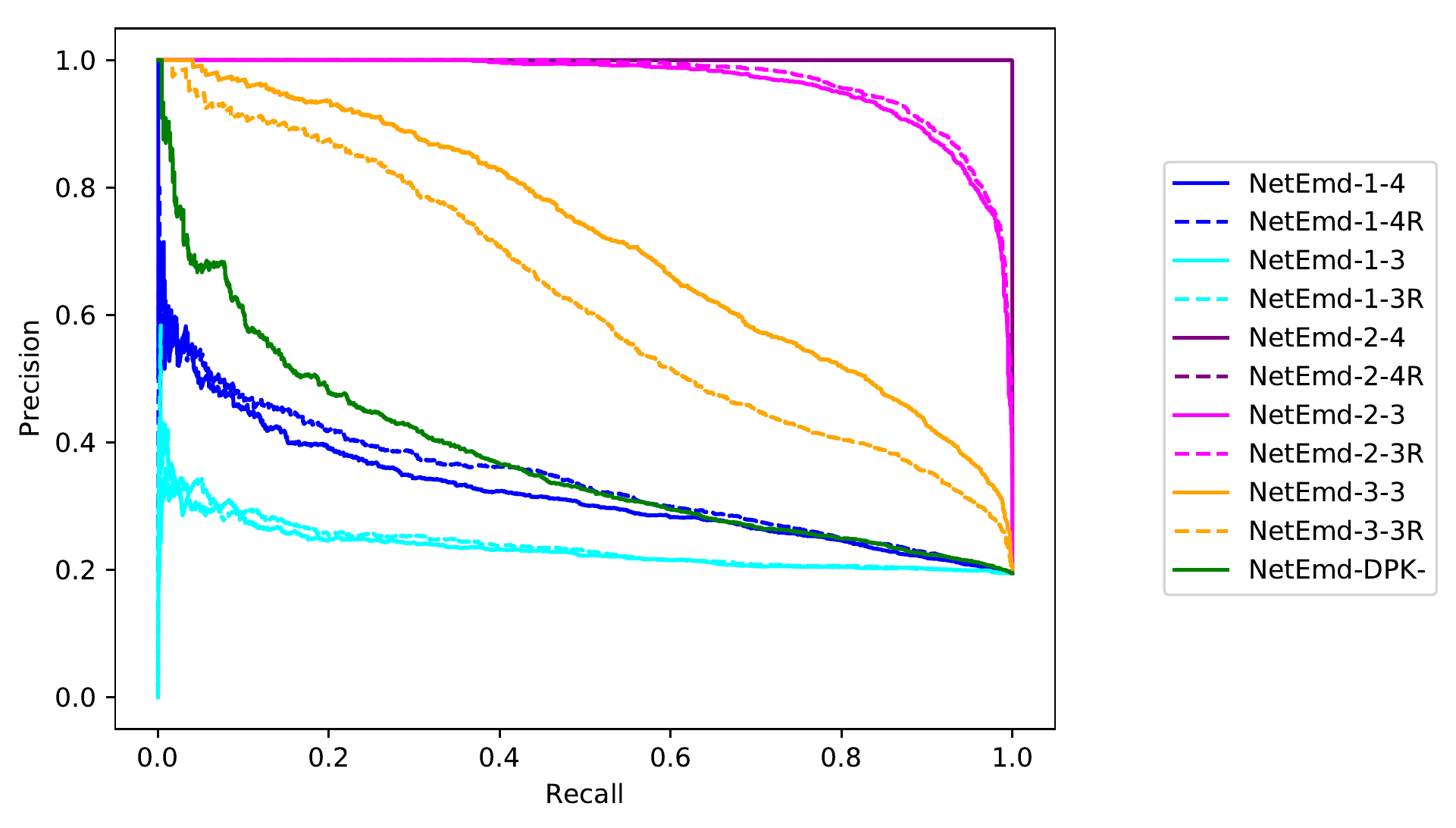}
\end{center}
\caption{NetEmd 4-node-2-layer graphlet insertion precision-recall}
\label{fig:netemd-graphlet42-prerec}
\end{figure}

\begin{figure}[H]
\begin{center}
\includegraphics[width=0.8\textwidth]{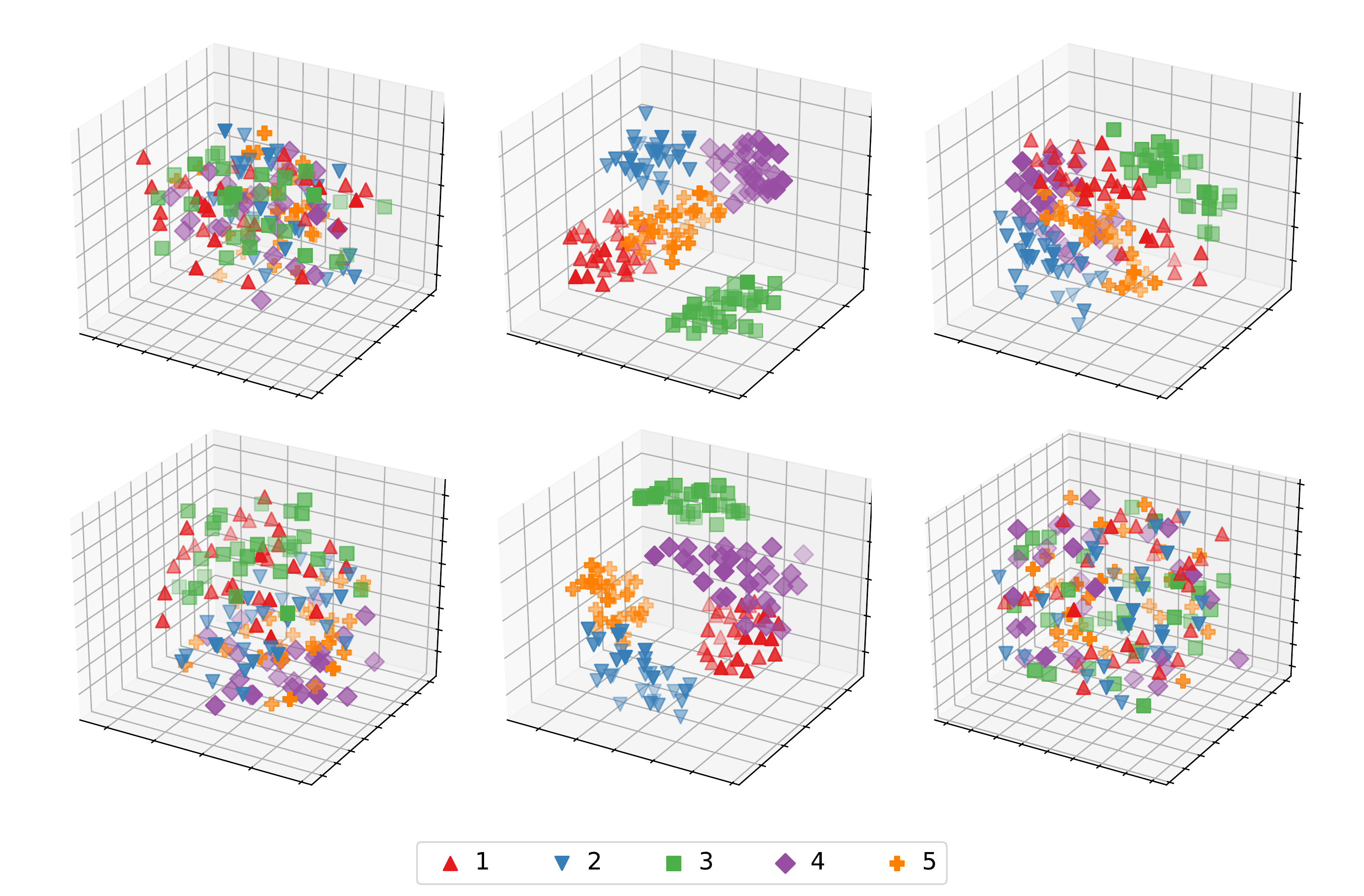}
\end{center}
\caption{NetEmd 4-node-2-layer graphlet insertion MDS}
\label{fig:netemd-graphlet42-mds}
\end{figure}

\begin{figure}[H]
\begin{center}
\includegraphics[width=0.8\textwidth]{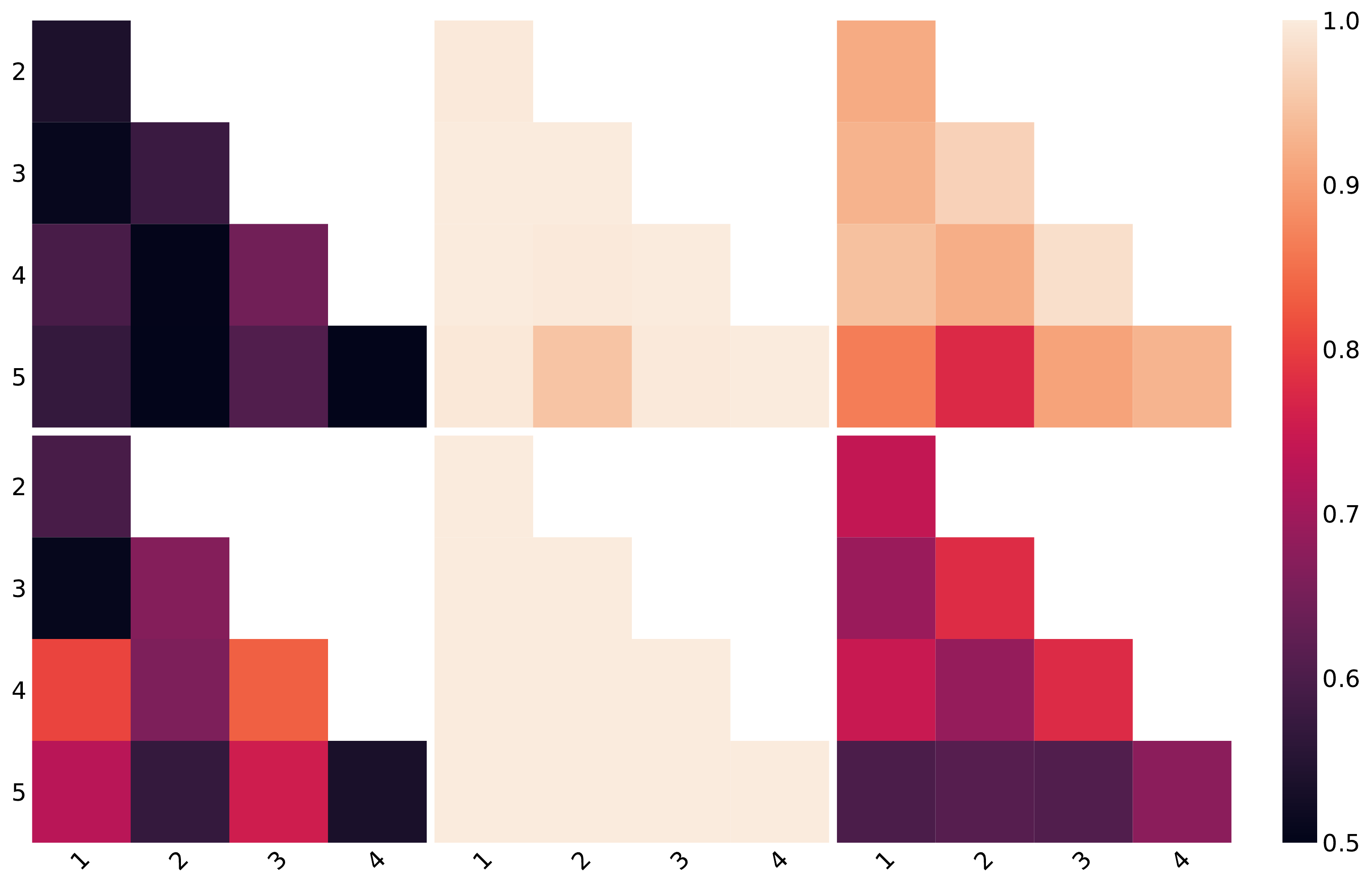}
\end{center}
\caption{NetEmd 4-node-2-layer graphlet insertion pairwise AUPRs}
\label{fig:netemd-graphlet42-auprs}
\end{figure}

\begin{figure}[H]
\begin{center}
\includegraphics[width=0.8\textwidth]{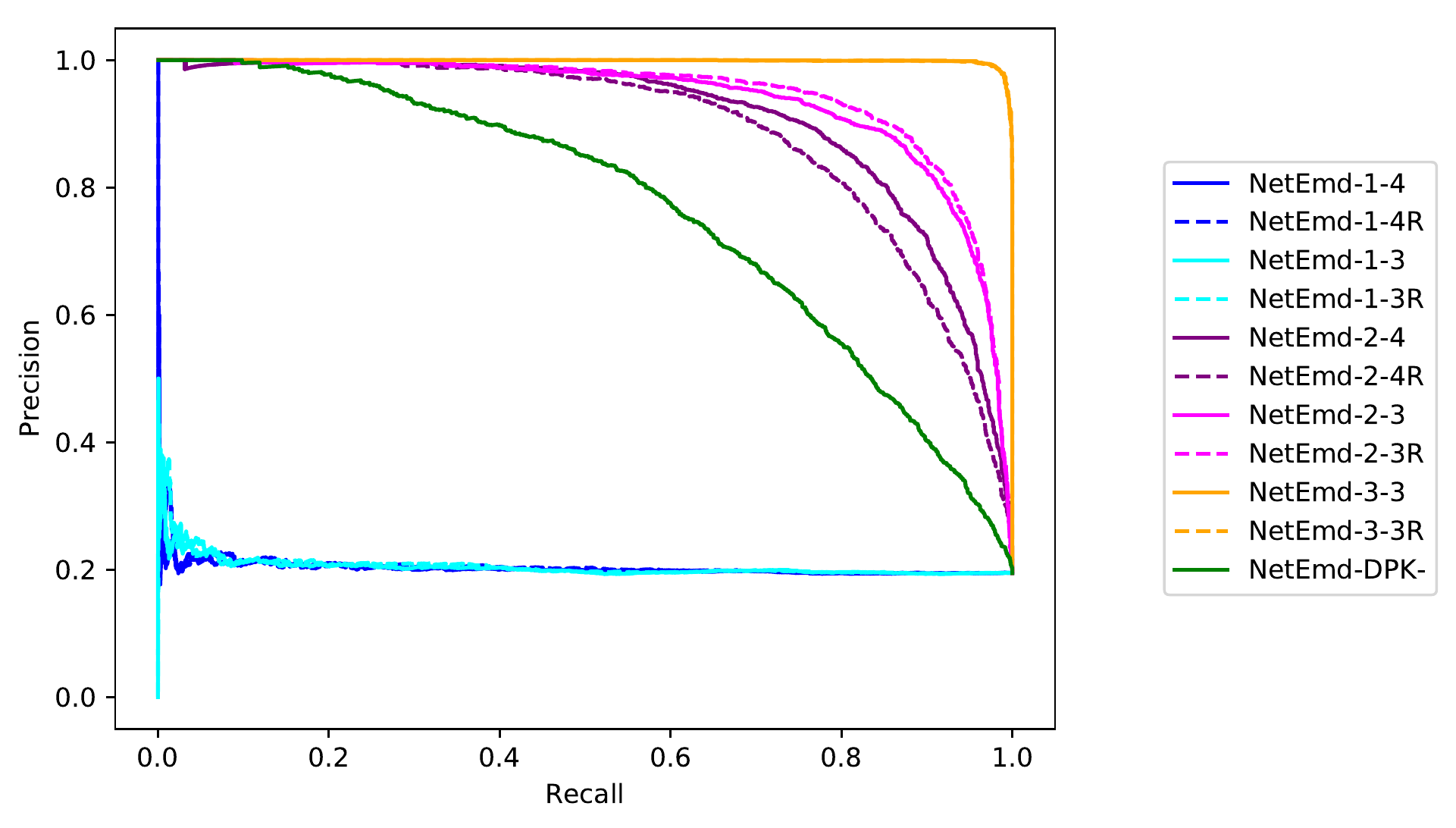}
\end{center}
\caption{NetEmd 3-node-3-layer graphlet insertion precision-recall}
\label{fig:netemd-graphlet33-prerec}
\end{figure}

\begin{figure}[H]
\begin{center}
\includegraphics[width=0.8\textwidth]{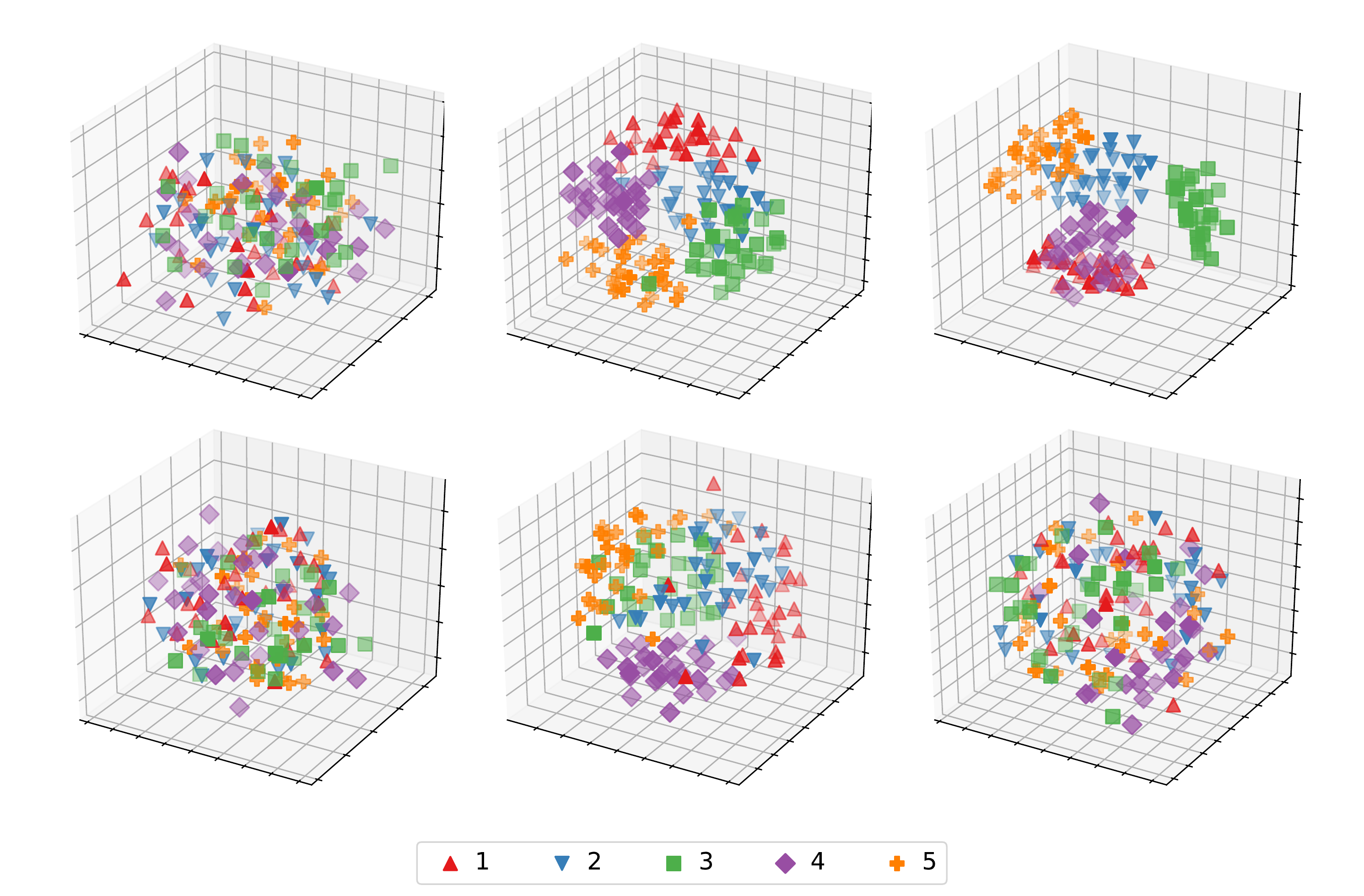}
\end{center}
\caption{NetEmd 3-node-3-layer graphlet insertion MDS}
\label{fig:netemd-graphlet33-mds}
\end{figure}

\begin{figure}[H]
\begin{center}
\includegraphics[width=0.8\textwidth]{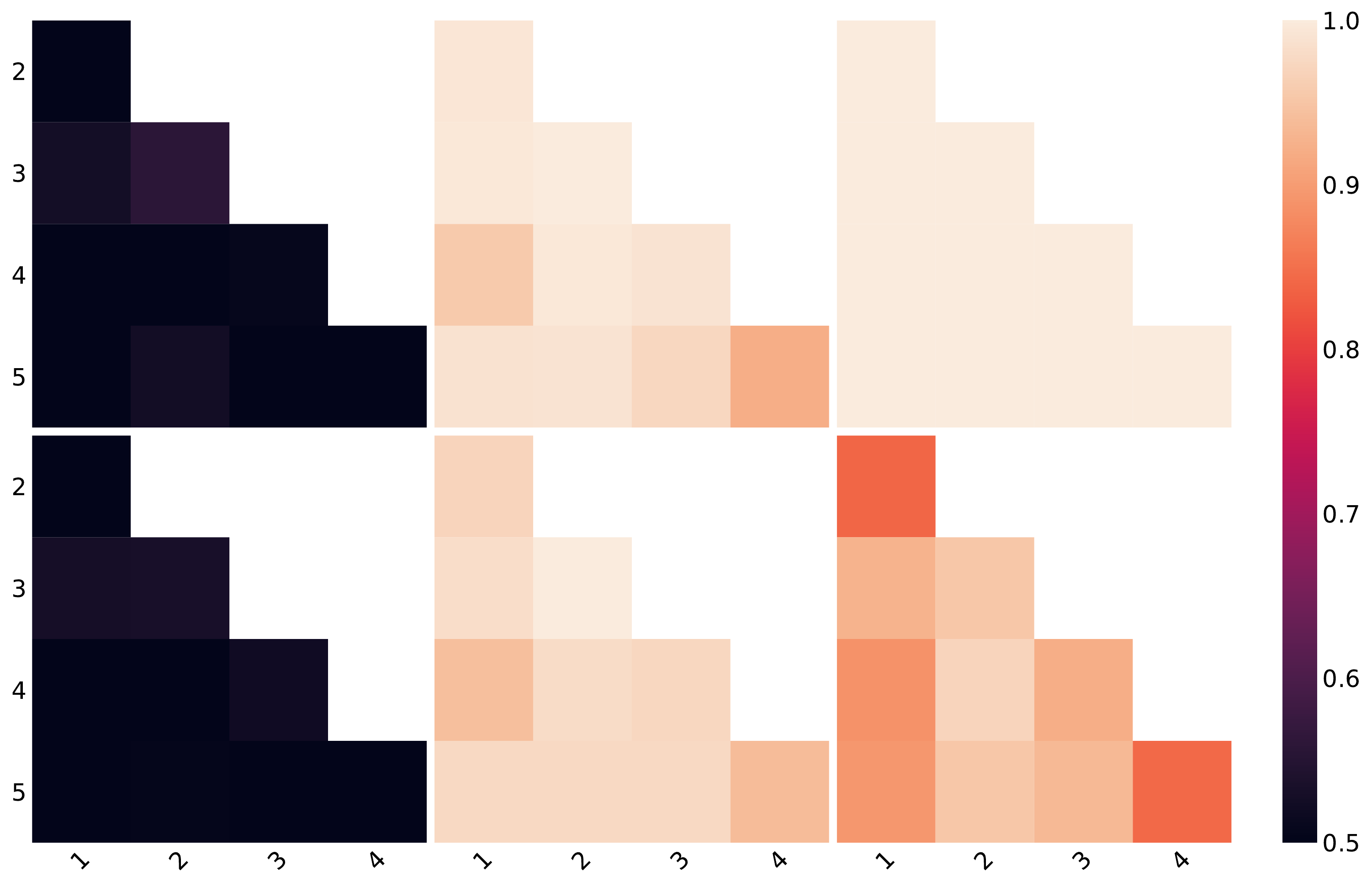}
\end{center}
\caption{NetEmd 3-node-3-layer graphlet insertion pairwise AUPRs}
\label{fig:netemd-graphlet33-auprs}
\end{figure}

The results using NetEmd (Figures \ref{fig:netemd-fixed-prerec}--\ref{fig:netemd-graphlet33-auprs}) are in line with the results using graphlet correlation distance (GCD). Using multilayer graphlet degree distributions yields better precision-recall curves than using single-layer graphlet degree distributions of aggregated networks.

In Figures \ref{fig:netemd-fixed-mds}, \ref{fig:netemd-fixed-auprs}, \ref{fig:netemd-degprog-mds}, \ref{fig:netemd-degprog-auprs}, \ref{fig:netemd-graphlet42-mds}, \ref{fig:netemd-graphlet42-auprs}, \ref{fig:netemd-graphlet33-mds}, and \ref{fig:netemd-graphlet33-auprs} we use the following convention for subfigures: \textbf{top left:} NetEmd-1-3; \textbf{top middle:} NetEmd-2-3; \textbf{top right:} NetEmd-3-3; \textbf{bottom left:} NetEmd-1-4; \textbf{bottom middle:} NetEmd-2-4; \textbf{bottom right:} NetEmd-DPK.

\subsection{Graphlet degree distribution agreement}
Graphlet degree distribution agreement \cite{prvzulj2007biological} (GDDA) is a distance measure which is based on calculating distances between graphlet degree distributions. It is defined as follows:

Let $d^t_G(k)$ be the $t$th graphlet degree distribution of network $G$, \textit{i.e.} it is the distribution of the number of nodes in $G$ on the $t$th automorphism orbit for a total of $k$ times. Then,
\begin{align*}
S^t_G(k) &= \frac{d^t_G(k)}{k} \\
T^t_G &= \sum^\infty_{k=1}S^t_G(k) \\
N^t_G(k) &= \frac{S^t_G(k)}{T^t_G}
\end{align*}
and the distance between the $t$th orbits/graphlet degree distributions (GDDs) of networks $G$ and $G'$ is defined as:
\[
D^t(G,G') = \frac{1}{\sqrt{2}}\left(\sum^\infty_{k=1}\left[N^t_G(k)-N^t_{G'}(k)\right]^2\right)^{\frac{1}{2}}
\]
and the $t$th GDD agreement is defined as:
\[
A^t(G,G') = 1 - D^t(G,G')
\]
and the agreement between networks is then defined as an average of the graphlet degree distribution agreements. Here, the arithmetic mean of the agreements is used, so that the agreement (GDDA) between $G$ and $G'$ is:
\[
A(G,G') = \frac{1}{m}\sum^{m}_{t=1}A^t(G,G')
\]
if we index the graphlet degree distributions from $0$ to $m-1$ instead of $1$ to $m$, this becomes:
\[
A(G,G') = \frac{1}{m}\sum^{m-1}_{t=0}A^t(G,G')
\]

Since $A(G,G')$ is an agreement value, it is larger for networks that are more similar than for networks that are more dissimilar (and $A(G,G) = 1$). Because the other measures, GCD and NetEmd, are distances which are smaller for networks that are more similar than for networks that are more dissimilar (and $GCD(G,G) = NetEmd(G,G) = 0$), we define the GDDA distance $Adist(G,G')$ and use it in our analysis for consistency:
\[
Adist(G,G') = 1 - A(G,G')
\]
This is equal to taking the average of the distances $D^t(G,G')$:
\begin{align*}
Adist(G,G') &= 1 - A(G,G') \\
&= 1 - \frac{1}{m}\sum^{m}_{t=1}A^t(G,G') \\
&= 1 - \frac{1}{m}\sum^{m}_{t=1}\left(1 - D^t(G,G')\right) \\
&= 1 - \frac{1}{m}m + \frac{1}{m}\sum^m_{t=1}D^t(G,G') \\
&= \frac{1}{m}\sum^m_{t=1}D^t(G,G')
\end{align*}

\begin{figure}[H]
\begin{center}
\includegraphics[width=0.8\textwidth]{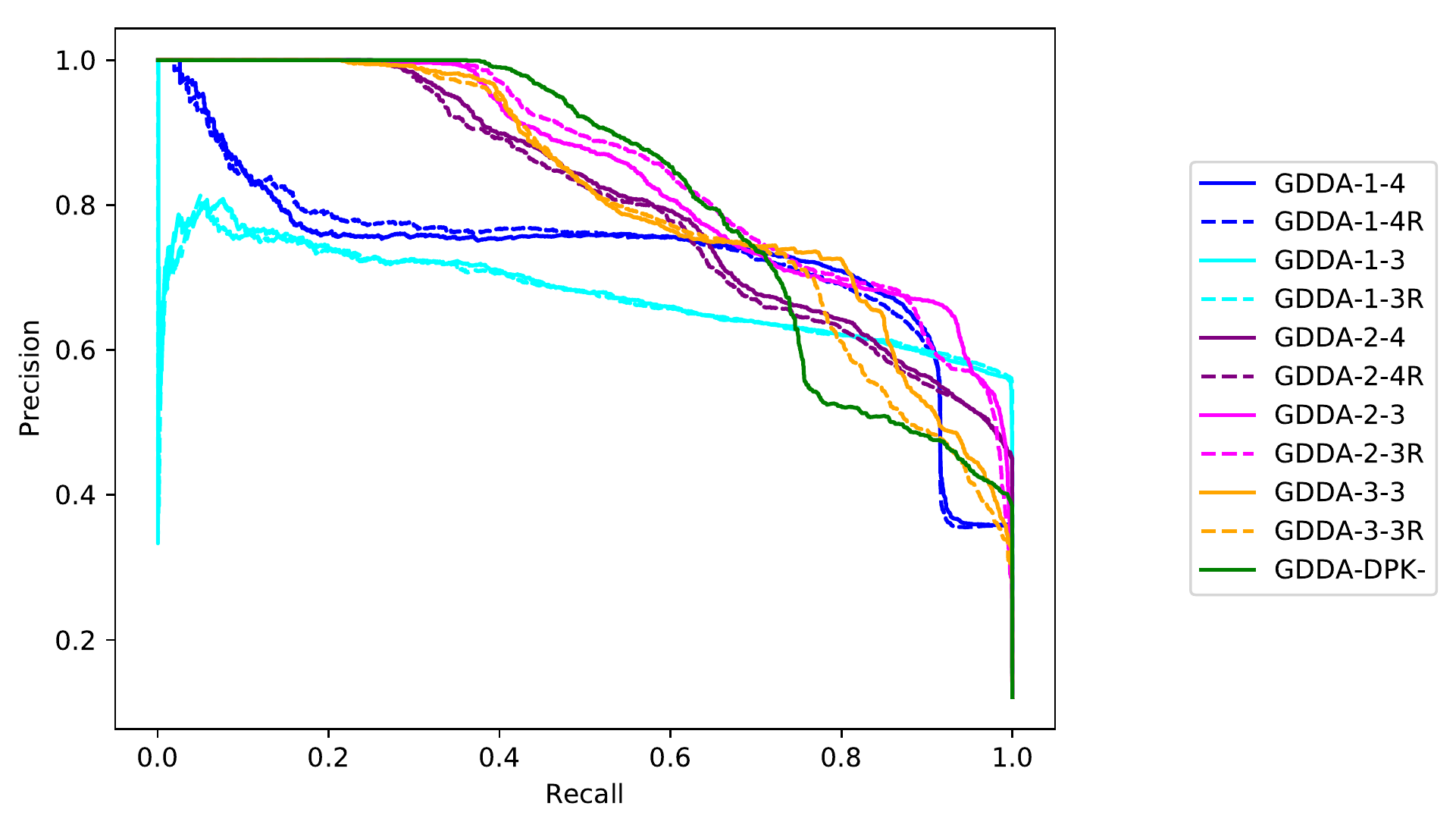}
\end{center}
\caption{GDDA Constant degree precision-recall}
\label{fig:gdda-fixed-prerec}
\end{figure}

\begin{figure}[H]
\begin{center}
\includegraphics[width=0.8\textwidth]{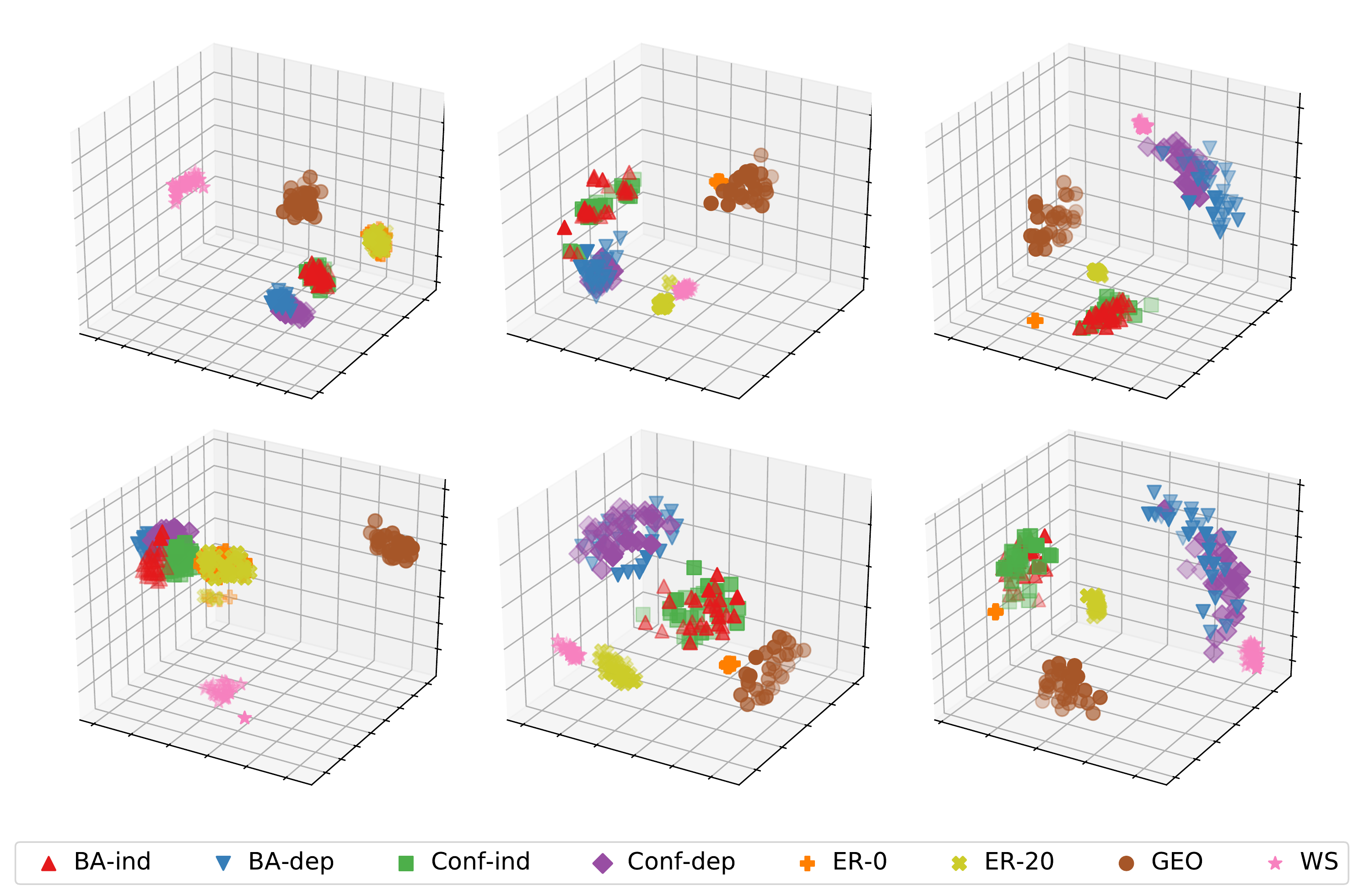}
\end{center}
\caption{GDDA Constant degree MDS}
\label{fig:gdda-fixed-mds}
\end{figure}

\begin{figure}[H]
\begin{center}
\includegraphics[width=0.8\textwidth]{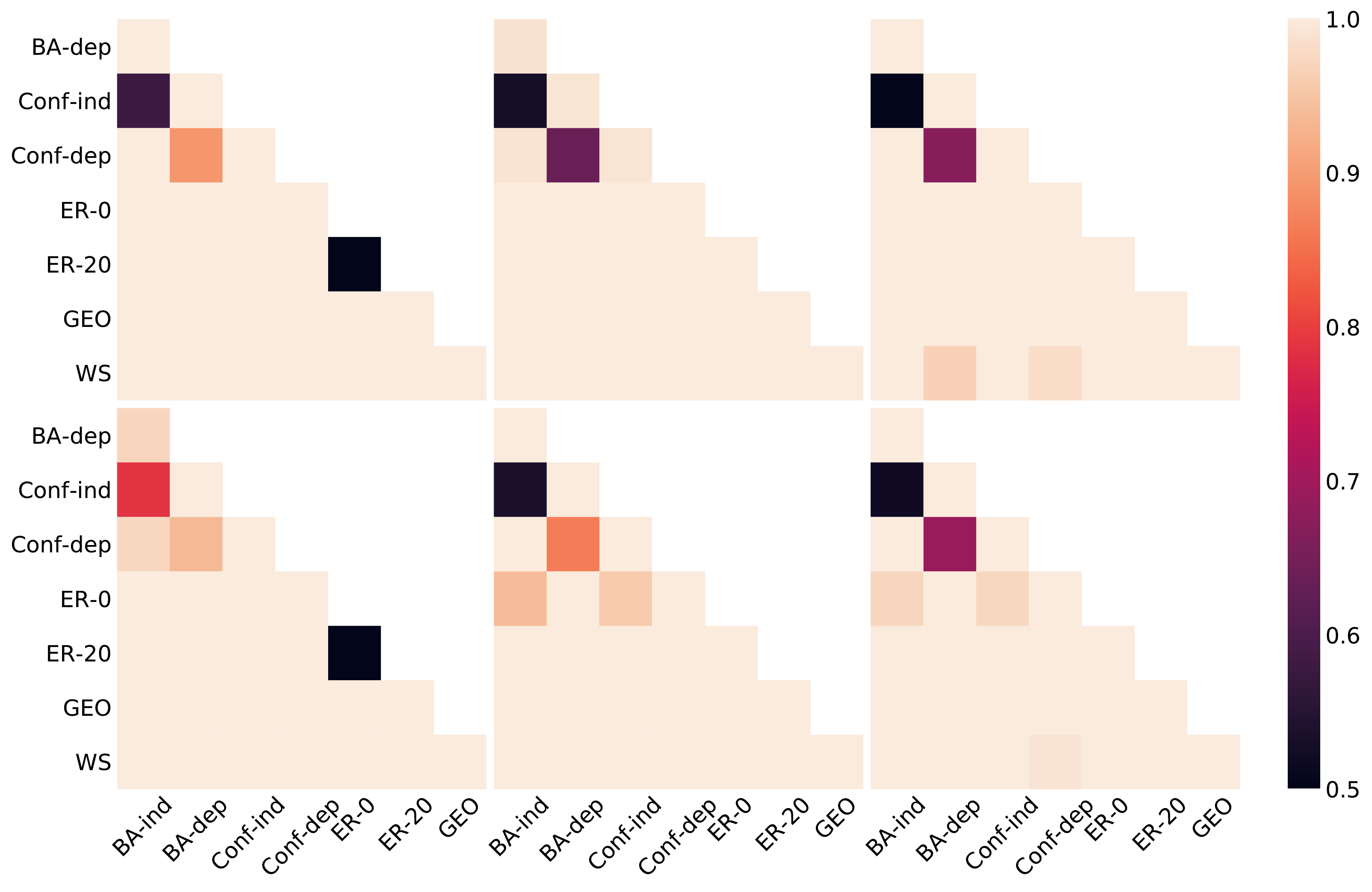}
\end{center}
\caption{GDDA Constant degree pairwise AUPRs}
\label{fig:gdda-fixed-auprs}
\end{figure}

\begin{figure}[H]
\begin{center}
\includegraphics[width=0.8\textwidth]{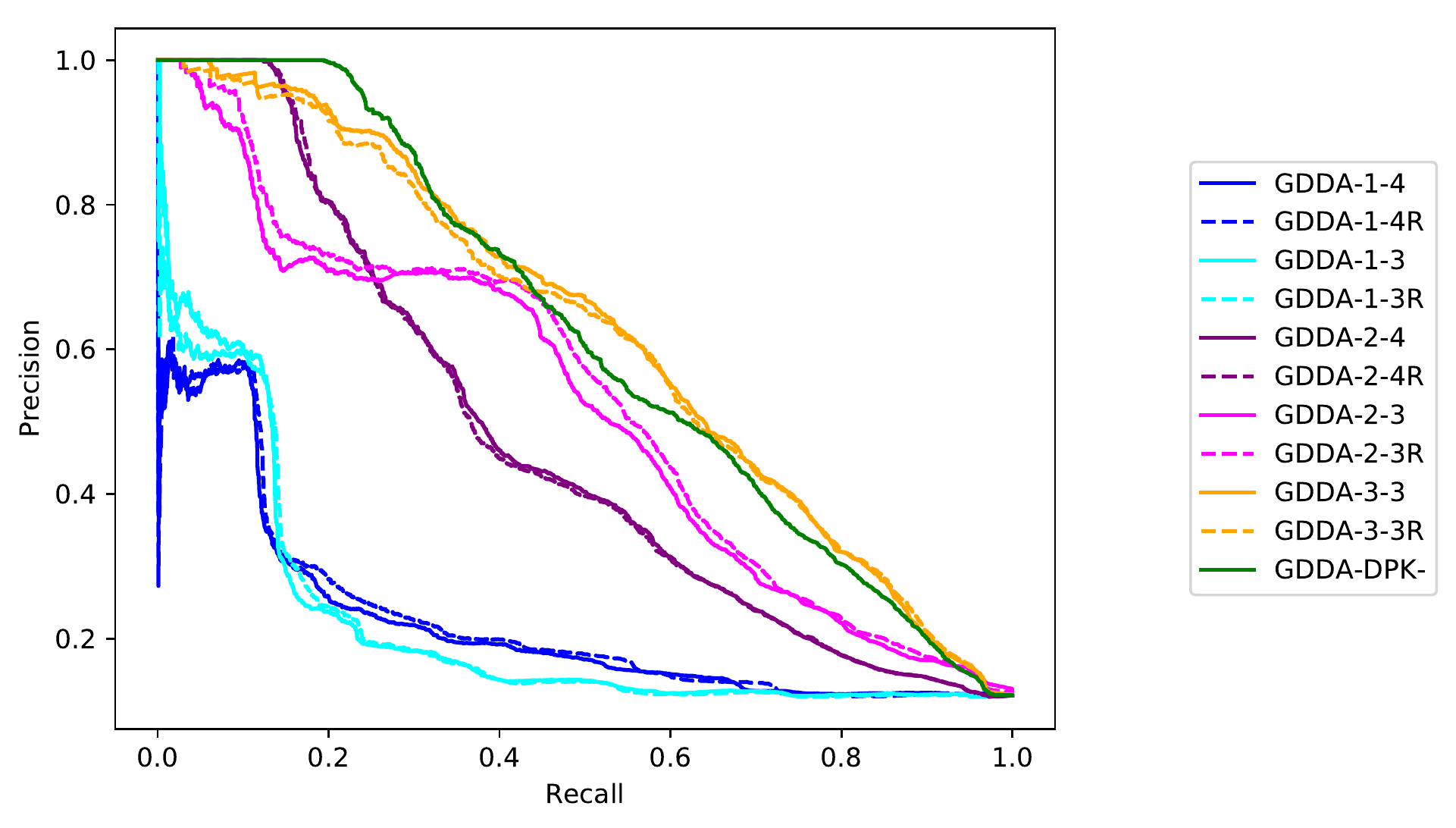}
\end{center}
\caption{GDDA Degree progression precision-recall}
\label{fig:gdda-degprog-prerec}
\end{figure}

\begin{figure}[H]
\begin{center}
\includegraphics[width=0.8\textwidth]{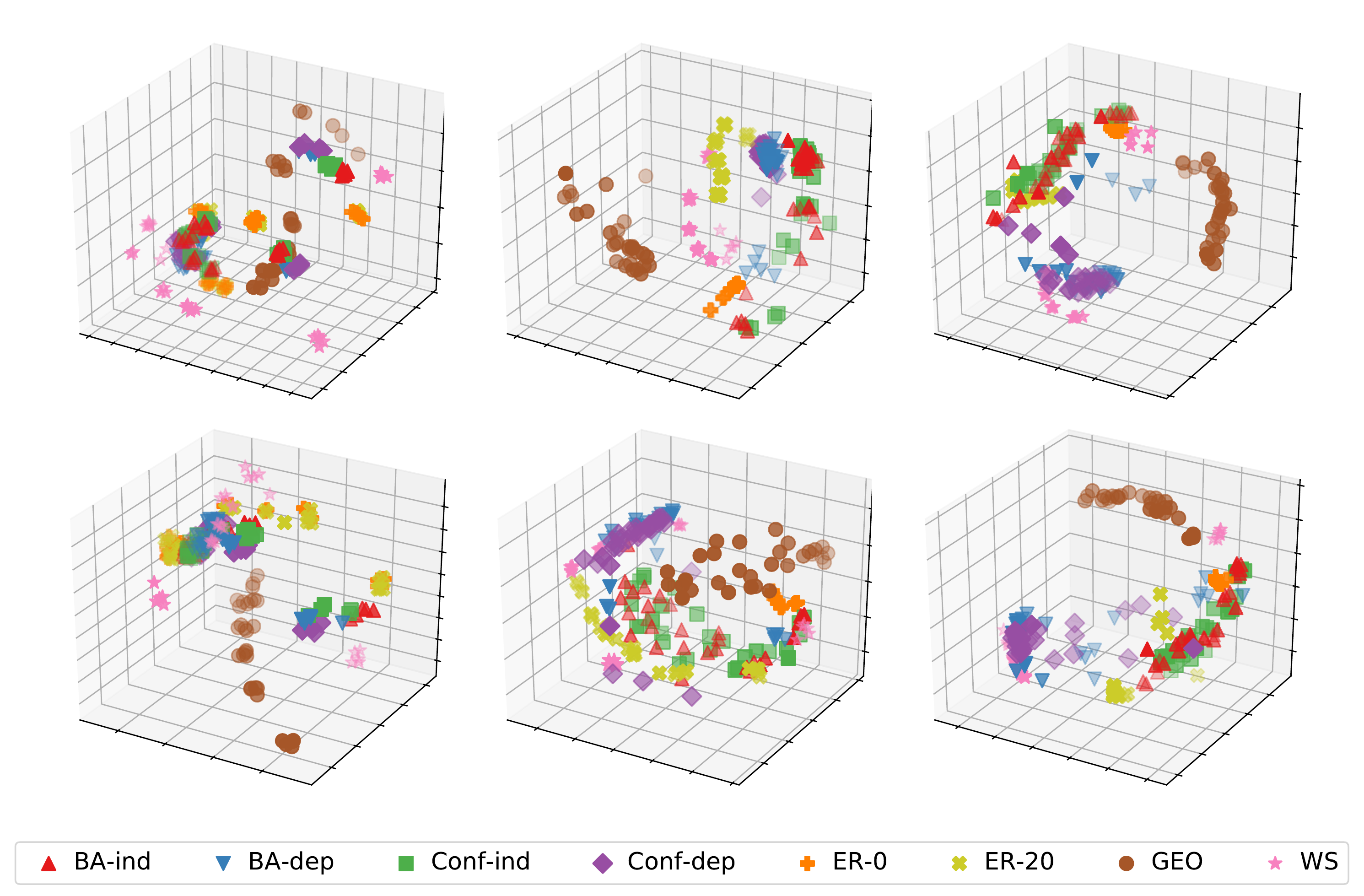}
\end{center}
\caption{GDDA Degree progression MDS}
\label{fig:gdda-degprog-mds}
\end{figure}

\begin{figure}[H]
\begin{center}
\includegraphics[width=0.8\textwidth]{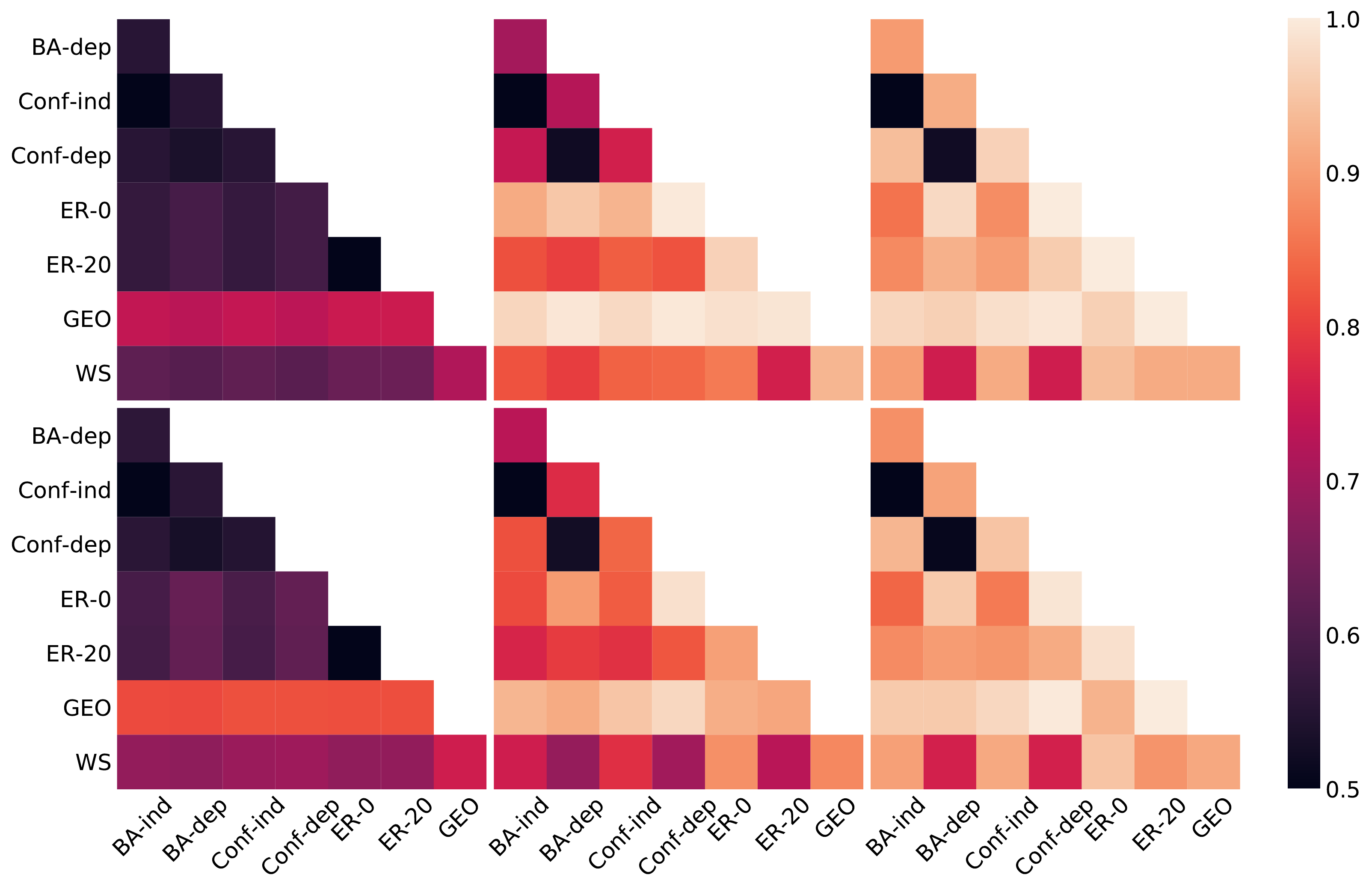}
\end{center}
\caption{GDDA Degree progression pairwise AUPRs}
\label{fig:gdda-degprog-auprs}
\end{figure}

\begin{figure}[H]
\begin{center}
\includegraphics[width=0.8\textwidth]{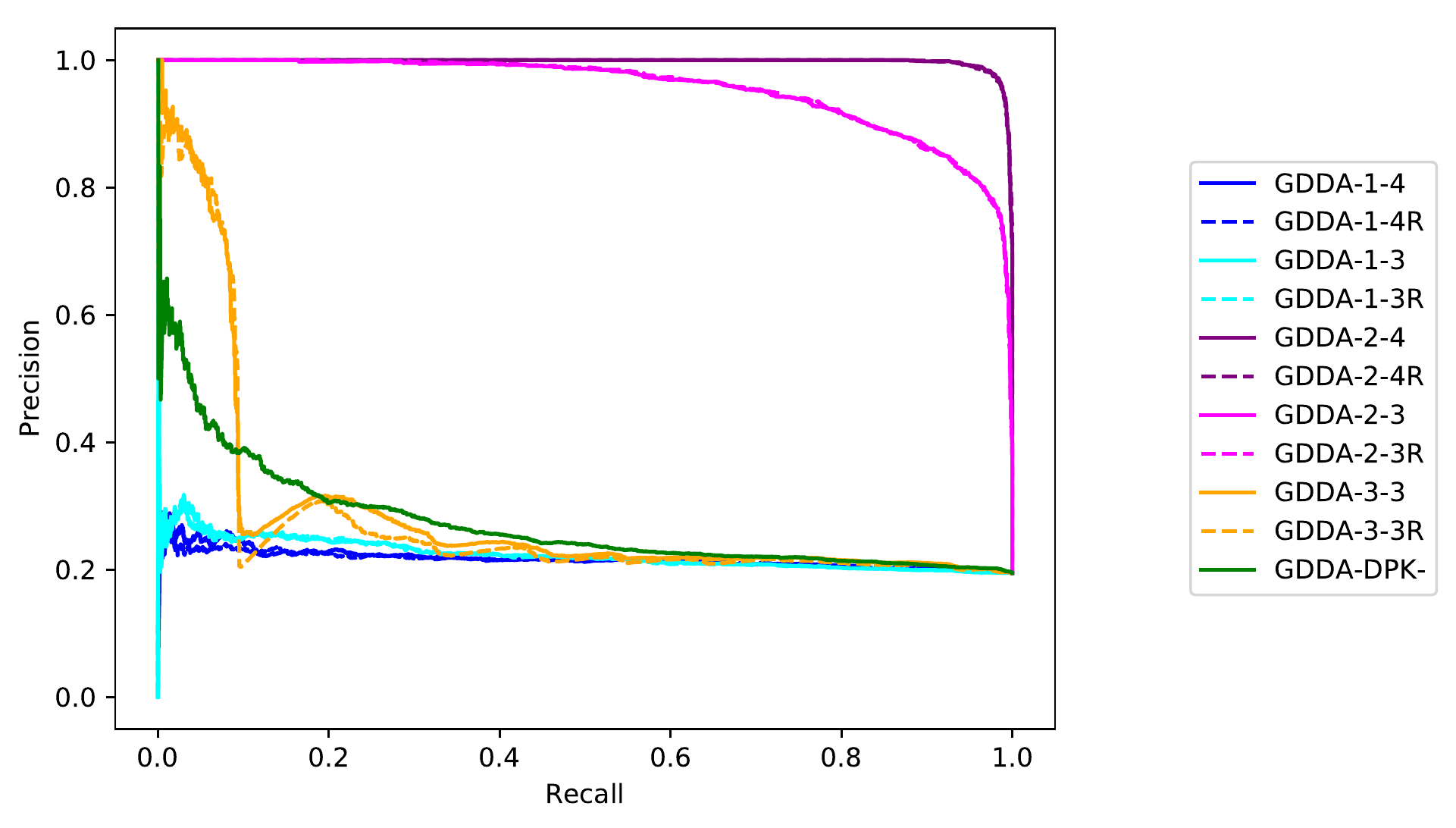}
\end{center}
\caption{GDDA 4-node-2-layer graphlet insertion precision-recall}
\label{fig:gdda-graphlet42-prerec}
\end{figure}

\begin{figure}[H]
\begin{center}
\includegraphics[width=0.8\textwidth]{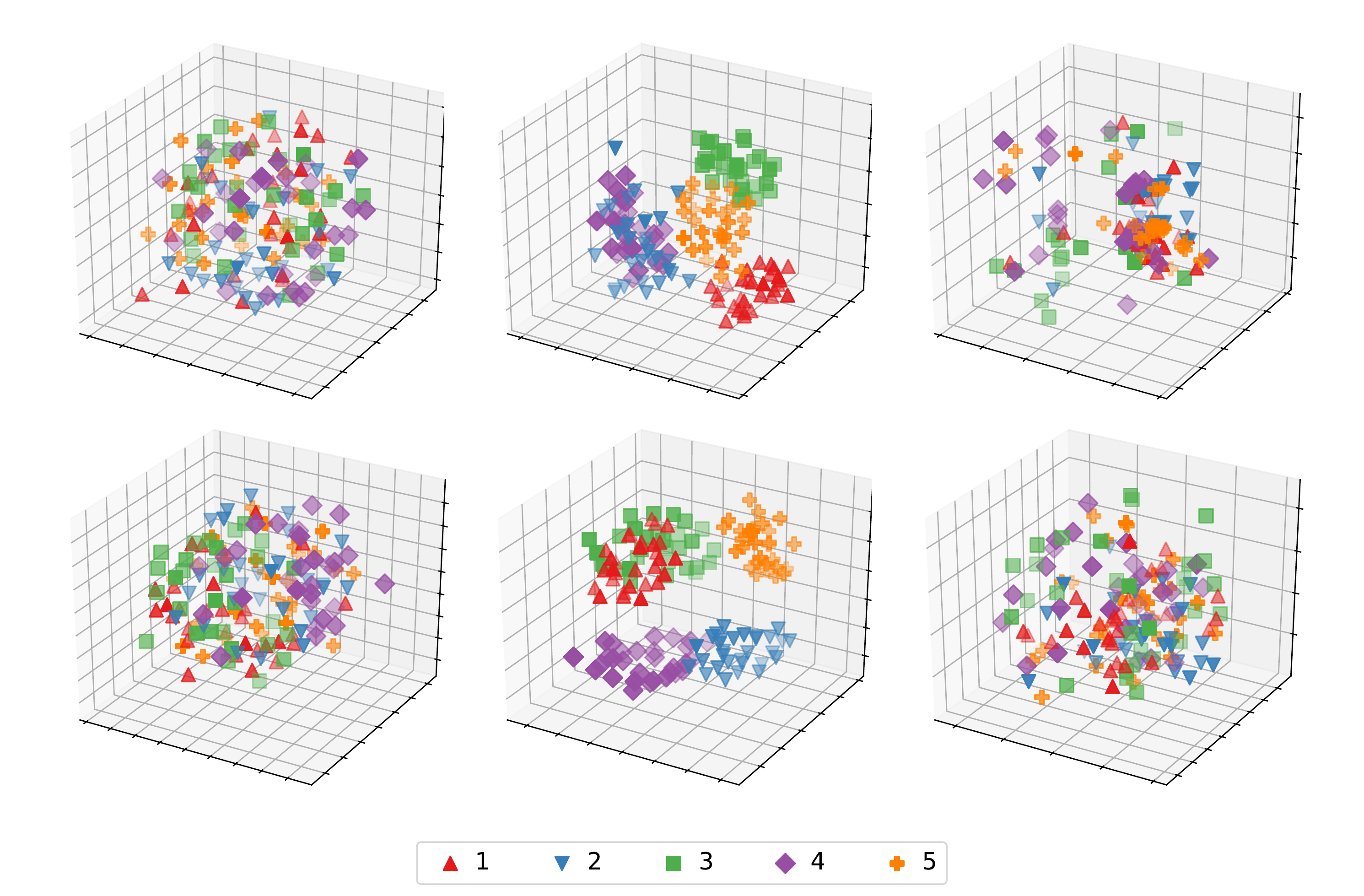}
\end{center}
\caption{GDDA 4-node-2-layer graphlet insertion MDS}
\label{fig:gdda-graphlet42-mds}
\end{figure}

\begin{figure}[H]
\begin{center}
\includegraphics[width=0.8\textwidth]{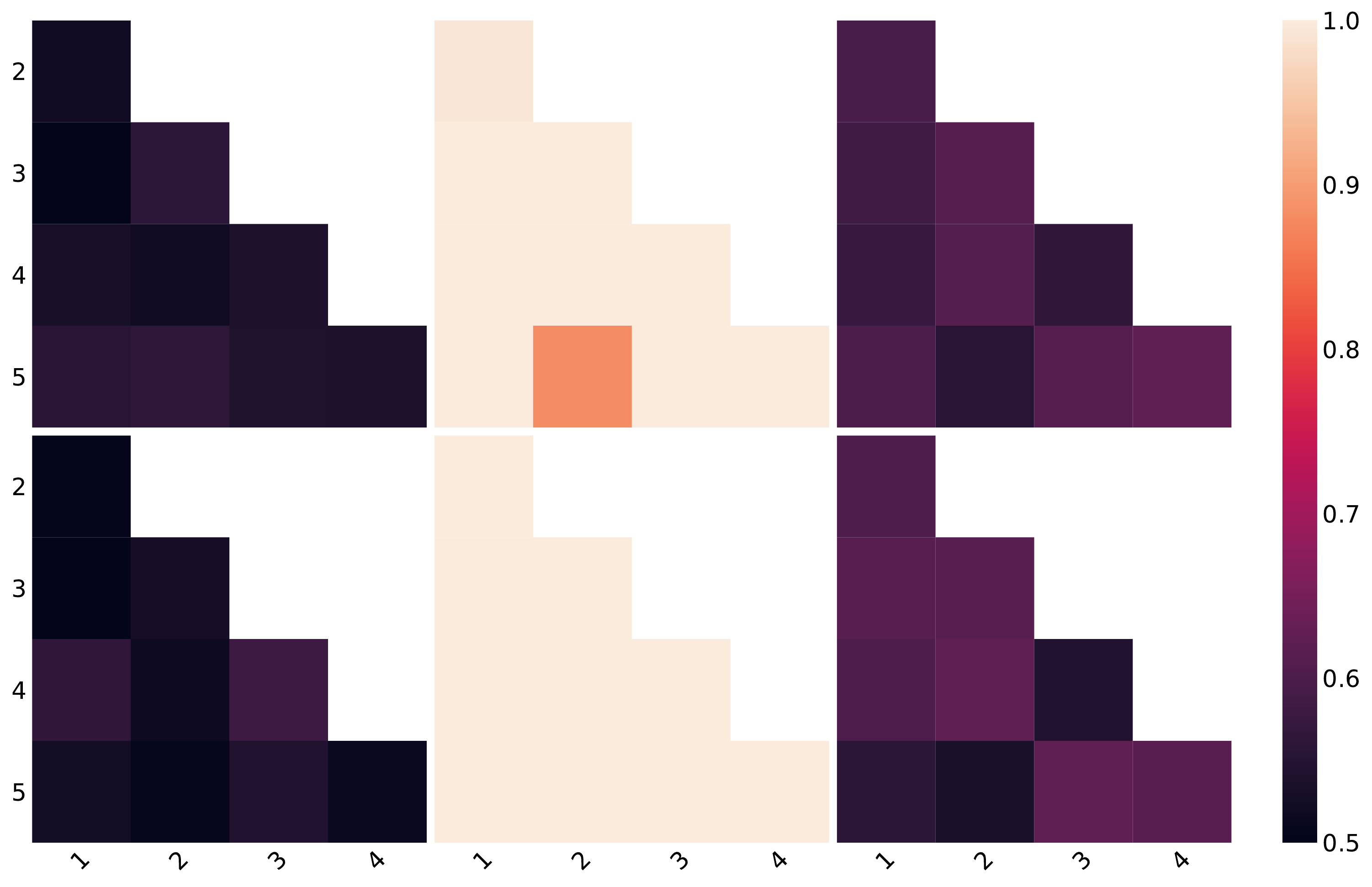}
\end{center}
\caption{GDDA 4-node-2-layer graphlet insertion pairwise AUPRs}
\label{fig:gdda-graphlet42-auprs}
\end{figure}

\begin{figure}[H]
\begin{center}
\includegraphics[width=0.8\textwidth]{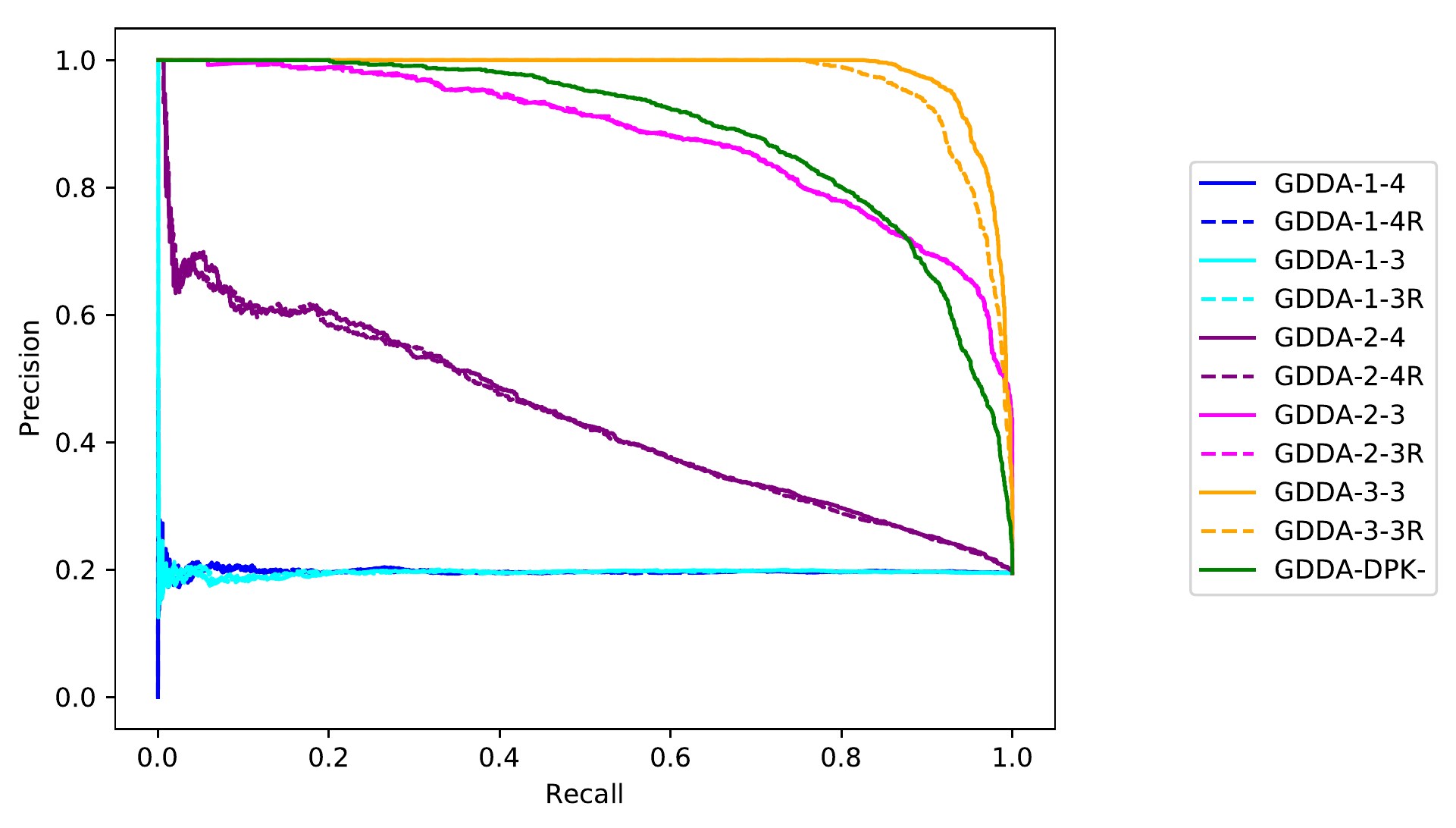}
\end{center}
\caption{GDDA 3-node-3-layer graphlet insertion precision-recall}
\label{fig:gdda-graphlet33-prerec}
\end{figure}

\begin{figure}[H]
\begin{center}
\includegraphics[width=0.8\textwidth]{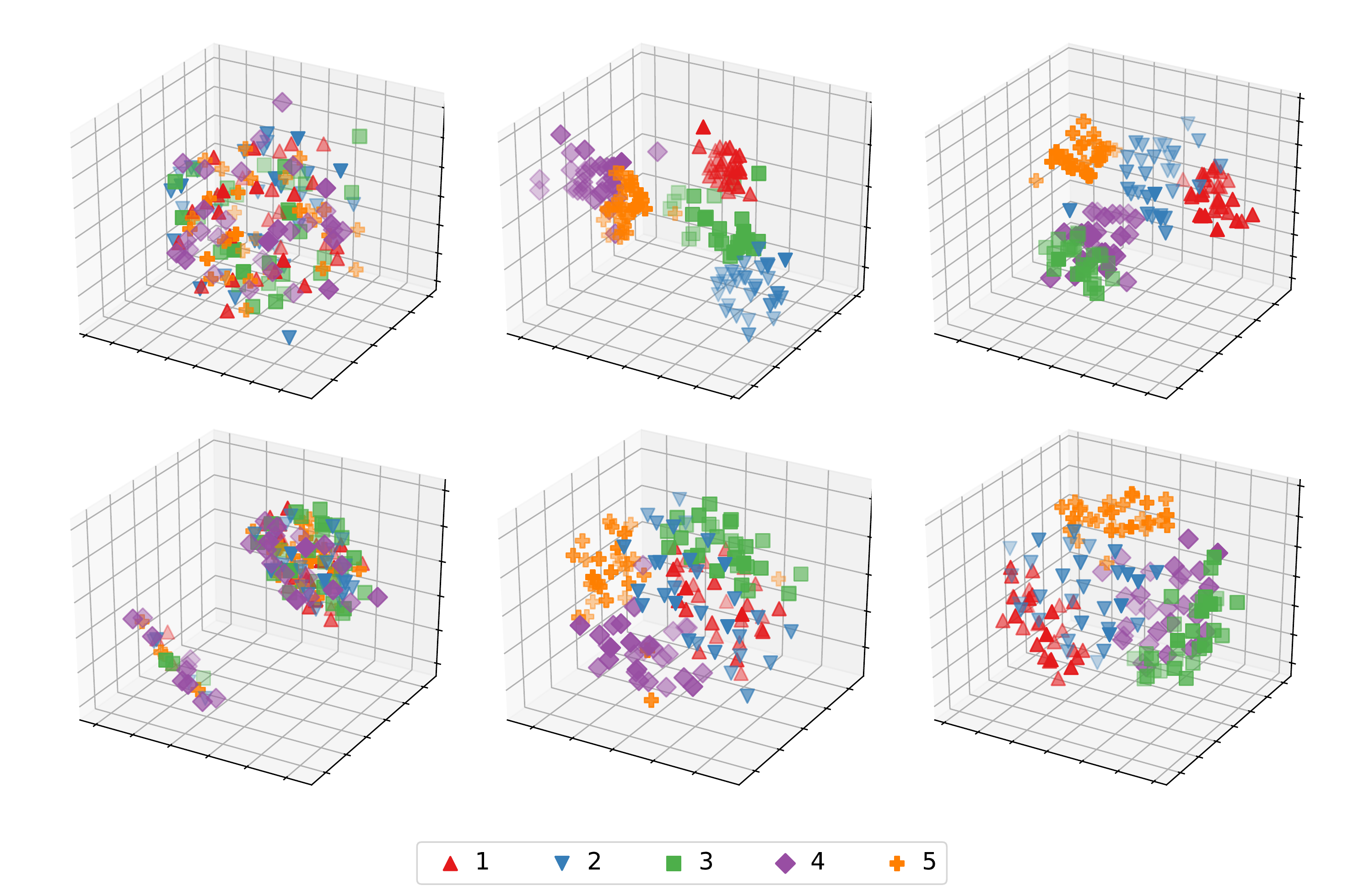}
\end{center}
\caption{GDDA 3-node-3-layer graphlet insertion MDS}
\label{fig:gdda-graphlet33-mds}
\end{figure}

\begin{figure}[H]
\begin{center}
\includegraphics[width=0.8\textwidth]{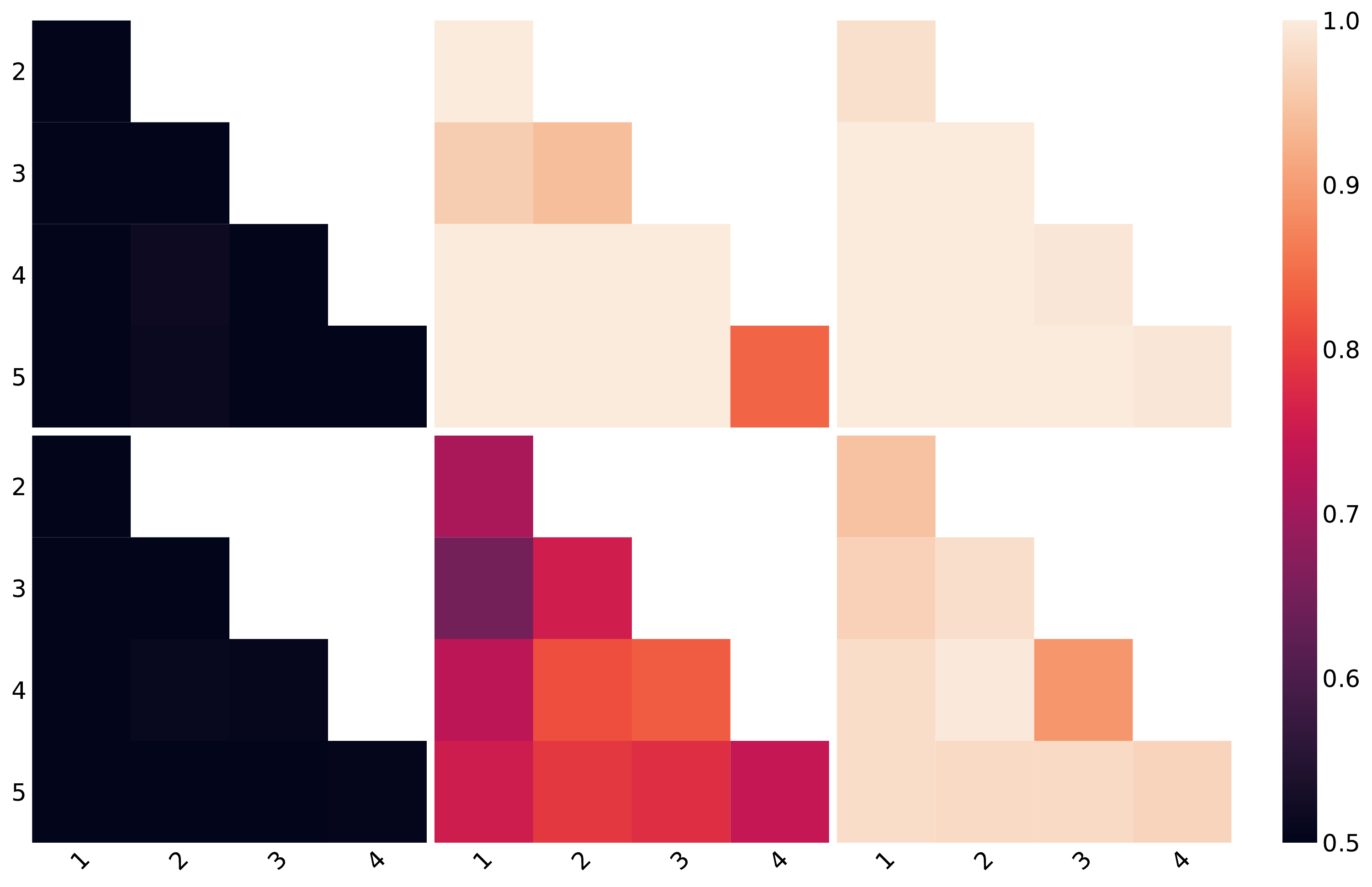}
\end{center}
\caption{GDDA 3-node-3-layer graphlet insertion pairwise AUPRs}
\label{fig:gdda-graphlet33-auprs}
\end{figure}

The results using GDDA (Figures \ref{fig:gdda-fixed-prerec}--\ref{fig:gdda-graphlet33-auprs}) are in line with the results using graphlet correlation distance (GCD). Using multilayer graphlet degree distributions yields better precision-recall curves than using single-layer graphlet degree distributions of aggregated networks.

In Figures \ref{fig:gdda-fixed-mds}, \ref{fig:gdda-fixed-auprs}, \ref{fig:gdda-degprog-mds}, \ref{fig:gdda-degprog-auprs}, \ref{fig:gdda-graphlet42-mds}, \ref{fig:gdda-graphlet42-auprs}, \ref{fig:gdda-graphlet33-mds}, and \ref{fig:gdda-graphlet33-auprs} we use the following convention for subfigures: \textbf{top left:} GDDA-1-3; \textbf{top middle:} GDDA-2-3; \textbf{top right:} GDDA-3-3; \textbf{bottom left:} GDDA-1-4; \textbf{bottom middle:} GDDA-2-4; \textbf{bottom right:} GDDA-DPK.

\clearpage

\section{Computational considerations}

A noteworthy remark is that the running times required to compute the orbit counts increases as the number of nodes in the graphlets is increased.
Another matter to consider is the number of layers to be included in the graphlets.
As the number of layers in the studied networks increases, the number of layer combinations for which one needs to compute the orbit counts increases.
If the number of layers in the graphlets is smaller than in the analyzed network, the summed graphlet degree vectors from different layer combinations can become difficult interpret especially if the layers depict very diverse relationships between the nodes.
However, the advantage of summing the vectors from different layer combinations is that one is able to compare networks with different numbers of layers. Clearly, this summation only makes sense in a context where one uses isomorphism in which layer labels are allowed to be permuted.

\end{document}